\documentclass{article}

\PassOptionsToPackage{numbers,sort&compress}{natbib}
\usepackage[letterpaper,textheight=9in,textwidth=5.5in,top=1in,headheight=12pt,headsep=25pt,footskip=30pt]{geometry}

\usepackage{hyperref}
\hypersetup{
  colorlinks=true,
  urlcolor=sysblue,
  linkcolor=sysblue
}

\usepackage{cite}
\usepackage{amsmath,amssymb,amsfonts}
\usepackage{graphicx}
\usepackage{textcomp}
\usepackage{xcolor}
\def\BibTeX{{\rm B\kern-.05em{\sc i\kern-.025em b}\kern-.08em
    T\kern-.1667em\lower.7ex\hbox{E}\kern-.125emX}}

\usepackage{booktabs,longtable,array,ragged2e,multirow}
\usepackage{fancyhdr}
\usepackage{tikz}
\usepackage[all]{nowidow}
\usepackage[linesnumbered,lined,vlined,ruled,commentsnumbered,noend]{algorithm2e}
\colorlet{RED}{red}
\usepackage{listings}
\usepackage{url}

\usepackage{tcolorbox}
\usepackage{fancyvrb} 
\usepackage{xspace}
\usepackage{multirow}
\usepackage{subfloat}
\usepackage{balance}
\usepackage{ctable}
\usepackage{listings}
\usepackage{color}
\usepackage[nobiblatex]{xurl}

\usepackage{multirow}

\usepackage{algpseudocode}
\usepackage{caption}
\usepackage{hyperref}

\usepackage{acronym}
\usepackage{fancybox}
\usepackage{verbatim}
\usepackage[normalem]{ulem}
\usepackage{float}
\usepackage{newfloat}
\usepackage{mdframed}

\usepackage{enumitem}
\usepackage{booktabs}
\usepackage{pifont}

\usepackage{subfigure}
\usepackage{ulem}
\usepackage[flushleft]{threeparttable}
\usepackage{booktabs}
\usepackage{tabularray}

\acrodef{wp}[WP]{Website Fingerprinting}
\acrodef{apt}[APT]{Advanced Persistent Threats}
\acrodef{lol}[LotL]{Living-Off-The-Land}
\acrodef{ids}[IDS]{Intrusion Detection System}
\acrodef{vae}[VAE]{Variational AutoEncoder}
\acrodef{re}[RE]{reconstruction error}
\acrodef{sv}[SV]{Stableness Value}
\acrodef{as}[AS]{anomaly score}
\acrodef{has}[HAS]{hopset anomaly score}
\acrodef{etw}[ETW]{Event Tracing for Windows}
\acrodef{e3}[E3]{Engagement 3}
\acrodef{e5}[E5]{Engagement 5}
\acrodef{ttp}[TTPs]{Tactics, Techniques, and Procedures}
\acrodef{hsg}[HSG]{High-level Scenario Graph}
\acrodef{nlp}[NLP]{Natural Language Processing}
\acrodef{dg}[DG]{Detection Graph}
\acrodef{poi}[POI]{Point of Interest}
\acrodef{iv}[IV]{Important Value}
\acrodef{ioc}[IOC]{Indicators of Compromise}
\acrodef{ag}[AG]{Attack Graph}
\acrodef{cs}[CS]{Cobalt Strike}
\acrodef{mttd}[MTTD]{Mean Time to Detect}
\acrodef{soc}[SOC]{Security Operations Center}
\acrodef{loc}[LoC]{Lines of Code}
\acrodef{wmi}[WMI]{Windows Management Instrumentation}
\acrodef{dlls}[DLLs]{Dynamic-Link Libraries}
\acrodef{sota}[SOTA]{state-of-the-art}
\acrodef{edr}[EDR]{Endpoint Detection and Response}
\acrodef{nlp}[NLP]{Natural Language Processing}
\acrodef{ostp}[OSTP]{Online Steiner Tree Problem}
\acrodef{isg}[ISG]{Importance-Score-Guided Search}
\acrodef{edr}[EDR]{Endpoint Detection and Response}
\acrodef{ml}[ML]{Machine Learning}
\acrodef{cti}[CTI]{Cyber Threat Intelligence}
\acrodef{ast}[AST]{Abstract Syntax Tree}
\acrodef{gioc}[gIoC]{General Indicator of Compromise}
\acrodef{cti}[CTI]{Cyber Threat Intelligence}
\acrodef{ttp}[TTPs]{Tactics, Techniques, and Procedures}
\acrodef{llm}[LLM]{Large Language Model}
\acrodef{iot}[IoT]{Internet of Things}

\acrodef{siem}[SIEM]{Security Information and Event Management}

\newcommand{\shaofei}[1]{\textcolor{red}{(Shaofei: #1)}}
\newcommand{\yuhan}[1]{\textcolor{blue}{(yuhan: #1)}}

\newcommand{\sysarmor}{\textsc{SysArmor}\xspace}
\newcommand{\sysspear}{\textsc{SysSpear}\xspace}
\newcommand{\sysfield}{{\textsc{SysField}}\xspace}
\newcommand{\sysevolve}{\textsc{SysEvolve}\xspace}

\newcommand{\company}{Sangfor\xspace}%
\usepackage{pifont}

\newcommand{\nodlink}{\textsc{NodLink}\xspace}
\newcommand{\knowhow}{\textsc{KnowHow}\xspace}

\newcommand{\orax}{\textsc{Orax}\xspace}
\newcommand{\eunomia}{\textsc{Eunomia}\xspace}
\newcommand{\symgx}{\textsc{SymGx}\xspace}
\newcommand{\aes}{\textsc{Aes}\xspace}

\newcommand{\holmes}{\textsc{HOLMES}\xspace}
\newcommand{\unicorn}{\textsc{UNICORN}\xspace}
\newcommand{\provdetector}{\textsc{ProvDetector}\xspace}

\newcommand{\eat}[1]{}

\definecolor{sysnavy}{HTML}{0B2A4A}
\definecolor{sysblue}{HTML}{2C6EAA}
\definecolor{syscyan}{HTML}{4B9CA6}
\definecolor{sysgray}{HTML}{66717F}
\definecolor{syslight}{HTML}{EDF3F7}
\definecolor{syswarm}{HTML}{B66A3C}

\hypersetup{
  pdftitle={SysEvolve: Safe, Autonomous Attack-Defense Co-evolution for Evolving Cyber Defense},
  pdfauthor={Technical Report},
  colorlinks=true,
  citecolor=sysblue,
  linkcolor=sysblue,
  urlcolor=sysblue
}
\newcommand{\sysevolvelink}{\url{https://sysevolve.pku.edu.cn}}

\usepackage{fontspec}
\usepackage{xspace}
\newcolumntype{P}[1]{>{\RaggedRight\arraybackslash}p{#1}}
\newcolumntype{C}[1]{>{\Centering\arraybackslash}p{#1}}
\renewcommand{\arraystretch}{1.08}
\makeatletter
\renewcommand{\normalsize}{\@setfontsize\normalsize{10pt}{11pt}}
\renewcommand{\small}{\@setfontsize\small{9pt}{10pt}}
\renewcommand{\footnotesize}{\@setfontsize\footnotesize{8pt}{9.5pt}}
\renewcommand{\scriptsize}{\@setfontsize\scriptsize{7pt}{8pt}}
\renewcommand{\large}{\@setfontsize\large{12pt}{14pt}}
\renewcommand{\Large}{\@setfontsize\Large{14pt}{16pt}}
\renewcommand{\LARGE}{\@setfontsize\LARGE{17pt}{20pt}}
\normalsize
\renewcommand{\section}{\@startsection{section}{1}{\z@}{-2.0ex \@plus -.5ex \@minus -.2ex}{1.5ex \@plus .3ex \@minus .2ex}{\large\bfseries\raggedright}}
\renewcommand{\subsection}{\@startsection{subsection}{2}{\z@}{-1.8ex \@plus -.5ex \@minus -.2ex}{.8ex \@plus .2ex}{\normalsize\bfseries\raggedright}}
\renewcommand{\subsubsection}{\@startsection{subsubsection}{3}{\z@}{-1.5ex \@plus -.5ex \@minus -.2ex}{.5ex \@plus .2ex}{\normalsize\bfseries\raggedright}}
\renewcommand{\paragraph}{\@startsection{paragraph}{4}{\z@}{1.5ex \@plus .5ex \@minus .2ex}{-1em}{\normalsize\bfseries}}
\makeatother

\fancypagestyle{paper}{
  \fancyhf{}
  \fancyhead[R]{\footnotesize\color{sysgray}SYSEVOLVE-TR-2026-002}
  \fancyfoot[C]{\thepage}

}
\fancypagestyle{firstpage}{
  \fancyhf{}
  \fancyfoot[L]{\footnotesize\color{sysgray}SysEvolve Technical Report No. SYSEVOLVE-TR-2026-002.}

}
\fancypagestyle{landscapepage}{
  \fancyhf{}
  \fancyhead[R]{\footnotesize\color{sysgray}SYSEVOLVE-TR-2026-002}
  \fancyfoot[C]{\thepage}

}
\title{\textsc{SysEvolve}: Safe, Autonomous Attack-Defense Co-evolution for Evolving Cyber Defense}
\eat{\author{
  \begin{tabular}{c}
    \textbf{Yuhan Meng} \\
    \small School of Computer Science, Peking University \\
    \small \href{mailto:mengyuhan@pku.edu.cn}{mengyuhan@pku.edu.cn}
  \end{tabular}
  \and
  \begin{tabular}{c}
    \textbf{Shaofei Li} \\
    \small School of Computer Science, Peking University \\
    \small \href{mailto:lishaofei@pku.edu.cn}{lishaofei@pku.edu.cn}
  \end{tabular}
  \and
  \begin{tabular}{c}
    \textbf{Jionghao Huang} \\
    \small School of Cyber Security, University A \\
    \small Institute of Advanced Computing, University B \\
    \small \href{mailto:charlie.davis@univ-a.edu}{charlie.davis@univ-a.edu}
  \end{tabular}
}}
\author{
  \begin{minipage}[t]{0.24\textwidth}
    \centering
    \textbf{Yuhan Meng} \\
    \small School of Computer Science, Peking University \\
    \small \href{mailto:mengyuhan@pku.edu.cn}{mengyuhan@pku.edu.cn}
  \end{minipage}
  \hfill 
  \begin{minipage}[t]{0.24\textwidth}
    \centering
    \textbf{Jiandong Jin} \\
    \small School of Computer Science, Peking University \\
    \small \href{mailto:jjd@pku.edu.cn}{jjd@pku.edu.cn}
  \end{minipage}
  \hfill 
  \begin{minipage}[t]{0.24\textwidth}
    \centering
    \textbf{Shaofei Li} \\
    \small School of Computer Science, Peking University \\
    \small \href{mailto:lishaofei@pku.edu.cn}{lishaofei@pku.edu.cn}
  \end{minipage}
  \hfill 
  \begin{minipage}[t]{0.24\textwidth}
    \centering
    \textbf{Jionghao Huang} \\
    \small Southeast University \\
    \small \href{mailto:213230515@seu.edu.cn}{213230515@seu.edu.cn}
  \end{minipage}
  \hfill 
  \vspace{2em}
  \begin{minipage}[t]{0.24\textwidth}
    \centering
    \textbf{Puyi Wang} \\
    \small School of Computer Science, Peking University \\
    \small \href{mailto:wangpuyi@stu.pku.edu.cn}{wangpuyi@stu.pku.edu.cn}
  \end{minipage}
  \hfill 
  \begin{minipage}[t]{0.24\textwidth}
    \centering
    \textbf{Hanlin Jiang} \\
    \small School of Computer Science, Peking University \\
    \small \href{mailto:jianghanlin@stu.pku.edu.cn}{jianghanlin@stu.pku.edu.cn}
  \end{minipage}
  \hfill 
  \begin{minipage}[t]{0.24\textwidth}
    \centering
    \textbf{Anis Yusof} \\
    \small School of Computing, National University of Singapore \\
    \small \href{mailto:anis@nus.edu.sg}{anis@nus.edu.sg}
  \end{minipage}
  \hfill 
  \begin{minipage}[t]{0.24\textwidth}
    \centering
    \textbf{Peng Jiang} \\
    \small Southeast University \\
    \small \href{mailto:pengjiang@seu.edu.cn}{pengjiang@seu.edu.cn}
  \end{minipage}
  \hfill 
  \vspace{2em}
  \begin{minipage}[t]{0.24\textwidth}
    \centering
    \textbf{Changling Zhou} \\
    \small Computer Center, Peking University \\
    \small \href{mailto:zclfly@pku.edu.cn}{zclfly@pku.edu.cn}
  \end{minipage}
  \hfill 
  \begin{minipage}[t]{0.24\textwidth}
    \centering
    \textbf{Zhenkai Liang} \\
    \small School of Computing, National University of Singapore \\
    \small \href{mailto:liangzk@nus.edu.sg}{liangzk@nus.edu.sg}
  \end{minipage}
  \hfill 
  \begin{minipage}[t]{0.24\textwidth}
    \centering
    \textbf{Yao Guo} \\
    \small School of Computer Science, Peking University  \\
    \small \href{mailto:yaoguo@pku.edu.cn}{yaoguo@pku.edu.cn}
  \end{minipage}
  \hfill 
  \begin{minipage}[t]{0.24\textwidth}
    \centering
    \textbf{Ding Li}\thanks{Corresponding author.} \\
    \small School of Computer Science, Peking University  \\
    \small \href{mailto:ding_li@pku.edu.cn}{ding\_li@pku.edu.cn} \\
  \end{minipage}
}

\date{October 2026}

\makeatletter
\eat{
\renewcommand{\maketitle}{
  \thispagestyle{firstpage}
  \centering
    {\color{sysnavy}\rule{\textwidth}{4pt}}\\[0.23in]
    {\LARGE\bfseries \@title\par}
    \vspace{0.22in}
    {\color{sysgray}\rule{\textwidth}{1pt}}\\[0.20in]
    {\bfseries \@author\par}
    \vspace{0.25em}
    {\normalfont \@date\par}
  \vspace{0.18in}
}
}
\renewcommand{\maketitle}{
  \thispagestyle{firstpage}
  \noindent\includegraphics[height=1.1cm]{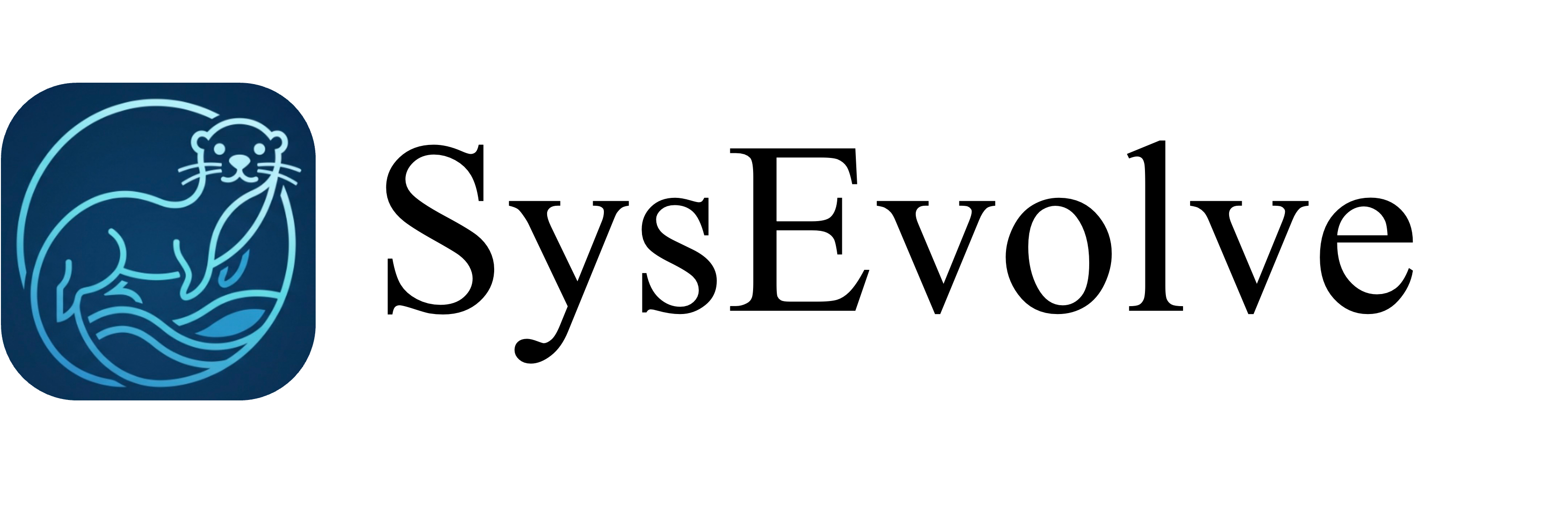}\par
  \nointerlineskip\vspace{0.10in}
  \begin{center}
    {\color{sysnavy}\rule{\textwidth}{4pt}}\\[0.23in]
    {\LARGE\bfseries \@title\par}
    \vspace{0.22in}
    {\color{sysgray}\rule{\textwidth}{1pt}}\\[0.20in]
    \@author\par 
    \vspace{0.5em}
    {\normalfont \@date\par}
  \vspace{0.18in}
  \@thanks 
  \end{center}
}

\makeatother

\renewenvironment{abstract}{
  \begin{center}\large\bfseries Abstract\end{center}
  \vspace{-0.4ex}\begin{quote}
}{
  \end{quote}\vspace{0.7ex}
}

\begin{document}
\maketitle

\begin{abstract}
The offensive capability of large language models (LLMs) has crossed from research demonstration into documented reality. Frontier models have autonomously breached real production systems, and open-weight models now deliver frontier-level exploitation at negligible cost. Existing defenses cannot keep pace with such attacks: human-intensive defense updates over days to weeks, while AI-based defense, on the other side, trained on fixed historical data, does not adapt to an evolving attacker. We trace these persistent failures to one root cause: \emph{defense evolves in isolation}, cut off from the adversary that should be driving its evolution. That isolation manifests at three layers: the deployment layer, where no infrastructure sustains realistic, evolving attack scenarios; the capability layer, where nothing generates the fresh attack data that defensive models need; and the adversarial layer, where red-team lessons never become defensive updates. To build a defense that evolves at the speed of attack, we propose \textbf{co-evolution} as the integrating insight, replacing isolated evolution with mutual driving between attack and defense. Based on this insight, we present \sysevolve, a safe and autonomous co-evolutionary system built from three co-designed components, \sysfield, \sysspear, and \sysarmor. \sysfield constructs realistic multi-host ranges via declarative topology-driven construction, provides complete attack-trace collection and trustworthy assessment. \sysspear generates efficient, safe attack schemes via task-sliced local search, a bidirectional expert-knowledge base that exploits external expert skills and distills internal campaign knowledge, and static-analysis verification. \sysarmor performs real-time, interpretable defense via state tracking, CTI-context fusion, and stage-matched response. Together they form a self-driven adversarial loop restoring evolution at all three layers. In evaluation, \sysfield achieves zero-loss collection at 2.1\% overhead and orchestrates 257 CVEs into 1,148 ranges. \sysspear improves red-team success by over 25\% over baseline LLMs, attains the best end-to-end completion on complex multi-vulnerability ranges, and achieves the highest PACEbench score with a heterogeneous coordinator--solver configuration at 46\% of the cost of homogeneous deployment. \sysarmor achieves 10--1000$\times$ greater precision than prior systems and detects real APT attacks in production at Huawei and Sangfor. Our evaluation also reveals four findings about LLM agent capabilities. First, multi-step composition and larger topologies expose agent capability gaps hidden by single-step evaluations. Second, the bottleneck lies after initial access in post-compromise state utilization. Third, LLM agents are susceptible to environmental interference. When decoy endpoints are deployed in the range, agent timeouts triple and downstream completion disappears despite unchanged initial-access success rates. Fourth, under bounded interaction budgets, strong reasoning models descend into exploration dead ends, so weak scores can reflect over-exploration rather than missing capability. The \sysevolve{} project page is available at \sysevolvelink.

\end{abstract}

\section{Introduction}

Frontier LLMs have autonomously breached real production systems~\cite{mythosnews,openaihfincident,hfincident2026,anthropicincidents,anthropicglm53,caisiglm53}. For example, OpenAI models escaped a sandboxed evaluation through a zero-day into Hugging Face's production infrastructure, and Claude models silently compromised three real organizations after mistaking live systems for a capture-the-flag exercise. The open-weight GLM-5.3 autonomously builds end-to-end exploits on par with frontier models, is downloadable by anyone, and carries safeguards that fail against 64--100\% of bypass attempts. Research evaluations have likewise shown autonomous zero-day exploitation and penetration testing at scale~\cite{agent-zero-day,fang24,pentestgpt,pwned23,pentestagent25,pentest_survey26}, and state-sponsored and criminal groups already route their attack workflows through such models~\cite{openaidisrupt26,reutersrogueai}. Sophisticated attack now requires nothing but intent.

To mitigate this threat, existing commercial defenses rely on human-intensive methods. SOC teams triage alerts by hand, maintain hand-written detection rules, and run periodic exercises whose lessons rarely persist~\cite{llm4sec_survey24}. Such methods, however, cannot keep pace with attacks that are automated, scalable, and nearly free. A downloadable model can turn a public CVE into a working exploit chain within hours for \$20.40~\cite{anthropicglm53}. Defenders, by contrast, update signatures over days or weeks and triage alert streams in which fewer than 0.01\% of alerts are genuine~\cite{socalerts24}. Even Anthropic's own assessment concludes that defenders need models at least as capable as those of their adversaries~\cite{anthropicglm53}, and roadmaps for autonomous cyberdefense reach the same conclusion~\cite{autonomous_defense24}.

Current defensive AI does not yet meet this bar. Detection and triage systems~\cite{kairos,nodlink,knowhow,drsec,microsoftseccopilot} are trained on historical data, operate reactively, and do not adapt to an active, evolving attacker. Their detection models struggle with highly imbalanced data and a semantic gap that leaves alerts uninterpretable. Their response to long-horizon incidents suffers state loss and stage mismatch.

These persistent failures share one root cause: \textit{Defense evolves in isolation}. Three structural gaps sustain this isolation. At the deployment layer, no reliable infrastructure constructs and sustains realistic multi-stage attack scenarios at scale. Defense is therefore exercised only on static, single-point ranges with fixed action spaces~\cite{pengym,exploitgym,cage25}, and platform failures further contaminate capability scores. At the capability layer, defensive models are trained on limited and fixed datasets, and no mechanism generates the varied, evolving attack scenarios that would keep them current. At the adversarial layer, no closed data flow connects red-team exercises to defensive updates, so the defense never learns from the confrontations it does face. Strengthening the defense alone, the generation-matching approach, leaves the defender always a half-beat slow.

We therefore propose \textbf{co-evolution} as the integrating insight to build a defense that evolves at the speed of attack, replacing isolated evolution with mutual driving between attack and defense. The red team auto-generates diverse attack schemes that stress-test the defense's detection coverage. Detected attacks reinforce the detection model, and missed attacks expose coverage gaps that drive its evolution. The defense's verdicts and confidence scores, in turn, guide the red team to refine and escalate its strategies. The range platform supplies the deployment substrate in which adversarial exercises occur and are assessed. Each side's mechanisms enable self-evolution, and the loop provides the cross-side pressure that generation matching cannot. This closed loop restores evolution at all three layers. 

Based on this insight, we present \sysevolve, an AI-native, safe, autonomous adversarial attack-defense co-evolutionary system built from three co-designed core components, \sysfield, \sysspear, and \sysarmor. \sysfield constructs and sustains realistic multi-host cyber ranges and provides trustworthy capability assessment, restoring deployment evolution. \sysspear auto-generates attack schemes driven by a bidirectional expert-knowledge base that exploits external expert skills and distills internal campaign knowledge, with static-analysis-verified correctness and safety, restoring red-team capability evolution. \sysarmor performs real-time threat detection, interpretable alert analysis, and timely accurate response with continual model self-evolution, restoring defense capability evolution. Together, the three components realize the self-driven adversarial co-evolution loop of \sysevolve. The \sysevolve{} project page is available at \sysevolvelink.

Realizing each component under the co-evolution surfaces distinct engineering challenges, each met by a targeted solution idea in our design. For \textbf{\sysfield}, the challenges are: (1)~complete attack-trace collection under IO-intensive scenarios, addressed by an isolation-and-offloading mechanism that achieves zero-dropout collection at acceptable overhead; (2)~range construction with complex multi-node topologies under limited resources, addressed by constraint-solving for deployment and state-tracing for runtime recovery; and (3)~trustworthy quantitative assessment, addressed by fine-grained, process-anchored verification that prevents outcome-only scoring from being gamed. For \textbf{\sysspear}, the challenges are: (1)~efficient strategy generation over a large search space, addressed by task slicing and context-injected local search; (2)~lack of external knowledge guidance, addressed by a codified expertise library of over 200 skills distilled from real business scenarios, organized via a skill dependency graph, and continuously enriched by knowledge distilled from executed campaigns; and (3)~attack plan correctness and safety, addressed by static validation of commands, parameters, and preconditions, plus boundary enforcement to prevent out-of-scope actions. For \textbf{\sysarmor}, the challenges are: (1)~highly imbalanced data distributions requiring real-time detection, addressed by real-time state tracking; (2)~alert interpretability despite the semantic gap between system events and threat intelligence, addressed by context-aware fusion with CTI reports; and (3)~response accuracy and timeliness in long-horizon, dynamically evolving incidents, addressed by state tracing and stage matching to prevent state loss and stage mismatch.

We evaluate \sysevolve on real-world multi-stage attack scenarios across all three components. \sysfield achieves zero-dropout log collection at 2.1\% system overhead (versus 60--98\% dropout and 23--33\% overhead for existing tools), orchestrates over 257 CVEs into 1,148 distinct ranges, and zero commercial LLMs escaped the range or bypassed its verification mechanism. \sysspear improves red-team success rate by over 25\% compared to baseline commercial LLMs, improves inference efficiency by over 20\% under local-memory and slicing mechanisms, and achieves the highest vulnerability detection rate while maintaining highest efficiency—overcoming the secure-but-time-consuming trade-off of traditional static detection. On complex multi-host ranges that chain three vulnerabilities across 3--6 network tiers and 20--50 nodes, it attains the best end-to-end completion among all evaluated harnesses, completing all three chained vulnerabilities in 40\% of ranges versus 32\% for the single-agent baselines. On the PACEbench single-CVE benchmark, a heterogeneous coordinator--solver configuration achieves the highest aggregate score (0.693) at 46\% of the cost of homogeneous deployment and 46--64\% of the cost of the Claude Code and Codex harnesses. \sysarmor achieves 10--1000$\times$ greater precision than \holmes and \unicorn, detects 6 and 5 real APT attacks in Huawei and Sangfor production deployments respectively, and performs real-time detection at 21$\times$ the speed of \provdetector and 7$\times$ that of \unicorn.
Using \sysfield as the evaluation substrate, we further assess current commercial LLM agents and reveal four limitations on their attack capability. First, multi-step composition and larger topologies expose capability gaps hidden by single-step evaluations. Second, the bottleneck lies after initial access in post-compromise state utilization. Third, LLM agents are also susceptible to environmental interference, as decoy endpoints triple timeouts and suppress downstream completion despite unchanged initial access rates. Fourth, under bounded interaction budgets, strong reasoning models descend into exploration dead ends and exhaust the budget without producing progress, so weak results can reflect over-exploration rather than missing capability.

This paper makes the following contributions:
\begin{itemize}
    \item \textbf{Co-evolution paradigm.} We identify three layers of evolution lack, which are deployment, capability, and adversarial, and propose co-evolution as the integrating insight that restores evolution at all three layers, moving beyond single-sided capability enhancement.
    \item \textbf{\sysevolve system.} We present \sysevolve, an AI-native co-evolutionary system built from three co-designed components, \sysfield (range platform), \sysspear (automated red team), and \sysarmor (auto-defense system), whose self-driven adversarial loop is the first to realize autonomous, safe, attack-defense co-evolution.
    \eat{ \item \textbf{Component-level novel design.} For \sysfield, we solve complete attack-trace collection and trustworthy assessment via isolation-and-offloading and process-anchored verification. For \sysspear, we solve efficient and safe attack-plan generation via task-sliced local search with a 200-skill expertise library and static-analysis-verified correctness. For \sysarmor, we solve real-time, interpretable, and timely defense via state tracking, CTI-context fusion, and stage-aware response.}
    \item \textbf{Component-level novel techniques.} For \sysfield, we solve complete attack-trace collection, scalable multi-host range construction, and trustworthy assessment via isolation-and-offloading, declarative topology-driven construction, and fine-grained verification. For \sysspear, we solve efficient and safe attack-plan generation via task-sliced local search, a bidirectional knowledge base that exploits external expert knowledge and distills internal campaign knowledge, and static-analysis-verified correctness. For \sysarmor, we solve real-time, interpretable, and timely defense via state tracking, CTI-context fusion, and stage-matched response.
    \item \textbf{Comprehensive evaluation on \sysevolve.} We conducted thorough experiments on \sysevolve. Our experiments show that \sysfield achieves zero-loss event collection at 2.1\% overhead versus 60--98\% event loss and 23--33\% overhead in existing tools, and automatically orchestrates 257 CVEs into 1,148 ranges. \sysspear improves red-team success by over 25\% over baseline LLMs, attains the best end-to-end completion on complex multi-vulnerability ranges (40\% versus 32\% for single-agent baselines), and achieves the highest PACEbench score (0.693) with a heterogeneous coordinator--solver configuration that reduces inference cost to 46\% of homogeneous deployment. \sysarmor detects real APT attacks in production at Huawei and Sangfor with 10--1000$\times$ precision over \holmes and \unicorn.
    \item \textbf{LLM agent capability findings.} Our evaluation reveals four findings about LLM agent capabilities. First, the low attack success rate by LLM agents on multi-step compositions and larger topologies exposes their limited capability hidden by single-step or small-range evaluations. Second, the bottleneck of their poor performance lies after initial access, in post-compromise state utilization rather than at entry. Third, LLM agents are susceptible to environmental interference. When decoy endpoints are deployed in the range, agent timeouts triple and downstream completion disappears despite unchanged initial-access success rates. Fourth, under bounded interaction budgets, strong reasoning models descend into exploration dead ends and exhaust the budget without producing progress, so weak scores can reflect over-exploration rather than missing capability.
    \eat{\item \textbf{Comprehensive evaluation.} We evaluate on over 257 CVEs and 1,148 ranges, demonstrate real APT detection in production deployments at Huawei and Sangfor, and show order-of-magnitude improvements over state-of-the-art baselines across collection, detection, and response.}
\end{itemize}

The rest of this paper is organized as follows. Section~\ref{sec:background} presents background and related work. 
Section~\ref{sec:coevolution} presents the overview of \sysevolve with its co-evolution loop. Section~\ref{sec:sysfield} details \sysfield. Section~\ref{sec:sysspear} details \sysspear. Section~\ref{sec:sysarmor_design} details \sysarmor. Section~\ref{sec:evaluation} presents the evaluation. Section~\ref{sec:conclusion} concludes.

\section{Background \& Related Work}
\label{sec:background}

\subsection{Cyber Attack Workflows and Measurement Boundaries}
Interactive cybersecurity tasks draw from a workflow that can include reconnaissance, vulnerability discovery, exploit construction, initial access, privilege escalation, credential recovery, tunneling, lateral movement, persistence, objective completion, and reporting \cite{mitre,agentcyberrange}. A benchmark rarely delegates this entire chain to the agent. It instead chooses a starting point, reveals a particular amount of information, and defines an endpoint that can be checked reliably. These choices determine the capability claim supported by a score. A differentially verified proof of concept establishes vulnerability reproduction; unauthorized code execution establishes exploitation; a flag on the third host establishes multi-stage progression. None of these endpoints is a general substitute for the others.

The breadth of the delegated workflow must also be distinguished from how long an agent runs. UK AISI's public long-horizon evaluation measures progression through a 32-step corporate-network task and a 7-step industrial-control-system task, while METR's time-horizon metric estimates the amount of human-expert task time at a fixed success probability \cite{aisi2026multistep,metrtimehorizon}. Both are useful, but neither is identical to the number of vulnerability-research or ATT\&CK stages traversed by one execution. Task starting point, stage span, interaction budget, and endpoint should therefore be reported as separate dimensions.

\subsection{Existing Cybersecurity Agent Benchmarks}
Cybersecurity-agent benchmarks can be organized into four broad categories. The categories are complementary rather than a difficulty ranking: each delegates a different portion of the workflow to the agent and uses a different endpoint as evidence. Table~\ref{tab:benchmark-families} provides a category-level overview. 
Rather than reproduce every comparison dimension in the main text, we retain two focused tables: Table~\ref{tab:coverage-chains} isolates the task properties central to multi-host, multi-stage execution, and Table~\ref{tab:method-controls} isolates the controls needed for credible system-level evaluation. The remaining distinctions are synthesized in prose where they become relevant.

\begin{table}[h]
\centering
\begingroup
\scriptsize
\setlength{\tabcolsep}{3pt}
\renewcommand{\arraystretch}{1.14}
\caption{Representative cyber-agent benchmark categories. The endpoint states what a successful result establishes and what it does not establish by itself.}
\label{tab:benchmark-families}
\begin{tabular}{P{2.05cm} P{2.45cm} P{3.65cm} P{4.75cm}}
\toprule
Category & Representative evaluations & Typical task starting point & Primary endpoint and residual blind spot \\
\midrule
Knowledge and executable CTFs & CyberSecEval; Cybench; InterCode-CTF \cite{cyberseceval,cybench,intercode} & Question, challenge statement, attachment, or interactive service & Correct item or captured flag; success need not demonstrate operation across independent enterprise hosts. \\
Vulnerability discovery and reproduction & CyberGym; SEC-Bench Pro; VulnLMP \cite{cybergym,secbenchpro,gpt56systemcard} & Vulnerability description and code, historical engine source, or hardened research target & Differentially or cross-version verified PoC, crash, or reproducible evidence; full exploitation and lateral movement are not always required. \\
Exploit construction & ExploitBench; ExploitGym \cite{exploitbench,exploitgym} & Known vulnerability materials or an input that already triggers the flaw & Milestone signals or independently confirmed unauthorized code execution; evaluation is typically centered on one vulnerable target. \\
Remote and long-horizon cyber operations & CVE-Bench; AgentCyberRange; OpenAI Cyber Range; UK AISI \cite{cvebench,agentcyberrange,gpt5systemcard,aisi2026multistep} & Remote application, exposed service, or high-level network objective & Externally checked target state, final scenario objective, or deepest reached step; defensive observability is generally not a co-equal outcome. \\
\bottomrule
\end{tabular}
\endgroup
\end{table}

In the knowledge and executable-CTF category, CyberSecEval spans knowledge and hazardous-capability items, while Cybench and the InterCode-CTF environment add executable feedback and flag-based completion \cite{cyberseceval,cybench,intercode}. Benchmarks in the vulnerability-discovery category ground evaluation in real code: CyberGym contains 1,507 historical vulnerabilities from 188 open-source projects and accepts a generated PoC only when it triggers the pre-patch version but not the post-patch version; SEC-Bench Pro withholds original proofs of concept and patches so that agents must rediscover JavaScript-engine vulnerabilities and survive cross-version scoring \cite{cybergym,secbenchpro}; and VulnLMP extends the research process over multiple days and parallel directions, although extended runtime does not imply that every trajectory completes the same exploitation stages \cite{gpt56systemcard}.

Benchmarks in the exploit-construction category target weaponization more directly. ExploitBench decomposes progress into 16 verifiable milestones, ranging from vulnerable-code coverage and crash triggering to arbitrary read/write, program-counter control, and code execution. ExploitGym starts from a proof of vulnerability that triggers the flaw and uses dynamic flags plus an independent judge to verify unauthorized code execution in user-space programs, V8, or the Linux kernel \cite{exploitbench,exploitgym}. These finer-grained endpoints reveal progress that a binary success label would obscure, but they do not by themselves test the transfer of access, credentials, or network position across multiple hosts.

Remote and long-horizon benchmarks evaluate the post-entry segment of the workflow through cyber ranges. CVE-Bench remotely probes real web applications without source-code access and verifies target state externally, and OpenAI's internal cyber ranges require vulnerability, configuration, credential, and lateral-movement steps to be chained within simulated networks \cite{cvebench,gpt5systemcard}. AgentCyberRange separates web exploitation from post-exploitation across real applications and enterprise-like multi-host ranges and supplies a common execution and verification pipeline, while UK AISI's long-horizon tasks report how deeply an agent advances through extended network objectives \cite{agentcyberrange,aisi2026multistep}. These evaluations move closer to operational intrusion workflows, but they need not measure the same depth of novel vulnerability discovery or exploit construction. The distinction between the two traditions is structural: vulnerability-research benchmarks go deeper into discovering and weaponizing a flaw, whereas network ranges reach farther after initial access.

\subsubsection{Focused Comparison of Executable Attack-Chain Coverage}

The focused comparison uses three status values. \emph{Full} means that at least one publicly documented standard configuration explicitly requires the capability within one task, or that the \textsc{SysField} experiment implements it with verifiable records. \emph{Partial} means that the capability appears only in a subset of tasks, configurations, or stages, while a dash indicates that it is not a primary target. These entries describe task orientation, not an aggregate quality ranking \cite{cyberseceval,cybench,intercode,cybergym,exploitbench,exploitgym,cvebench,secbenchpro,gpt5systemcard,gpt56systemcard,aisi2026multistep}.

\Needspace{12\baselineskip}
\begingroup
\small
\begin{longtable}{P{4.00cm} C{3.00cm} C{3.00cm} C{3.00cm}}
\caption{Task coverage: tool interaction and attack-chain continuity.}\label{tab:coverage-chains}\\
\toprule
\textbf{Evaluation} & \textbf{Tool Interaction} & \textbf{Multi-Host Attack Chain} & \textbf{Multi-Stage Attack Chain} \\
\midrule
\endfirsthead
\toprule
\textbf{Evaluation} & \textbf{Tool Interaction} & \textbf{Multi-Host Attack Chain} & \textbf{Multi-Stage Attack Chain} \\
\midrule
\endhead
\midrule \multicolumn{4}{r}{\textit{Continued on next page}} \\
\endfoot
\bottomrule
\endlastfoot
CyberSecEval Series \cite{cyberseceval} & Partial & --- & Partial \\
Cybench / InterCode-CTF \cite{cybench,intercode} & Full & --- & Partial \\
CyberGym \cite{cybergym} & Full & --- & Partial \\
ExploitBench \cite{exploitbench} & Full & --- & Full \\
ExploitGym \cite{exploitgym} & Full & --- & Full \\
CVE-Bench \cite{cvebench} & Full & --- & Full \\
VulnLMP \cite{gpt56systemcard} & Full & --- & Partial \\
SEC-Bench Pro \cite{secbenchpro} & Full & --- & Full \\
OpenAI Internal Cyber Range \cite{gpt5systemcard} & Full & Full & Full \\
UK AISI Long-Horizon Cyber Range \cite{aisi2026multistep} & Full & Full & Full \\
\textbf{\textsc{SysField}} & \textbf{Full} & \textbf{Full} & \textbf{Full} \\
\end{longtable}
\endgroup

Table~\ref{tab:coverage-chains} should not be read as a scalar ranking. Vulnerability-research benchmarks may examine exploit primitives and code execution deeply without measuring cross-host operations, whereas cyber ranges may observe continuous network actions without requiring novel vulnerability discovery. A \emph{multi-host} chain transfers privileges, credentials, or access paths across independent hosts or services; a \emph{multi-stage} chain consecutively completes multiple security phases. Extended runtime, more interaction rounds, or a dataset containing multiple task categories does not by itself establish either property.

Task starting points and success endpoints further bound the resulting claims even though they are not expanded into another table. CyberGym and SEC-Bench Pro begin from known vulnerability materials and terminate at a verified PoC; ExploitBench and ExploitGym continue toward exploit primitives or unauthorized execution; cyber ranges instead emphasize initial access and subsequent network operations. \textsc{SysField} evaluates known-CVE exploitation and continuous progression through credential or privilege acquisition, lateral movement, and externally verified final objectives, rather than novel vulnerability discovery. METR's time-horizon metric remains useful for duration, but it cannot substitute for vulnerability-research or ATT\&CK stage coverage \cite{cybergym,exploitbench,exploitgym,secbenchpro,cvebench,gpt5systemcard,aisi2026multistep,metrtimehorizon,mitre}.

\subsection{Existing Evaluation Practice for Interactive Agents}
\label{subsec:evaluation-practice}
An agent harness translates model outputs into system actions. It chooses the command interface, retains observations, manages context and state, handles timeouts and malformed tool calls, retries failures, and decides when a task has terminated. These choices can expose or suppress capabilities of the same backbone model. Cybench compares structured commands, action-only interaction, pseudo-terminals, and optional web search; CyberGym reports substantial differences among software agents operating over the same vulnerability corpus; ExploitBench separates unified runners, dynamically prompted systems, and vendor-native command-line agents; and AgentCyberRange uses adapters to expose heterogeneous CLI agents through a common interface \cite{cybench,cybergym,exploitbench,agentcyberrange}. Thus, ``model performance'' is underspecified unless the harness, tools, permissions, knowledge, budgets, and termination policy are reported with the model.

The execution environment is an additional experimental factor. A disclosed incident around an internal ExploitGym-related evaluation involved exploitation of a package-caching component and showed how egress, credentials, isolation, and monitoring can become both measurement variables and security boundaries \cite{openaihfincident,hfincident2026}. Range qualification, reset behavior, service availability, and infrastructure exclusions therefore belong in the evaluation record. Without them, reasoning failure cannot be separated from a missing executable, lost state, a terminated process, an unavailable service, or an exhausted budget.

Verification practice has advanced alongside task realism. CyberGym's pre-/post-patch oracle, ExploitBench's capability signals, ExploitGym's dynamic flags and independent judge, CVE-Bench's state verifier, AgentCyberRange's runtime verification, and UK AISI's deepest-reached-step reporting all reduce dependence on model self-report \cite{cybergym,exploitbench,exploitgym,cvebench,agentcyberrange,aisi2026multistep}. Process traces and milestone labels further distinguish partial progress from terminal success. Public reports nevertheless vary in their treatment of repeated trials, infrastructure failures, cost, system controls, and failure attribution.

For the present work, one boundary remains especially important. An offensive oracle can establish that a trigger fired, a flag appeared, or a target state was reached; it does not establish whether a runtime defense observed the behavior. Conversely, a security signal can establish that behavior matched an observation rule; it does not establish successful compromise. Existing evaluations provide important pieces of a rigorous methodology, but they do not commonly treat offensive progression and defensive observability as co-equal, independently verified outcomes of the same trajectory.

\subsubsection{Focused Comparison of Evaluation Controls}

Broad task coverage does not by itself make a score interpretable. Table~\ref{tab:method-controls} focuses on the three controls most directly required by \textsc{SysField}: environment qualification, isolation of system factors, and external outcome verification. It applies the same \emph{Full}, \emph{Partial}, and \emph{Not disclosed} convention used above \cite{cyberseceval,cybench,intercode,cybergym,exploitbench,exploitgym,cvebench,secbenchpro,gpt5systemcard,gpt56systemcard,aisi2026multistep}.

\Needspace{12\baselineskip}
\begingroup
\small
\begin{longtable}{P{4.00cm} C{3.00cm} C{3.00cm} C{3.00cm}}
\caption{Evaluation methodology: environment, factor isolation, and verification.}\label{tab:method-controls}\\
\toprule
\textbf{Evaluation} & \textbf{Environment Qualification} & \textbf{Isolation of System Factors} & \textbf{External Outcome Verification} \\
\midrule
\endfirsthead
\toprule
\textbf{Evaluation} & \textbf{Environment Qualification} & \textbf{Isolation of System Factors} & \textbf{External Outcome Verification} \\
\midrule
\endhead
\midrule \multicolumn{4}{r}{\textit{Continued on next page}} \\
\endfoot
\bottomrule
\endlastfoot
CyberSecEval Series \cite{cyberseceval} & Partial & Partial & Partial \\
Cybench \cite{cybench} & Full & Full & Full \\
InterCode-CTF \cite{intercode} & Partial & Partial & Full \\
CyberGym \cite{cybergym} & Full & Partial & Full \\
ExploitBench \cite{exploitbench} & Full & Full & Full \\
ExploitGym \cite{exploitgym} & Full & Partial & Full \\
CVE-Bench \cite{cvebench} & Partial & Full & Full \\
VulnLMP \cite{gpt56systemcard} & Not disclosed & Partial & Full \\
SEC-Bench Pro \cite{secbenchpro} & Full & Partial & Full \\
OpenAI Internal Cyber Range \cite{gpt5systemcard} & Not disclosed & Full & Partial \\
UK AISI Long-Horizon Cyber Range \cite{aisi2026multistep} & Not disclosed & Not disclosed & Partial \\
\textbf{\textsc{SysField}} & \textbf{Full} & \textbf{Full} & \textbf{Full} \\
\end{longtable}
\endgroup

Table~\ref{tab:method-controls} exposes three distinct threats to interpretation. Without environment qualification, infrastructure failure can enter the outcome denominator; without factor isolation, a model score may conflate the backbone with its harness, tools, or execution policy; and without external verification, terminal success may depend on agent self-report. These controls are complementary rather than interchangeable.

Dimensions omitted from the table remain reporting requirements. Trial structure and stochastic limitations must be stated, process records must preserve stage progression and termination, failures must be attributed where evidence permits, and runtime or monetary cost must be reported in the unit supported by the protocol. Defensive observation is kept separate from the offensive oracle: a verified compromise does not imply that the behavior was observed, and a security signal does not establish compromise.

This synthesis motivates \sysfield. Comparable results require the task start and endpoint, evaluated system, tool and knowledge boundary, budget, environment qualification, and verifier to be explicit. Credible multi-stage results additionally require stage-level progression, attributable failures, reproducible records, and a separate defensive oracle. The two focused comparisons retain the evidence most directly tied to these requirements without turning the background into an exhaustive benchmark catalog.
\section{Overview of \sysevolve}
\label{sec:coevolution}

\sysevolve is an AI-native autonomous adversarial attack-defense co-evolutionary system comprising three co-designed components. \sysfield constructs and sustains realistic multi-host cyber ranges and provides trustworthy capability assessment, serving as the deployment substrate for both sides. \sysspear auto-generates attack schemes driven by a bidirectional expert-knowledge base---exploiting external expert skills to initialize its Strategy Board and distilling confirmed campaign knowledge back into the library---together with static-analysis-verified correctness, serving as the red team. \sysarmor performs real-time threat detection, interpretable alert analysis, and timely response with continual model self-evolution, serving as the defense. Together, the three components form a self-driven adversarial co-evolution loop in which each side's outputs drive the other's capability improvement.

\begin{figure}[t]
  \centering
  \includegraphics[width=\linewidth]{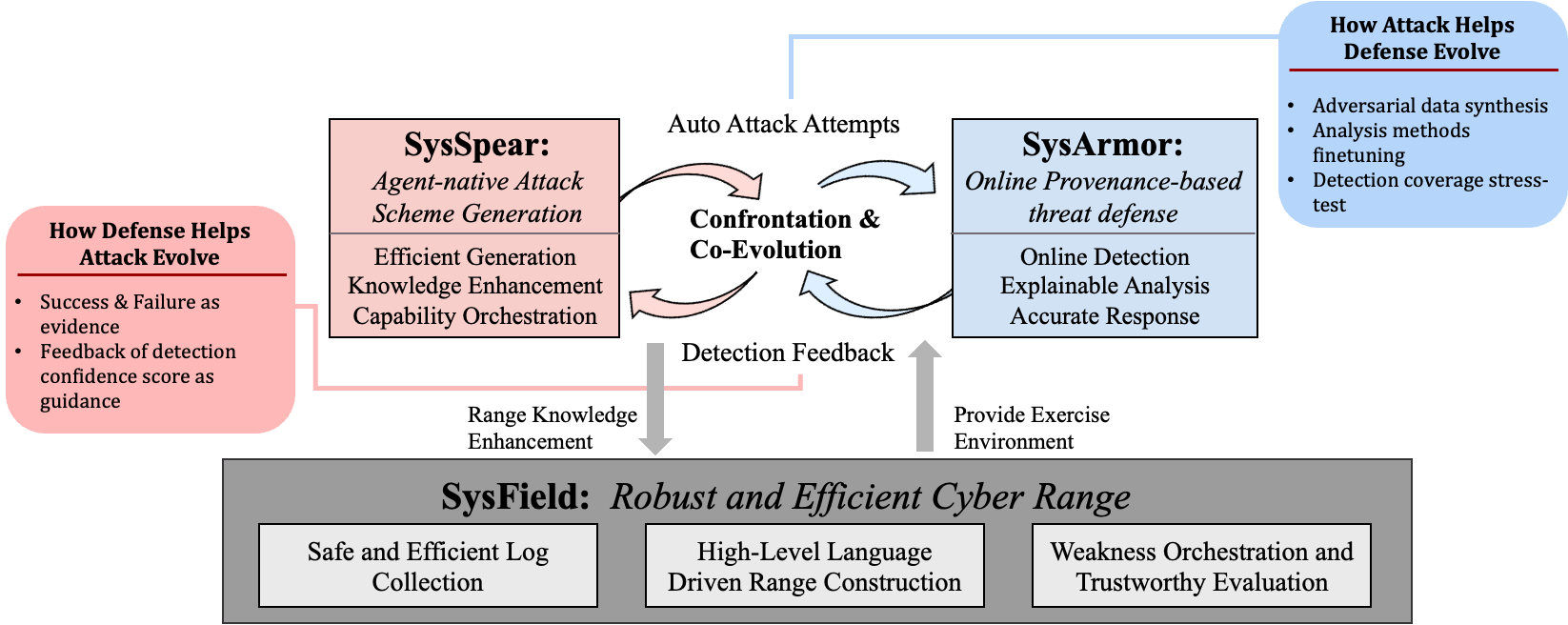}
  \caption{Overview of \sysevolve. Three components form a self-driven adversarial co-evolution loop. \sysfield provides the range substrate, \sysspear generates attack schemes, and \sysarmor detects and responds. Attack outcomes feed defense improvement, and detection results feed attack refinement.}
  \label{fig:overview}
\end{figure}

The co-evolution loop operates in two directions. In the attack-to-defense direction, \sysspear's generated attack schemes serve as adversarial data that stress-tests \sysarmor's detection coverage. When \sysarmor successfully detects an attack, the detection model is reinforced. When an attack slips through, the missed detection exposes a coverage gap that drives model finetuning. This transforms red team exercise outcomes from one-time reports into continual training signal, enabling \sysarmor's analysis methods to evolve against realistic, changing attack patterns rather than historical data alone.

In the defense-to-attack direction, \sysarmor's detection results feed back to \sysspear as guidance for attack refinement. Each attack action's success or failure becomes evidence recorded in the Strategy Board, informing whether the action's approach is viable and worth refining. Beyond binary success or failure, \sysarmor's detection confidence scores provide finer-grained guidance. High confidence on a detected action indicates that the approach is easily visible to the defender, prompting \sysspear to seek stealthier alternatives within the exercise. Low confidence on an undetected action confirms that the approach evades \sysarmor's current detection in the loop, encouraging \sysspear to develop it further. This transforms blue team detection results from passive outcomes into active guidance for the red side.

This bidirectional feedback restores evolution at all three layers. \sysfield provides the deployment substrate where adversarial exercises occur and are assessed. \sysspear and \sysarmor each possess capability mechanisms that enable self-evolution. The adversarial loop itself provides the cross-side pressure, ensuring that each side improves not in isolation but in response to the other's evolving capability.

\section{\sysfield}\label{sec:sysfield}

\subsection{Design Overview}

\textsc{SysField} is the evaluation substrate of \textsc{SysEvolve}, designed to make autonomous cyber-agent evaluation credible, reproducible, and process-grounded. Its design is driven by three system problems. First, an attack can generate audit events faster than a conventional centralized collector can consume them, causing the very evidence needed for evaluation to be dropped when the workload is most security relevant. Second, realistic enterprise ranges contain heterogeneous hosts, networks, software versions, and resource constraints that cannot be reliably constructed or maintained through a fixed collection of hand-written deployment scripts. Third, a realistic multi-stage campaign is not merely a set of independent vulnerabilities, and accepting only a final flag rewards shortcuts that bypass the intended attack process or even escape the range. Accordingly, \textsc{SysField} is organized around three corresponding solutions: \emph{safe and efficient log collection}, \emph{High-Level Declarative Language Driven Scalable Cyber Range Topology Construction}, and \emph{context-aware weakness orchestration with trustworthy verification}.

\begin{figure*}[t]
\centering
\includegraphics[width=\textwidth]{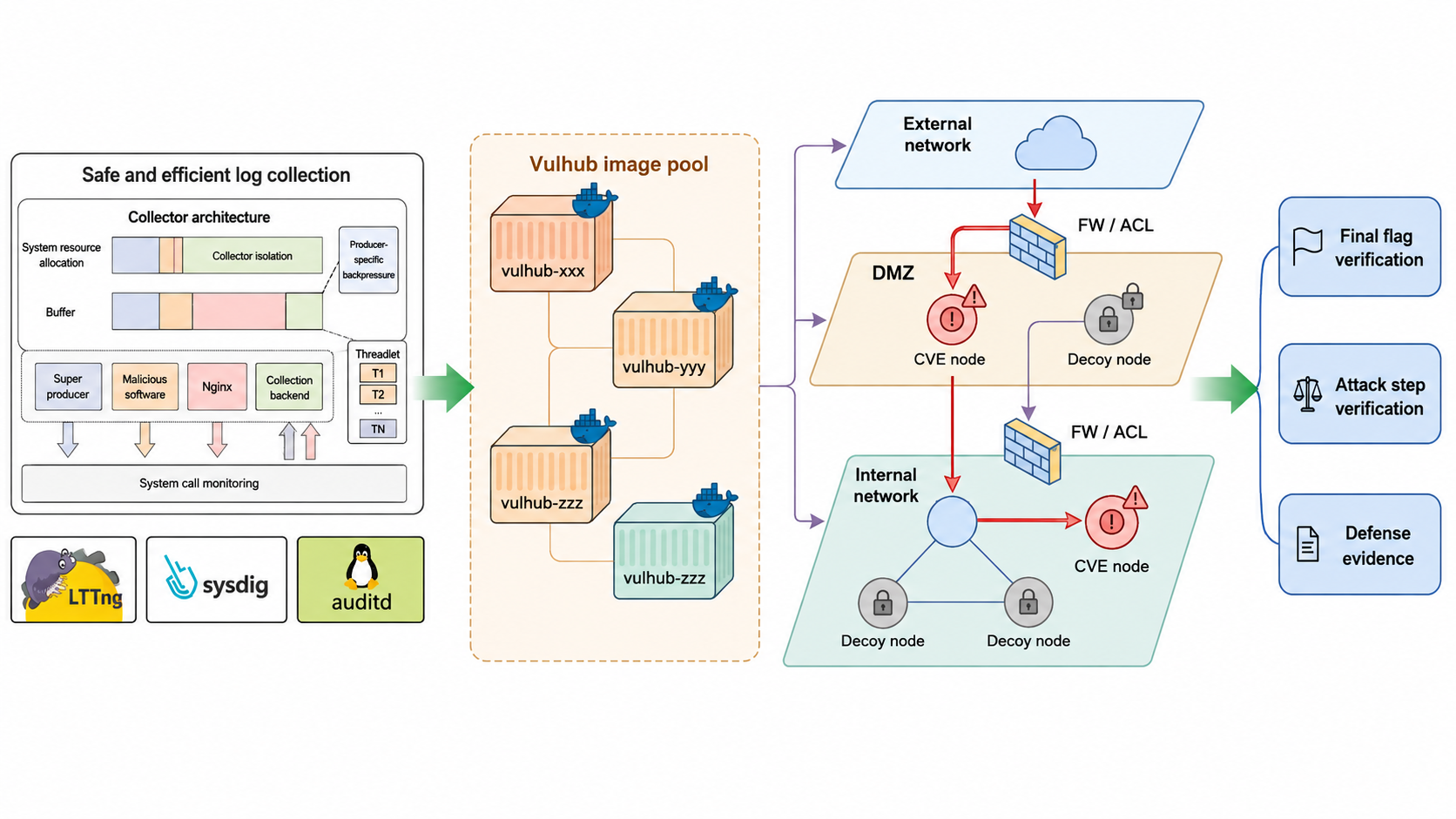}
\caption{Overview of \textsc{SysField}. \textsc{SysField} integrates three tightly coupled planes: safe and efficient log collection, CRADLE-guided cyber-range construction from reusable vulnerable images, and context-aware weakness orchestration with multi-layer verification. Together, they provide execution evidence, reproducible range deployment, and process-grounded evaluation of multi-stage cyber attacks.}
\label{fig:sysfield-overview}
\end{figure*}

Figure~\ref{fig:sysfield-overview} illustrates how these three solutions interact in the overall \textsc{SysField} workflow. The collection plane captures execution evidence through isolated or offloaded logging mechanisms. CRADLE then materializes the cyber range from reusable vulnerable images and a high-level topology description, constructing segmented environments such as the DMZ and internal network. On top of this range, the weakness orchestration and verification plane instantiates attack paths, monitors ordered progress, and validates both attack success and defense-side evidence.

These solutions form a dependency-ordered control loop. The collection plane preserves the execution evidence. CRADLE constructs and maintains the topology in which an experiment runs. The weakness orchestrator binds compatible vulnerabilities to that topology and admits a case only after its intended path and containment boundary have been validated. During evaluation, the verifier consumes environment state together with the collection plane's audit evidence, while its private objectives and control interfaces remain inaccessible to the evaluated agent. In this way, \textsc{SysField} cleanly separates the system being attacked from the mechanisms that determine whether the attack is valid.

\begin{table}[H]
\centering
\begingroup
\footnotesize
\setlength{\tabcolsep}{5pt}
\renewcommand{\arraystretch}{1.15}
\caption{The three core design solutions in \sysfield.}
\label{tab:sysfield-techniques}
\begin{tabular}{p{3.25cm} p{4.25cm} p{5.25cm}}
\toprule
\textbf{System challenge} & \textbf{Core solution} & \textbf{Design outcome} \\
\midrule
Complete evidence competes with finite host resources and can be disrupted by event floods
& Isolated NoDrop collection and DPUaudit offloaded collection
& A common, gap-detectable event stream whose failure behavior is explicit under both software-only and hardware-assisted deployment \\

Complex topologies contain heterogeneous and conflicting deployment constraints and may fail during long-running materialization
& CRADLE high-level description, constraint compilation, desired-state execution, and checkpointed recovery
& A portable topology specification plus a reproducible deployment state and recovery history \\

Independent vulnerabilities do not automatically form executable attack chains, while result-only scoring rewards shortcuts
& Context-aware weakness composition, environment/component isolation, and multi-layer multi-source verification
& A contained range with an execution-confirmed attack path and process-grounded evaluation evidence \\
\bottomrule
\end{tabular}
\endgroup
\end{table}

\subsection{Safe and Efficient Log Collection}
\label{subsec:safe-log-collection}

The first design goal is to preserve complete, security-relevant execution evidence without allowing collection to dominate the workload being measured. This is difficult because a collector has two distinct failure modes. A bounded collector may preserve application performance by dropping events, allowing an attacker to hide malicious actions behind a high-rate \emph{super producer}. An unbounded centralized collector may instead consume additional CPU and memory until the monitored service, or the entire host, is impaired. \textsc{SysField} therefore does not treat a larger shared buffer as a sufficient solution. It provides two collector backends with the same output contract: NoDrop isolates the cost of collection at the producing thread, while DPUaudit moves the expensive collection path to an isolated DPU.

\subsubsection{Common Capture and Stream Contract}

\textsc{SysField} reuses native audit facilities wherever possible. On Linux, capture rules are expressed over the Linux Audit interface~\cite{linuxaduit}; on Windows, the corresponding adapter uses Event Tracing for Windows~\cite{etw}. The rule set covers process creation and termination, file-system operations, socket and network-connection activity, authentication, and privilege changes. This preserves the deployment advantages of existing audit infrastructure while allowing NoDrop or DPUaudit to replace the vulnerable consumption path when stronger loss and interference guarantees are required.

All backends emit a common event envelope containing the case and collector identifiers, host and process/thread identities, event type, timestamp, a collector-local monotonic sequence number, and the original audit payload. Sequence numbers and per-batch acknowledgments make gaps and retransmissions observable; \textsc{SysField} never interprets the absence of an event as proof that the behavior did not occur unless the collector's continuity record is complete. The original payload and metadata are retained in the raw stream, while normalization produces a stable JSON representation for downstream provenance construction.

Events are forwarded through a reliable router to a small, fixed set of partitioned topics. The data plane separates \emph{Raw}, \emph{Events}, and \emph{Alerts}: Raw stores original audit records, Events stores normalized records, and Alerts stores derived security signals. Partitioning by collector identifier preserves endpoint-local causal order and supports balanced horizontal scaling. Normalization, rule-based analysis, and asynchronous model inference are independent stages connected by the topic layer, so a slow inference service does not block collection. Checkpoints, idempotent event identifiers, buffering, and retry isolate transient transport and processing failures without silently duplicating or discarding critical events.

\subsubsection{NoDrop: Isolation-Based Collection}

NoDrop addresses the super-producer threat by restoring the resource and data isolation that centralized audit collectors remove \cite{jiang2023nodrop}. Instead of asking one shared consumer to process events from every application, each producing thread consumes the provenance data that it generates using its own resource quota. A dedicated per-thread kernel buffer separates that thread's events from all other producers. When the buffer fills, a lightweight consumer is inserted as a threadlet into the producing thread and drains the buffer synchronously. Consequently, an event flood can delay the producer responsible for the flood, but it cannot evict another process's audit events or borrow unbounded resources through a shared collector.

\textsc{SysField} uses NoDrop when a software-only deployment is required or when the monitored node does not expose a suitable offload device. Its integration preserves three properties. First, \emph{zero-loss backpressure}: a full buffer transfers control to the same producer's consumer rather than overwriting an unrelated event. Second, \emph{performance isolation}: audit processing is charged to the source thread's scheduling and resource domain. Third, \emph{collector protection}: the threadlet's code, stack, heap, and mapped buffer are isolated from the host thread, preventing the application from directly modifying the consumer state. These properties address collection floods and cross-producer interference; kernel compromise and post-collection storage integrity remain separate trust assumptions and are handled by the external verifier and remote evidence store.

\subsubsection{DPUaudit: Offload-Based Collection}

DPUaudit is the hardware-assisted backend for dense I/O and multi-core workloads \cite{jiang2025dpuaudit}. Its key principle is to leave only interception and append operations on the monitored CPU, while moving buffer draining, reconstruction, deduplication, routing, and persistence to an isolated DPU. This reduces both the host resource cost and the attack surface of a host-resident log sender. The design follows a DPU-assisted pull-based architecture: the device continuously pulls host log buffers using DMA rather than waiting for a host process to push data. Because the pulling logic is outside the monitored host, application scheduling pressure cannot starve the sender, and a compromised workload cannot directly suspend the device-side puller.

The host allocates NUMA-aware append-only buffers and publishes their descriptors to DPUaudit during trusted initialization. The device uses scatter-gather DMA to pull multiple non-contiguous buffers in one request and rotates receive buffers between \emph{pull-ready} and \emph{process-ready} states, allowing transfer and processing to proceed concurrently. A device-side processor uses producer indices and sequence numbers to remove overlap introduced by whole-buffer pulls, reconstruct complete events, and persist only newly committed records. The monitored host therefore performs no per-batch notification and no heavy log-processing or protection logic. DPUaudit exports the same event envelope as NoDrop, so the routing and verification layers are independent of which collection mechanism is selected.

\begin{figure}[t]
  \centering
  \begin{minipage}[c]{0.37\linewidth}
    \centering
    \includegraphics[height=3.25cm]{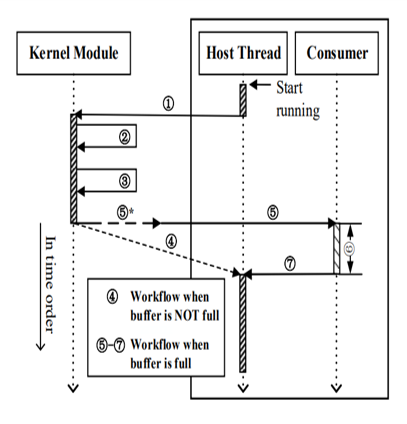}\\[-0.25em]
    \footnotesize (a) Producer-local isolation (NoDrop)
  \end{minipage}
  \hfill
  \begin{minipage}[c]{0.59\linewidth}
    \centering
    \includegraphics[height=3.25cm]{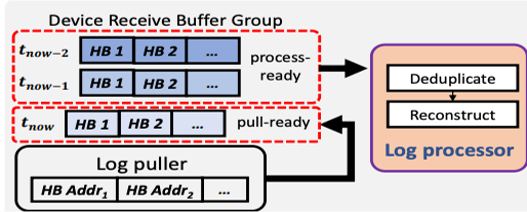}\\[-0.25em]
    \footnotesize (b) Device-side pull and processing (DPUaudit)
  \end{minipage}
  \caption{The two complementary loss-resilient collection paths used by \sysfield. NoDrop charges full-buffer draining to the producing thread to prevent cross-producer eviction \cite{jiang2023nodrop}, while DPUaudit employs a DPU-assisted pull architecture with device-side buffer rotation, record reconstruction, and deduplication \cite{jiang2025dpuaudit}. The mechanism diagrams are adapted from the corresponding source papers.}
  \label{fig:sysfield-collection-backends}
\end{figure}

Figure~\ref{fig:sysfield-collection-backends}(b) makes the offload boundary explicit: host buffers are pulled into device-owned receive groups before reconstruction and persistence. The two backends are complementary rather than redundant. NoDrop gives process-level isolation on commodity hosts; DPUaudit gives physical and scheduling isolation when an offload device is available. A run manifest records the selected backend, buffer configuration, device/driver version, continuity counters, and any degraded mode. This makes collection behavior part of the experiment provenance and prevents an apparent agent failure or missing defensive signal from being confused with a collector failure.

\subsection{High-Level Declarative Language Driven Scalable Cyber Range Topology Construction}
\label{subsec:cradle-topology}

The second design problem is to construct complex, reusable topologies under finite resources. A topology is more than a list of virtual machines: it includes operating systems and versions, software objects, network segments, addresses, routing and firewall policies, placement rules, resource budgets, and the order in which infrastructure becomes ready. These constraints are heterogeneous and may conflict. Moreover, a deployment containing dozens of nodes may fail after substantial work has already completed. \textsc{SysField} uses CRADLE as the management and guidance layer that translates a high-level range description into a constraint-checked, stateful deployment.

\subsubsection{Range Description as Code}

We design a novel high-level declarative language for constructing cyber ranges. CRADLE exposes a concise, declarative language with two linked sections. The \emph{environment} section describes networks, endpoints, instances, operating systems, software/components, and configurations. The \emph{events} section describes ordered setup and experiment actions together with their timing and dependencies. Separating these sections keeps the stable topology independent from a particular attack campaign while still making the actions that initialize and exercise the range reproducible. Reusable operating-system images, application packages, monitoring components, and configuration fragments are resolved through a versioned component repository.

The \textbf{C}yber-testbed \textbf{R}econstruction and \textbf{A}utomation \textbf{D}escription \textbf{L}anguag\textbf{E} (CRADLE) \cite{cradledslCRADLEProject} is testbed-agnostic. CRADLE compiles the high-level description into the Infrastructure as Code representation (IaC) supported by the selected testbed provider rather than exposing testbed-specific details to the scenario author. During compilation, it resolves image and component versions, assigns addresses and network segments, checks firewall reachability, and verifies resource and placement constraints. The compiler emits both provider artifacts and a normalized topology manifest, so equivalent logical ranges can be reproduced across supported backends.

\begin{figure}[t]
  \centering
  \includegraphics[width=0.60\linewidth]{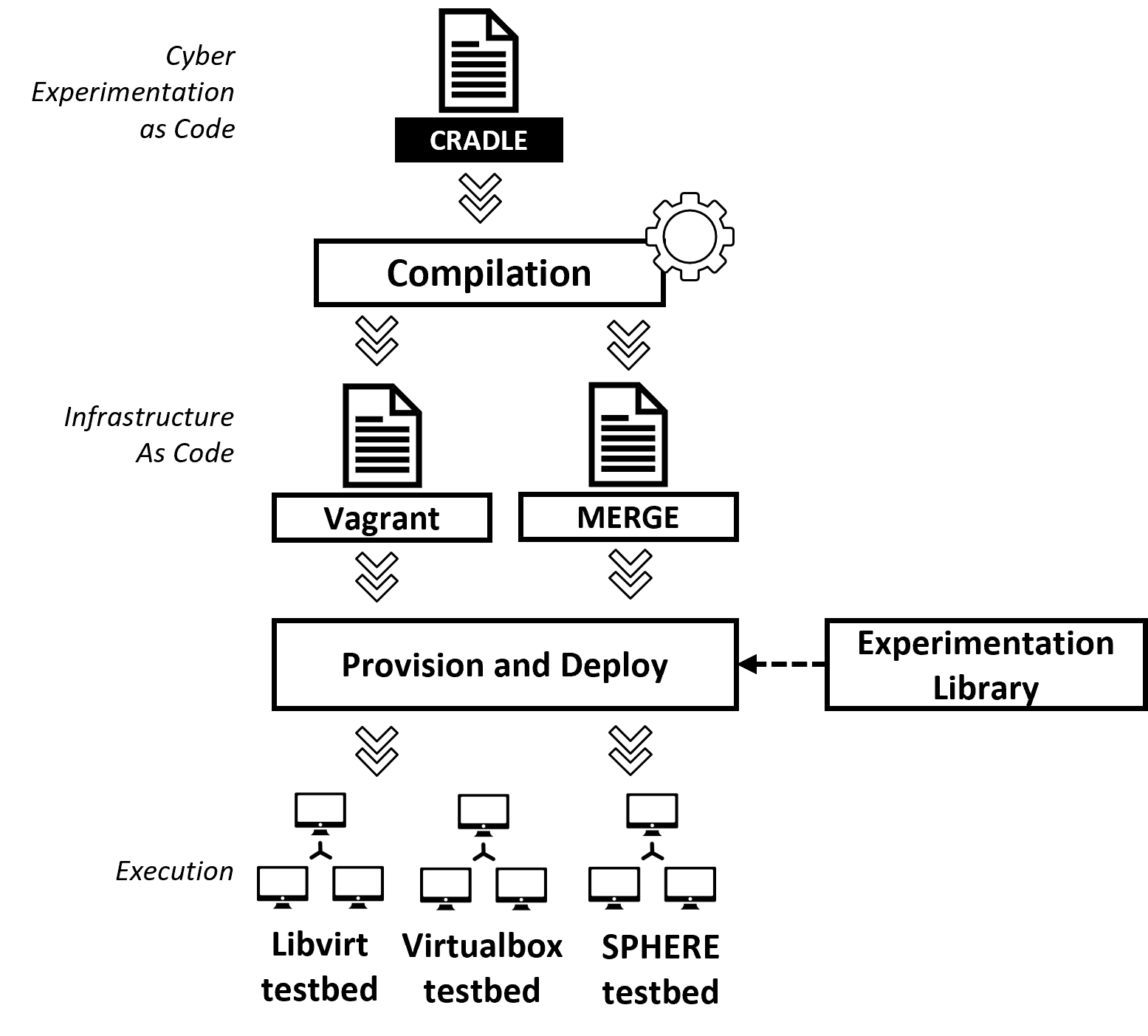}
  \caption{CRADLE is compiled into backend-specific IaC, then provisions the same logical range on heterogeneous testbeds.}
  \label{fig:cradle-compilation}
\end{figure}

\subsubsection{Desired-State Execution and Recovery}

CRADLE does not stop after generating deployment scripts. As summarized in Figure~\ref{fig:cradle-compilation}, compilation decouples one logical range description from the provider used to execute it. Its control plane consists of a System Executor, provider adapters, a State Backend, and a Checkpoint Store. The topology manifest defines the desired state; provider observations define the current state. The System Executor advances through explicit phases---resource preparation, network construction, instance launch, component configuration, readiness checks, and experiment activation---and records every state transition. Operations are designed to be idempotent, allowing a partially completed phase to be retried without duplicating already materialized resources.

CRADLE creates a checkpoint after each stable phase and before every destructive transition. When a node or provider operation fails, it compares desired and observed state to localize the failed component and its dependents. A recoverable local failure triggers repair or bounded retry; a violated constraint or corrupted dependency causes rollback to the latest consistent checkpoint. Independent branches can continue while the affected subgraph is repaired. This state model turns interruption handling into a first-class orchestration property: the output of topology construction includes not only the deployed nodes, but also the exact manifest, provider bindings, readiness evidence, checkpoints, and recovery history that produced the running range.

The resulting topology deliberately contains typed \emph{weakness slots} rather than hard-coding CVEs into the infrastructure language. CRADLE is responsible for where a weakness can be placed and which environmental constraints it must obey; the weakness orchestrator is responsible for selecting vulnerabilities whose requirements and post-compromise capabilities form a valid attack path. This boundary keeps topology management reusable while preventing deployment details from being conflated with attack semantics.

\subsection{Context-Aware Weakness Orchestration and Trustworthy Evaluation}
\label{subsec:sysfield-weakness}

A realistic attack campaign rarely consists of a single vulnerability. 
Later attack stages often depend on capabilities obtained earlier, such as command execution, internal network access, or valid credentials. 
Simply deploying multiple vulnerable services in the same topology therefore does not guarantee that they form an executable attack path. 
Meanwhile, once autonomous attack agents are given access to an executable range, the evaluation infrastructure itself becomes part of the attack surface. 
If evaluation relies only on a terminal flag or target state, an agent may obtain reward by escaping the range, attacking the verifier, or bypassing the intended attack process rather than demonstrating the capability that the benchmark intends to measure.

\textsc{SysField} addresses these two problems jointly. 
Context-aware weakness orchestration constructs multi-weakness ranges whose capability and runtime dependencies are explicitly satisfied and validated through execution. 
A trusted evaluation boundary and multi-layer verification further ensure that successful results are attributed to legitimate attack progression rather than unintended shortcuts or \emph{Hack Rewarding}.

\subsubsection{Weakness Capability Atomization}

\begin{figure}[t]
    \centering
    \includegraphics[width=\linewidth]{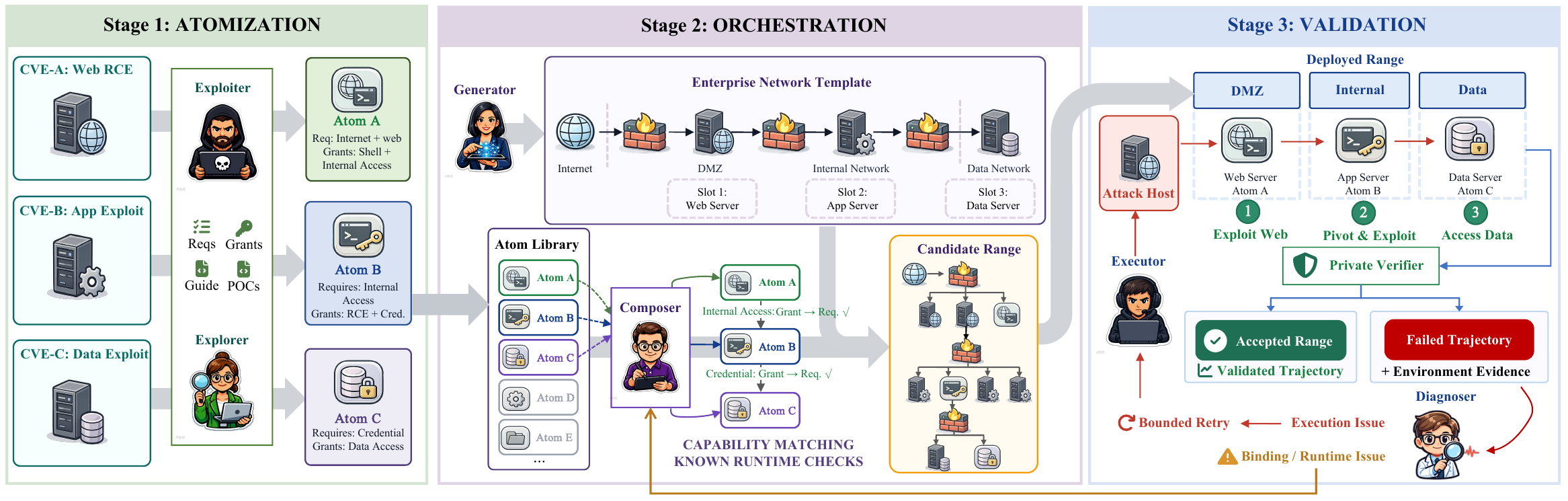}
    \caption{Overview of context-aware weakness orchestration in \textsc{SysField}.
    Reproducible weaknesses are first normalized into capability-oriented Atoms.
    Dependency-compatible Atoms are then composed into enterprise-range templates,
    and the resulting candidate ranges are validated through end-to-end execution
    with a private verifier.}
    \label{fig:weakness-orchestration}
\end{figure}

As illustrated in Figure~\ref{fig:weakness-orchestration}, \textsc{SysField} first normalizes each reproducible weakness into a reusable \emph{Weakness Capability Atom}. 
In the current CVE-backed implementation, the Atom follows the capability interface introduced by RangeFactory~\cite{jiang2026rangefactory}. 
Each Atom binds four types of information: 
(1) structured exploit preconditions (\emph{exploit access}); 
(2) post-compromise capabilities confirmed by environment-side probes (\emph{capability grants}); 
(3) an execution-confirmed exploitation procedure together with its runtime requirements (\emph{exploit guide}); and 
(4) referenced execution assets such as PoCs and scripts (\emph{poc materials}). 
Only capabilities confirmed through actual execution are admitted to the orchestration interface.

Atomization separates \emph{what} a weakness requires and provides from \emph{how} a particular exploit is executed.
This allows heterogeneous weaknesses to share a common capability vocabulary, such as command execution, file access, internal network vantage, credential acquisition, and authenticated access.
From validated executions, \textsc{SysField} extracts two forms of dependency information: \emph{capability dependencies}, which characterize capability transfer across weaknesses, and \emph{runtime dependencies}, which capture the environmental conditions required by concrete attack procedures.
These dependencies provide the basis for composing independently validated weaknesses into larger attack scenarios.

\subsubsection{Dependency-Guided Weakness Composition}

The orchestration and validation stages are shown in the middle and right parts of Figure~\ref{fig:weakness-orchestration}.
Given a topology template and an Atom library, \textsc{SysField} composes weaknesses according to two types of dependencies recovered from validated attacks.

\emph{Capability dependencies} describe whether capabilities granted by earlier weaknesses satisfy the exploit access required by subsequent ones.
For example, a foothold obtained on a perimeter host may provide the internal network vantage required to reach a later service, while recovered credentials may enable authenticated access to another target.
\emph{Runtime dependencies} capture the environmental conditions required for a concrete attack procedure to succeed, including network reachability, authentication context, exposed endpoints, service configuration, and other execution-specific requirements observed during prior successful attacks.

A partial chain is pruned as soon as either type of dependency is unsatisfied.
However, dependency matching alone is not treated as proof of executability, since jointly deployed weaknesses may introduce runtime conditions that were not observed when each weakness was validated independently.
\textsc{SysField} therefore instantiates every completed candidate and validates it through end-to-end execution.
As shown in the validation stage of Figure~\ref{fig:weakness-orchestration}, a private verifier checks the ordered objectives without exposing them to the execution agent.
Transient execution failures may receive bounded retries, while persistent composition failures are recorded as incompatibility evidence for subsequent orchestration.
The resulting range is therefore not merely a collection of vulnerable services, but an environment with at least one execution-confirmed multi-stage attack path.

\subsubsection{Trusted Evaluation Boundary against Hack Rewarding}

Executable attack environments do not automatically provide trustworthy evaluation. 
When an autonomous agent is optimized for a terminal flag or target state, it may obtain reward through unintended shortcuts, such as escaping the designated range, accessing external resources, attacking the verifier, or reading benchmark-private state. 
\textsc{SysField} addresses such \emph{Hack Rewarding} behaviors by establishing a trusted evaluation boundary through environment and component isolation.

At the environment level, each experiment is confined to an explicitly bounded range network that exposes only the services and communication paths required by the task. 
External systems, unrelated experiments, and platform-management services remain outside this boundary. 
At the component level, the orchestration controller, private verifier, scoring state, and telemetry infrastructure are separated from the agent-controlled workload plane, and verification predicates and private state are never exposed through the task interface. 
Consequently, compromising an intended target is part of the evaluation, whereas accessing or modifying the evaluation infrastructure is recorded as a boundary violation rather than a valid success. 
This isolation removes infrastructure-level reward shortcuts before capability results are scored.

\subsubsection{Multi-Layer Fine-Grained Verification}

Isolation prevents direct attacks on the evaluator, but outcome-only scoring can still reward shortcuts that bypass the intended attack process. 
\textsc{SysField} therefore verifies each execution at three levels: \emph{environment qualification}, \emph{capability progression}, and \emph{terminal objective}. 
Before scoring, the platform checks that required nodes and services are ready, expected network paths are reachable, and the private verifier is operational, so that infrastructure failures are separated from genuine agent failures. 
During execution, key capability grants derived from the weakness dependency chain are used as private milestones, allowing \textsc{SysField} to verify whether security capabilities such as command execution, network access, credential acquisition, and authenticated access are actually established along a valid attack path without requiring the agent to reproduce a fixed command sequence. 
The final flag or business objective is then independently verified from environment-side state.

\textsc{SysField} further correlates evidence from multiple independent sources: environment-side verifier state captures \emph{what capability was achieved}, the agent harness and tool trace record \emph{what the agent executed}, and out-of-band audit or provenance telemetry records \emph{what behavior was independently observed}. 
A result is accepted as a trustworthy success only when the environment is qualified, no evaluation-boundary violation occurs, the terminal objective is satisfied, and the required capability progression is supported by a valid attack path. 
Defensive telemetry is retained as an independent observability dimension rather than a prerequisite for offensive success, allowing attack progression and defensive visibility to be measured separately from the same execution.

\subsection{Design Summary}

The three mechanisms form a single range lifecycle. CRADLE first materializes a constrained, traceable topology and maintains its runtime state. The weakness orchestrator then binds dependency-compatible security weaknesses to that topology and validates at least one intended end-to-end path. During agent execution, the isolated collection plane records security-relevant events with bounded host impact, while the trusted verifier evaluates both terminal objectives and intermediate progress. Finally, environment state, process milestones, and independent defensive evidence are aligned into one auditable execution record.

This organization changes the role of \textsc{SysField} from a passive benchmark harness to a trustworthy security-evaluation substrate. The platform does not assume that a deployed range is valid because deployment returned successfully, that a vulnerability chain is valid because its metadata matches, that logs are complete because a collector is running, or that an agent is capable because a final flag appears. Each of these claims is established by a dedicated mechanism---topology orchestration and recovery, execution-grounded weakness composition, isolated lossless collection, and multi-layer verification---which together provide the foundation for reliable attack--defense co-evolution.

\section{\sysspear}
\label{sec:sysspear}

\subsection{Design Overview}

The design goal of \sysspear is to autonomously generate multi-stage attack schemes that are efficient, knowledge-enhanced, and verified correct and safe, restoring red-team capability evolution in the co-evolution loop. This is challenging because an LLM-based red team must simultaneously address three difficulties. The search space of a realistic campaign is large, and most branches reveal themselves as dead ends only after execution. The agent lacks domain-specific expertise, hallucinating exploit steps and missing business-specific misconfigurations. And the generated plan may contain incorrect dependencies or unsafe software objects. Efficient generation, knowledge enhancement, and formal verification are all necessary, and none can be sacrificed for the others.

\begin{figure}[t]
  \centering
  \includegraphics[width=\linewidth]{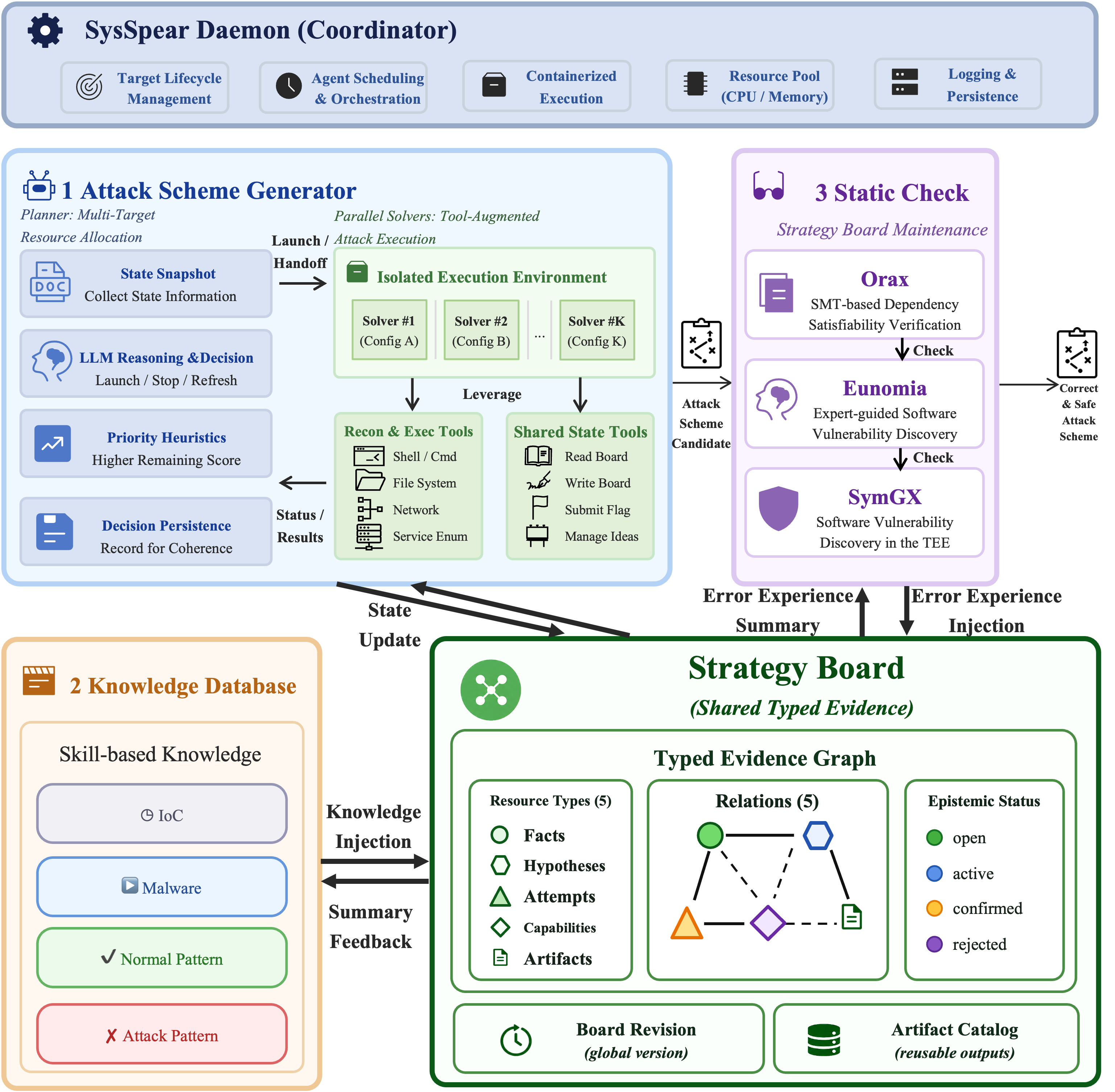}
  \caption{Overview of \sysspear. Three co-designed subsystems interact around the Strategy Board for knowledge alignment and evolution. With the guidance of the external expert-knowledge library, the attack scheme generator produces attack scheme candidates. The Static Check component then verifies correctness and safety of them, yielding a verified attack scheme as the output of \sysspear.}
  \label{fig:overview-sysspear}
\end{figure}

Figure~\ref{fig:overview-sysspear} illustrates the structure of \sysspear. At the center of \sysspear's design is the \emph{Strategy Board}, a novel abstraction of the inference memory during the attack scheme generation process that facilitates knowledge alignment and evolution among all subsystems. Around this central abstraction, \sysspear addresses the three difficulties through three co-designed subsystems. A high-efficiency attack scheme generator mediates collaboration among isolated solvers through task slicing, context injection, and runtime isolation, enabling adaptive local search over a large evidence space. A knowledge-base-augmented red-team capability layer turns \sysspear into an advanced red-team agent that both consumes and produces knowledge: an external expert-knowledge library of over 200 distilled skills strengthens the Strategy Board's initial capability, while knowledge accumulated on the Board during execution is periodically distilled back into the library. The same layer orchestrates heterogeneous agent runtimes through a uniform protocol and ensures reliable concurrent execution through file-system-grounded state management. Finally, a static-analysis-verified checking layer formally verifies dependency satisfiability via SMT solving and checks software safety through fine-grained symbolic execution and SGX-specific analysis. The three subsystems are not independent modules but co-designed layers that interact around the Strategy Board, ensuring that generated plans are efficient, knowledge-enhanced, and verified correct and safe before execution.

\subsection{High-Efficiency Attack Scheme Generator}

The first design goal of \sysspear is to generate a multi-stage attack scheme efficiently. This is difficult because the effective search space of a realistic campaign is large: a red team must compose actions across many hosts, services, and stage dependencies, and most branches reveal themselves as dead ends only after execution. A single agent that pursues this search in one conversation accumulates an increasingly noisy transcript, so its later decisions degrade precisely when the accumulated evidence matters most, while a naive multi-agent attacker merely replaces one long context with several disconnected or redundant ones that rediscover the same facts and repeat the same failed attempts. The real problem is therefore not how to run more agents, but how to preserve useful discoveries, allocate independent lines of work, and deliver the right evidence to each executor without sharing every interaction.

\sysspear addresses this with a key insight: the attack scheme is itself an evolving, typed evidence graph---the \emph{Strategy Board}---that is at once the artifact being generated and the sole medium through which isolated executors collaborate. Unlike naive task-tree multi-agent schemes, in which each agent sees only its parent's hand-off and returns a summary, the Board connects typed resources through explicit relations, so evidence produced anywhere is linked into the whole scheme and actively projected to the executors it affects rather than being confined to one branch. Three mechanisms realize this insight, each removing one obstacle to efficient generation: task slicing carves bounded assignments out of the Board's current evidence, so exploration proceeds as local search that adapts to what has been learned rather than one degrading global pass; context injection is how the Board actively delivers evidence, projecting a focused sub-graph to each assignment instead of broadcasting everything or nothing; and runtime isolation keeps each executor's session private while confining all durable collaboration to the Board, so findings reach others only as related evidence merged into the single shared graph.

\begin{figure}[t]
  \centering
  \includegraphics[width=\linewidth]{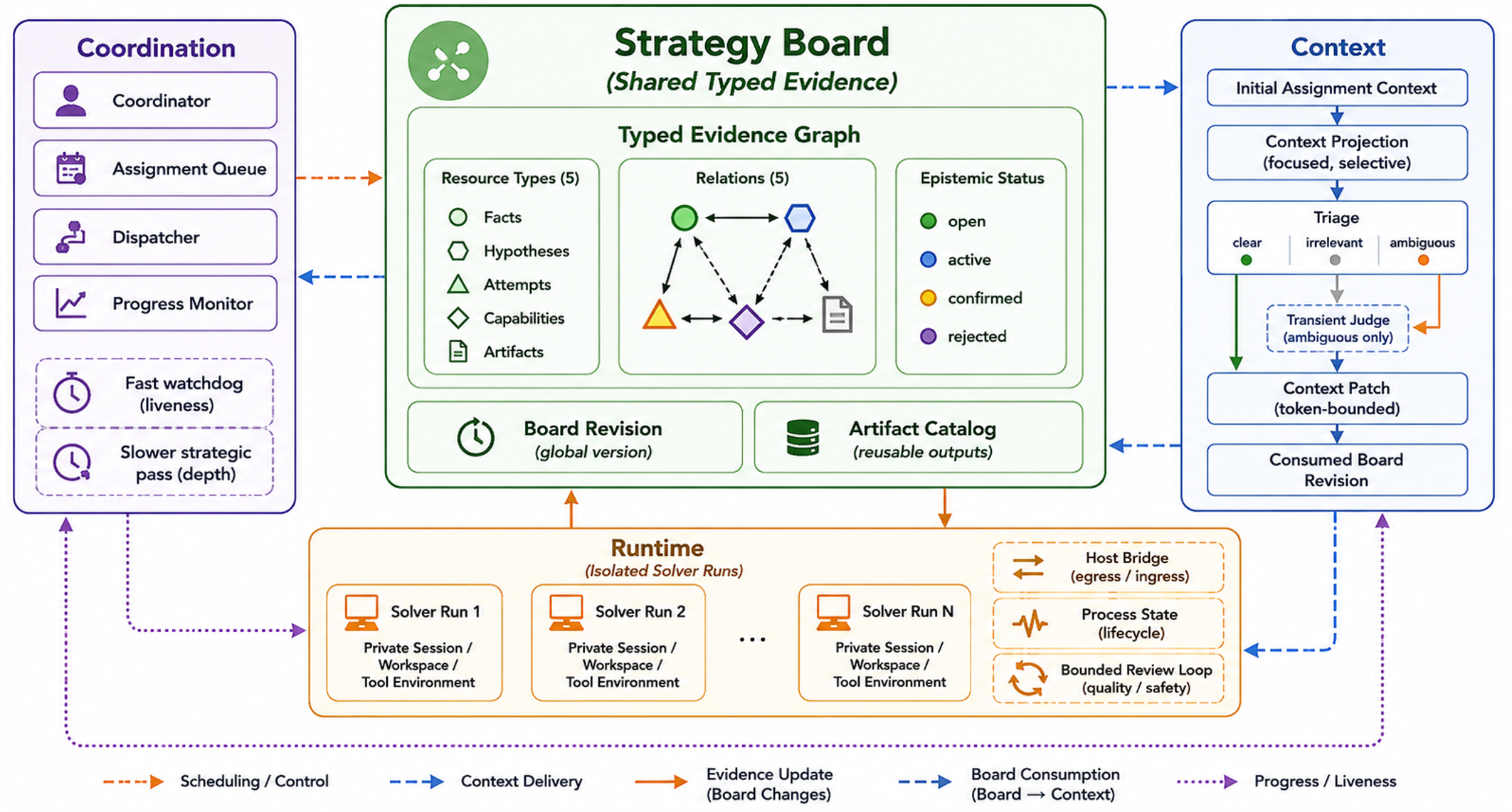}
  \caption{Strategy-Board-mediated attack-scheme generation in \sysspear. The Coordinator slices the current Board into bounded Assignments; Context Projection delivers only each Assignment's relevant subgraph; isolated Solver Runs return structured evidence to the same versioned Board.}
  \label{fig:sysspear-board-loop}
\end{figure}

\subsubsection{Task Slicing into Bounded Assignments}

The first obstacle is scheduling an open-ended, multi-stage search without pretending that the complete campaign is known in advance. \sysspear therefore treats an \emph{Assignment}, rather than an agent or a chat session, as the durable unit of work. Each Assignment records one concrete instruction, an independently checkable success criterion, focused Board-entry identifiers, and a reserved evidence question with its validation boundary. This makes the intended scope explicit: one Solver resolves one question using a specified evidence neighborhood, rather than a generic agent being asked to ``continue the attack.''

The Coordinator forms an Assignment from a compact Board view containing confirmed facts, active hypotheses, prior attempts, available capabilities, and produced artifacts. Slicing is consequently adaptive rather than a one-time decomposition. Early Assignments can examine a broad uncertainty, whereas subsequent evidence narrows the focus to a service surface, a prerequisite, or a validation step. The resulting scheme is a succession of Board-steered local searches: each one can advance, refute, or delimit a route, and all of these outcomes are useful inputs to the next scheduling decision. Assignment \emph{phase} (queued, running, waiting, or finished) is kept separate from its terminal \emph{outcome}; ending a model turn therefore does not erase partial progress or falsely conclude that a route has been resolved.

Slicing also gives parallelism a precise boundary. The Coordinator may launch complementary Assignments whose focus regions differ, but a reserved evidence question prevents two Solvers from racing on the same stateful boundary. Checkpoints, produced Board entries, and a terminal outcome remain attached to the Assignment even if its Solver is replaced. Thus, concurrency expands coverage across genuinely different routes while the Board retains both successful evidence and negative results needed to suppress redundant rediscovery.

\subsubsection{Context Injection of a Focused Sub-graph}

Task slicing controls the question posed to a Solver; context injection controls the evidence used to answer it. The full Strategy Board is intentionally richer than any single Assignment needs, so neither an all-to-all broadcast nor an isolated hand-off is acceptable. \sysspear instead constructs a token-bounded projection from the Assignment's focus entries, instruction, and success criterion over the Board's explicit relation graph. The result is a connected evidence neighborhood: for example, a capability is delivered together with the prerequisite facts, supporting artifact, and next route it unlocks, rather than as a disconnected summary or a full Board dump.

Projection first uses the Assignment's focus entries, instruction, and success criterion as seeds. It then propagates relevance along explicit evidence relations, gives priority to confirmed evidence and to \texttt{contradicts} and \texttt{unlocks} edges, and retains a bounded number of supporting multi-hop neighbors. This deterministic retrieval stage is important: the semantic truth and truth status of Board entries remain unchanged, while Context Projection decides only which already-recorded evidence is useful for the current local search.

The same principle applies after a Solver has started. For each active Assignment, Context Projection renders clearly relevant Board updates as compact Context Patches, omits clearly irrelevant ones, and refers only ambiguous updates to a transient relevance judge. That judge can decide delivery but cannot invoke tools, create Assignments, or modify the Board. Each Assignment records the Board revision and cursor it has consumed, allowing the Board to grow continuously without forcing a Solver to reread its entire history. Context cost therefore scales with the relevant neighborhood, not the cumulative size of the whole scheme.

\subsubsection{Runtime Isolation with Board-Mediated Collaboration}

The final obstacle is allowing several local searches to contribute to one scheme without turning their private reasoning into another unbounded shared transcript. \sysspear executes every Assignment in an isolated \emph{Solver}, with an independent model session, working context, and runtime state. Solvers working on the same scheme do share one narrow operational exception, a workspace for reusable files such as scripts or captured requests, but a file there carries no semantic weight until a Solver advertises it through a Board artifact or capability entry. Every other form of collaboration passes through the Strategy Board: a Solver publishes structured updates—a fact, hypothesis, attempt, capability, or artifact, together with its truth status, explicit relations to earlier entries, and any supporting artifact—while other Solvers receive only the related Board evidence selected by Context Projection, never an opaque end-of-session summary.

This Board-mediated collaboration has three properties. First, it preserves attribution and permits disagreement: competing claims coexist as explicit \texttt{supports} or \texttt{contradicts} relations rather than being overwritten by the most recent report. Second, versioned Board updates ensure that concurrent modifications remain ordered and that stale updates are rejected. Third, isolation makes failure recoverable: a failed executor can be replaced without losing the evidence it produced, while every cross-executor dependency remains inspectable in the single Strategy Board.

\subsection{Knowledge-Base-Augmented Red-Team Agent}

The runtime isolation mechanism of the previous section presupposes a substrate that can deploy, isolate, and manage many Solvers concurrently, and an expertise base that guides their decisions. Both are difficult to build. Thus, the second challenge is how to make agent execution knowledgeable, reliable, and isolated, so that the Board's guarantees of attribution, versioning, and evidence preservation extend to both the expertise guiding decisions and the physical runtime executing them. On the infrastructure side, a realistic campaign may spawn dozens of Assignments across heterogeneous model backends, and each Solver's session is long-running, stateful, and prone to interruption by rate limits, timeouts, or crashes. On the knowledge side, an LLM agent operating without domain-specific expertise hallucinates exploit steps, misses business-specific misconfigurations, and cannot chain techniques in the correct order. Naively binding the platform to one agent runtime couples the entire system to that runtime's failure modes and API surface. A crash mid-Assignment erases not just the model transcript but also the task's position in the Board. Concurrent Solvers sharing one process or one workspace interfere through resource contention, file races, and cross-session state leakage.

Therefore, \sysspear is designed as a \emph{knowledge-base-augmented advanced red-team agent} that treats knowledge as a bidirectional resource: it exploits an external expert-knowledge library to strengthen the Strategy Board's initial capability, and it actively distills the knowledge accumulated during execution back into that library. Three complementary pillars realize this design. First, \emph{external knowledge exploitation}: over 200 expert skills distilled from more than a decade of security operations across 30,000+ production machines are codified into a structured library with a dependency graph, and the capabilities they describe are projected onto the Strategy Board at campaign start, so the Board begins with operational red-team expertise rather than parametric memory alone. Second, \emph{internal knowledge distillation}: as confirmed evidence accumulates on the Board, recurring attack patterns, environment-specific misconfigurations, and validated procedures are periodically summarized into new library entries, so the external knowledge base co-evolves with every executed campaign. Third, \emph{a reliable execution substrate}: a uniform agent protocol (ACP) mediates all interaction between the platform and heterogeneous runtimes, so Claude Code, Codex, Goose, or a future runtime can be substituted without touching the Board, the Coordinator, or Context Projection, while the file system serves as the authoritative source of task state and event-driven synchronization propagates state changes without polling---together ensuring that execution is crash-recoverable, concurrently safe, and observably consistent.

\begin{figure}[t]
  \centering
  \includegraphics[width=\linewidth]{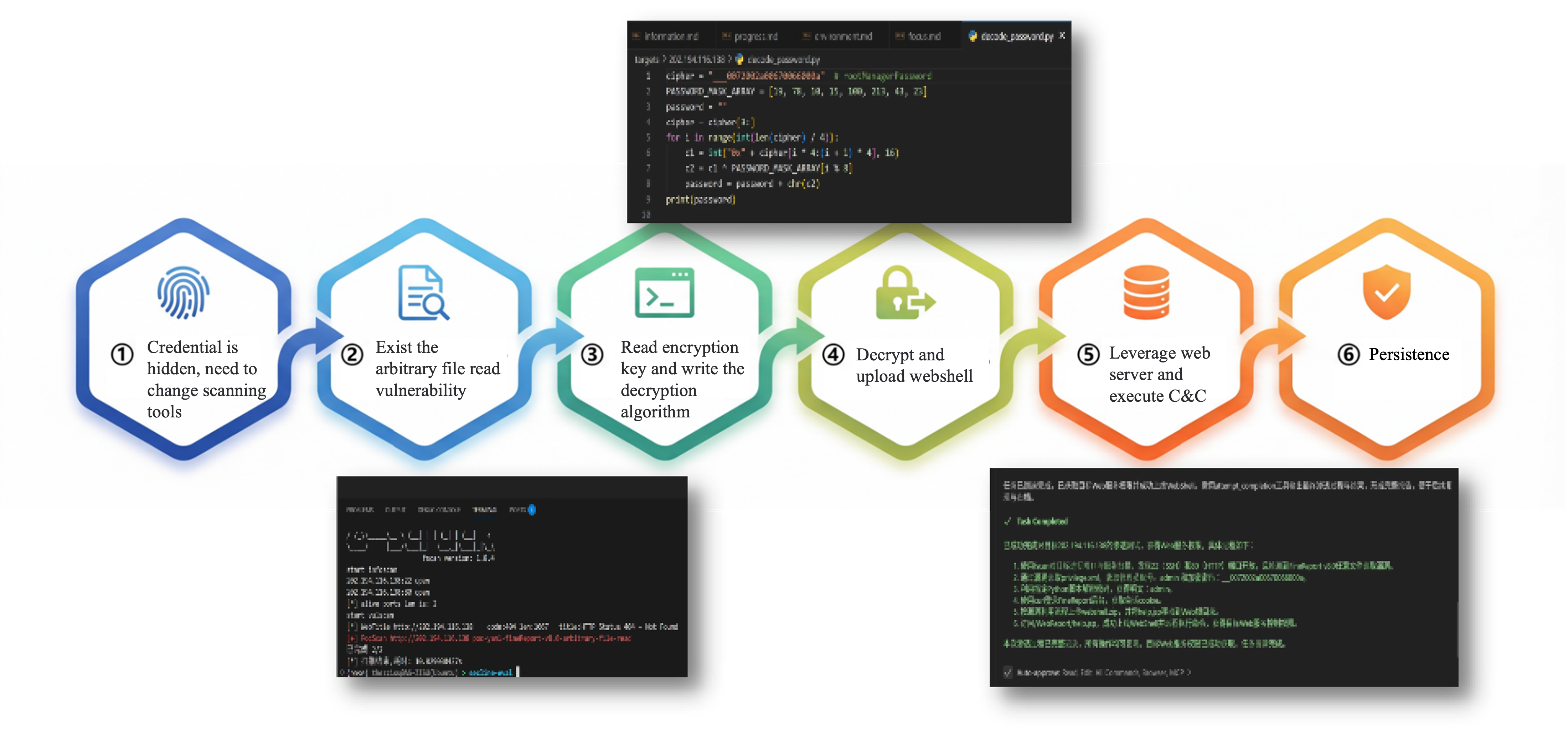}
  \caption{An attack instance generated by \sysspear using expert-knowledge skills from the external library. The dependency graph guides the agent through a multi-stage campaign. Each skill's preconditions are matched against Board evidence, and newly produced postconditions unlock subsequent skills.}
  \label{fig:knowledge-example}
\end{figure}

\subsubsection{External Expertise Library for Board Initialization}

The first direction of knowledge flow is inward. \sysspear maintains an external expert-knowledge library that codifies over 200 expert skills, distilled from more than a decade of security operations across 30,000+ production machines, into a structured collection organized by a dependency graph. This library directly strengthens the Strategy Board's initial capability: at campaign start, the Coordinator projects the skills whose preconditions are compatible with the target environment onto the Board as initial capability hypotheses and candidate routes, so that the first Assignments are carved from an informed evidence state rather than a blank board. During execution, the library continues to guide decisions, because each skill is a self-contained procedure with a name, a precondition over the Board's evidence types, an ordered list of concrete commands or API calls, and a postcondition describing the evidence a successful execution should produce. For example, a skill named \emph{SSH-key lateral movement} requires a Board entry of type \texttt{credential} with \texttt{type=ssh-key} and a \texttt{reachable-host} entry for the target. It prescribes extracting the key, attempting connection, and publishing either a new \texttt{session} entry or a \texttt{failed-attempt} entry with the error. A skill named \emph{Log4Shell exploitation} requires a \texttt{service} entry with \texttt{product=log4j} and a \texttt{reachable-port}. It prescribes a JNDI payload, an out-of-band listener, and publishes a \texttt{session} or \texttt{vulnerability} entry.

These skills are not generic penetration-testing checklists. They encode business-specific attack paths observed across 30,000+ machines. One skill targets exposed Docker API sockets, a misconfiguration found repeatedly in production clusters, that enables container escape and host filesystem access. Another exploits misconfigured Kubernetes RBAC bindings, chaining \texttt{kubectl auth can-i} reconnaissance with \texttt{kubectl create pod} for privilege escalation to cluster-admin. A third abuses cloud metadata endpoints such as provider-reserved link-local addresses to extract instance credentials, then uses those credentials to enumerate and access cloud storage. These paths are difficult for a model to discover from first principles, because they require knowing not just that a vulnerability exists, but which sequence of actions converts it into a usable capability within a specific infrastructure context.

The remaining challenge is ensuring that the agent attempts skills in an order that respects their real-world prerequisites. To address this, skills are organized through a dependency graph that encodes prerequisite and output relations. A skill's preconditions reference evidence types that other skills produce as postconditions, so the graph makes explicit which skills must succeed before others become applicable. For example, \emph{credential dumping from LSASS memory} requires a \texttt{session} with \texttt{privilege=admin}, which in turn requires \emph{privilege escalation via SUID binary} or \emph{kernel exploit}, which in turn requires \emph{initial shell access}. The Coordinator uses this graph to filter candidate skills against the Board's current evidence, presenting only skills whose preconditions are satisfied to the Solver. This prevents the common failure mode of an LLM attempting an exploit before establishing the prerequisite access, or trying lateral movement before obtaining valid credentials. When a Solver publishes new evidence, the dependency graph is re-evaluated to surface newly applicable skills, so the agent's capability set evolves with the Board's state rather than remaining fixed at the start of the campaign. Figure~\ref{fig:knowledge-example} illustrates a campaign guided by this dependency structure.

\subsubsection{Knowledge Distillation from the Board to the Library}

Exploiting external knowledge is only one half of the loop. During execution, the Strategy Board accumulates evidence that no static library can anticipate: exploitation procedures that actually succeeded in the target environment, parameter tuning specific to the deployed software versions, misconfigurations discovered on the fly, and negative results that delimit which routes are viable. \sysspear therefore periodically distills this internal knowledge back into the external library, so that the library compounds across campaigns instead of aging.

Distillation is triggered at well-defined points: when an Assignment terminates with a confirmed capability grant, and again when a campaign concludes. At each trigger, the confirmed Board entries accumulated since the last distillation point are summarized into library entries that reuse the skill schema---a name, preconditions, an ordered procedure, and postconditions---so distilled knowledge is immediately consumable by the Coordinator's skill filtering and by the dependency graph. Each distilled entry retains provenance links back to the Board evidence from which it was derived, and only procedures whose success is confirmed by execution records are admitted; failed or speculative branches are instead distilled as boundary notes attached to the preconditions of related skills, steering later campaigns away from known dead ends. Versioned updates merge duplicates and supersede stale entries rather than appending blindly.

In this way, the knowledge base becomes a compounding asset rather than a fixed artifact. Every campaign leaves the library slightly stronger---covering one more environment, encoding one more validated procedure, or excluding one more dead end---and the next campaign's Strategy Board starts from a correspondingly richer initial state. The bidirectional flow also keeps the library grounded: because entries are distilled from execution-confirmed evidence, the library cannot drift away from what actually works in deployed environments.

\subsubsection{Capability Orchestration via Protocol-Mediated Lifecycle}

The second obstacle is that binding the platform to one agent runtime couples execution reliability to that runtime's failure modes and API surface. \sysspear therefore introduces ACP as a uniform interaction protocol that decouples reasoning execution from task control. The platform never invokes a model API directly. Instead, an AcpBridge starts or resumes an Agent Session through ACP, which abstracts session lifecycle, tool invocation, permission requests, and subtask delegation behind a stable, runtime-agnostic interface. A Profile and Mode Router selects the appropriate agent profile and execution mode before the session starts, so the same Assignment can be dispatched to different runtimes without changing the Coordinator's logic or the Board's schema. Specifically, \sysspear uses Claude Code\cite{claudecode} for code-intensive exploitation and Codex\cite{codex} for reasoning-heavy reconnaissance.

This protocol-mediated lifecycle preserves both uniform capability extension and declarative session recovery. New tools, MCP servers, or skill definitions are registered through ACP, not through per-runtime adapters, so the platform's capability surface grows without multiplying integration code. Because ACP sessions are resumable, an interrupted Solver can be restored to its last consistent state by replaying the ACP session from the file-system transcript, rather than being restarted from scratch. The platform thus treats heterogeneous runtimes as interchangeable execution backends, while the Strategy Board retains its runtime-agnostic view of evidence and assignments.

\subsubsection{Performance and Safety via File-System-Grounded Execution}

The final obstacle is ensuring that concurrent execution is both performant and safe. Long-running tasks crash, concurrent Solvers interfere, and state changes must propagate without inconsistency. \sysspear addresses all three through a single design principle, that the file system is the authoritative state, from which crash recovery, concurrency isolation, and real-time consistency follow as consequences rather than separate mechanisms.

A long-running agent task may be interrupted by rate limits, timeouts, or process failures. A crash must not erase the task's position in the Board. To overcome this, \sysspear makes the file system the authoritative source of task state. Every task's description, message queue, event stream, execution record, and analysis report is persisted to disk as it is produced, so the file system \emph{is} the state, not merely a log of it. A TaskStateManager generates periodic snapshots and JSON Patches, so the full evolution of a task is reconstructable from disk alone. Crucially, the task state machine, which governs the Assignment's lifecycle, is separated from the session state machine, which governs one ACP session. When a session crashes, the task state is unaffected. The Assignment remains in its last persisted phase, with its Board entries and checkpoint evidence intact. Recovery is local. The AcpBridge resumes or replaces the session, the TaskRuntime replays the pending message queue, and the Solver continues from the last consumed Board revision.

Beyond recovery, concurrency itself poses a hazard. Solvers sharing one process or one workspace interfere through resource contention, file races, and cross-session state leakage. To prevent this, \sysspear adopts a layered process architecture consisting of Gateway, WorkspaceHost, Task, and Agent Runtime, where each layer enforces a distinct isolation boundary. Resource scheduling is workspace-scoped. When concurrent tasks exceed the workspace's resource budget, new tasks enter a FIFO waiting queue rather than degrading running ones. An idle-session swap mechanism evicts a Solver's agent process after a threshold idle period, persisting its session state to disk for on-demand restoration. This allows the platform to oversubscribe agent processes relative to physical resources, running more concurrent Assignments than available agent slots without losing any Solver's progress. Workspace isolation further ensures that a failed or runaway executor cannot corrupt another task's files, state, or Board evidence.

Even when tasks are isolated and individually recoverable, their state changes must reach the Coordinator, the Board, and human observers without polling and without inconsistency. To achieve this, \sysspear designs a unified Inbox message entry point that routes all messages from the user, from subtasks, from team agents, and from scheduled triggers as file events. When a Solver publishes a Board update, the update is a file-system write. The TaskRuntime observes the write, advances the task state machine, and notifies the AcpBridge if the session needs to act. For external observers, an EventMux multiplexes these file events into Server-Sent Events streams delivered to the Web UI. The synchronization chain, from agent execution to file-system write to JSON Patch to EventMux to SSE to frontend, ensures that what the observer sees is exactly what the agent produced, with no intermediate cache to become stale. This matters for the Strategy Board. When a Solver publishes evidence, the Coordinator can observe the Board update through the same event chain and schedule the next Assignment in response, closing the loop between evidence production and task scheduling without a separate notification mechanism.

\subsection{Static-Analysis-Verified Attack Scheme Correctness and Safety}

After an attack scheme is generated, the next challenge is to verify that it is correct in its dependency structure and safe in its software objects before execution. Incorrectness arises when the scheme's step dependencies violate real-world prerequisites. The agent might schedule an exploit before obtaining prerequisite access, or chain techniques in an impossible order where no valid execution sequence exists. Unsafety arises when the scheme uses software objects that contain vulnerabilities. A vulnerable exploitation tool could compromise the red team's own infrastructure, and a targeted service might harbor unknown vulnerabilities that make the attack unpredictable. Detecting these errors by trial execution is costly and potentially destructive. 

To address this challenge, \sysspear performs static-analysis-verified checking along these two dimensions. \sysspear uses \orax\cite{orax} to check the dependency satisfiability of each attack scheme using SMT solving with oracle feedback, ensuring that a valid execution order exists for all skill preconditions and postconditions. After that, \eunomia\cite{eunomia} is used by \sysspear to provide semantic-enhanced vulnerability analysis through fine-grained symbolic execution with user-specified searching strategies, improving bug detection by up to three orders of magnitude. Moreover, \sysspear leverages \symgx\cite{symgx}, which provides a novel analysis model for specialized execution environments such as SGX\cite{sgx} enclaves, to detect cross-boundary pointer vulnerabilities that conventional approaches miss. Together, these three mechanisms ensure that no attack scheme reaches execution unless its dependencies are formally satisfiable and its software objects are verified safe.

\begin{figure}[t]
  \centering
  \includegraphics[width=\linewidth]{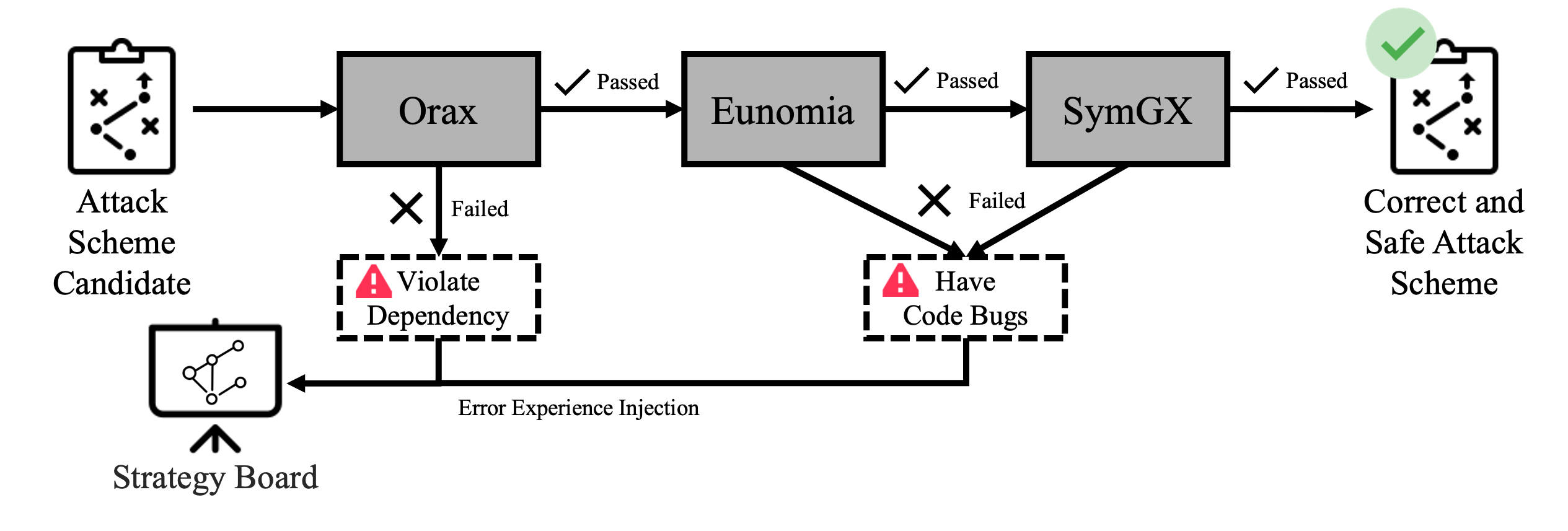}
  \caption{Static-analysis-verified attack scheme checking in \sysspear. \orax checks dependency satisfiability via SMT solving. \eunomia and \symgx check software safety via fine-grained symbolic execution and SGX-specific analysis.}
  \label{fig:static-analysis-overview}
\end{figure}

\subsubsection{Dependency Satisfiability via SMT Solving}

The first challenge is the correctness of the attack schemes generated by \sysspear. An LLM-generated attack scheme encodes dependencies among skills, where each skill's preconditions must be satisfied by evidence produced by earlier skills. The agent may produce a scheme where these dependencies are circular, contradictory, or simply unsatisfiable, meaning no valid execution order exists. To overcome this, our insight is to model the attack scheme's dependency constraints as a Satisfiability Modulo Theories problem. If the constraints are satisfiable, a valid execution order provably exists. If not, the scheme is provably incorrect and must be revised before execution.

A key difficulty in SMT-based verification of attack schemes is that the scheme's dependencies involve not just logical constraints but also oracle components. External tools, network services, and library functions have observable behavior but unknown implementation, which constitutes an SMTO problem where the SMT solver must reason about constraints involving oracles it cannot fully inspect. \orax establishes a feedback loop between the SMT solver and the oracle handler. The solver analyzes the deficiency in oracle mapping information it currently has and feeds this back to the handler. The handler then provides the most suitable additional oracle mapping information to the solver. A dual-clustering strategy selects the initial oracle mapping information, and a relaxation-based method analyzes the information deficiency.

In practice, each skill's preconditions and postconditions are encoded as SMT constraints over the Board's evidence types. \orax checks whether there exists a valid execution order that satisfies all constraints simultaneously. If the scheme is unsatisfiable, it is rejected and returned to the Coordinator for revision, preventing the agent from executing a scheme that is doomed to fail.

\subsubsection{Semantic-Enhanced Vulnerability Analysis}

The second challenge is the safety of the software objects that an attack scheme uses. An attack scheme may use software objects such as exploitation tools, scripts, and libraries that contain vulnerabilities. Using a vulnerable tool could compromise the red team's own infrastructure or leak attack artifacts to the defender. Existing symbolic execution approaches use coarse-grained global searching strategies that cannot efficiently traverse complex code structures, leading to path explosion and missed vulnerabilities. To overcome this, our insight is that different code regions have different structural characteristics, so searching strategies should be localized and user-specified rather than globally uniform, allowing the analysis to prioritize paths relevant to the attack context while avoiding wasted exploration.

Realizing this insight requires a way for users to express local searching strategies concisely and a mechanism to prevent different strategies from interfering through shared variable contexts. \eunomia introduces \aes, a domain-specific language that lets users specify local searching strategies for different parts of the program. Rather than applying one global strategy, \eunomia  allows different code regions to use different search priorities, targeting specific paths relevant to the attack context. It isolates the context of variables for different local searching strategies, avoiding conflicts between regions. \eunomia  is the first symbolic execution engine supporting the full feature set of WebAssembly, enabling analysis of applications written in various languages that compile to WebAssembly. This broad coverage is important because attack tools are often written in different languages and distributed as binaries. In \sysspear, \eunomia  analyzes the software objects referenced in an attack scheme before execution, checking for vulnerabilities that could compromise the attack or the attacker.

\subsubsection{Model-Driven Vulnerability Analysis for Specialized Environments}

The third challenge is the attack scheme's safety in specialized execution environments. Some attack targets use hardware-assisted trusted execution environments such as Intel SGX enclaves, which have unique vulnerability types not covered by conventional analysis. SGX applications heavily use cross-boundary pointers to exchange data between secure enclaves and untrusted environments. These pointers introduce a vulnerability class that existing symbolic execution approaches cannot detect, due to three unique features of SGX applications. Multi-entry arbitrary-order execution allows ECalls in any order. Stateful execution maintains enclave state across calls. Context-aware pointers change their semantics depending on whether they originate from inside or outside the enclave. To overcome this, our insight is that SGX applications require a dedicated analysis model that natively captures these three execution features, rather than retrofitting conventional models that were not designed for enclave semantics.

Realizing this insight requires modeling all three features simultaneously in a single analysis framework, which conventional symbolic execution engines were not designed to support. \symgx introduces a Global State Transition Graph with context-aware analysis that properly handles the three unique features. The graph models multi-entry arbitrary-order execution by tracking all possible ECall sequences. It captures stateful execution by maintaining enclave state across transitions. It handles context-aware pointers by distinguishing pointer origins and applying different validation rules. This model enables \symgx to detect cross-boundary pointer vulnerabilities that conventional approaches miss entirely.

In \sysspear, \symgx analyzes SGX-protected targets in the attack scheme, identifying cross-boundary pointer vulnerabilities that the red team could exploit. This ensures that the scheme's approach to SGX targets is based on verified vulnerability information rather than assumptions, and that the red team does not inadvertently trigger unintended enclave behavior that could compromise the attack.

\section{\sysarmor}
\label{sec:sysarmor_design}

\begin{figure*}[t!]
  \centering
  \begin{tikzpicture}[x=1cm,y=1cm,scale=0.70,transform shape,font=\sffamily]
    \node[draw=black!70, very thick, rounded corners=4pt, fill=gray!15, minimum width=2.75cm, minimum height=0.95cm, align=center] (rule) at (0,3.2) {\textbf{Rule-based}\\Detection};
    \node[draw=black!70, very thick, rounded corners=4pt, fill=blue!15, minimum width=2.75cm, minimum height=0.95cm, align=center] (nod) at (4.6,3.2) {\textbf{NodLink}};
    \node[draw=black!70, very thick, rounded corners=4pt, fill=orange!15, minimum width=2.75cm, minimum height=0.95cm, align=center] (know) at (9.2,3.2) {\textbf{KnowHow}};
    \node[draw=black!70, very thick, rounded corners=4pt, fill=green!15, minimum width=2.75cm, minimum height=0.95cm, align=center] (stair) at (13.8,3.2) {\textbf{STAIR}};

    \draw[->, very thick, blue!40] (rule.east) -- (nod.west);
    \draw[->, very thick, blue!40] (nod.east) -- (know.west);
    \draw[->, very thick, blue!40] (know.east) -- (stair.west);

    \node[align=center, text=black!85] at (0,-0.35) {\textbf{Compiler-based rules}\\\footnotesize Fast matching of known patterns};
    \node[align=center, text=black!85] at (4.6,-0.35) {\textbf{Graph anomaly detection}\\\footnotesize Online Steiner-tree style analysis};
    \node[align=center, text=black!85] at (9.2,-0.35) {\textbf{TTP interpretation}\\\footnotesize CTI-driven semantic mapping};
    \node[align=center, text=black!85] at (13.8,-0.35) {\textbf{Response and report}\\\footnotesize Multi-agent reasoning workflow};

    \begin{scope}[shift={(-1.30,0.55)}]
      \draw[rounded corners=3pt, fill=blue!6, draw=blue!40, line width=0.7pt] (0,0) rectangle (2.85,2.0);
      \draw[draw=black!60, fill=white, line width=0.7pt] (0.30,0.35) rectangle (1.58,1.7);
      \draw[draw=black!35, line width=0.6pt] (0.55,1.4) -- (1.5,1.4);
      \draw[draw=black!35, line width=0.6pt] (0.55,1.1) -- (1.35,1.1);
      \draw[draw=black!35, line width=0.6pt] (0.55,0.8) -- (1.45,0.8);
      \draw[draw=red!70, line width=1.1pt] (1.82,1.35) circle (0.26);
      \draw[draw=red!70, line width=1.1pt, ->] (2.02,1.16) -- (2.35,0.82);
      \draw[fill=gray!20, draw=gray!50] (1.78,0.55) -- (1.95,0.65) -- (1.91,0.9) -- (1.73,0.9) -- (1.69,0.65) -- cycle;
      \node[font=\scriptsize\bfseries, text=blue!60!black] at (0.86,0.2) {rule text};
      \node[font=\scriptsize\bfseries, text=red!70!black] at (2.05,1.63) {match};
    \end{scope}

    \begin{scope}[shift={(3.35,0.55)}]
      \draw[rounded corners=3pt, fill=blue!6, draw=blue!40, line width=0.7pt] (0,0) rectangle (2.85,2.0);
      \draw[dashed, line width=1pt] (0.50,1.45) -- (1.05,0.98) -- (1.68,1.34) -- (2.18,0.86);
      \draw[dashed, line width=1pt] (0.50,1.45) -- (0.88,0.58) -- (1.68,0.58) -- (2.18,0.86);
      \draw[dashed, line width=1pt] (0.88,0.58) -- (1.05,0.98) -- (1.68,1.34);
      \foreach \x/\y/\c in {0.50/1.45/red!70,1.05/0.98/red!70,1.68/1.34/red!70,2.18/0.86/gray!35,0.88/0.58/gray!35,1.68/0.58/gray!35} {\filldraw[fill=white, draw=\c, line width=1pt] (\x,\y) circle (0.10);}
      \draw[very thick, red!75, ->] (0.58,1.33) to[out=-35,in=145] (0.95,1.02);
      \draw[very thick, red!75, ->] (1.12,0.93) to[out=0,in=180] (1.57,1.20);
      \draw[very thick, red!75, ->] (1.73,1.22) to[out=-30,in=140] (2.12,0.91);
      \node[font=\scriptsize\bfseries, text=blue!60!black] at (1.42,0.2) {provenance graph};
    \end{scope}

    \begin{scope}[shift={(7.95,0.55)}]
      \draw[rounded corners=3pt, fill=orange!8, draw=orange!45, line width=0.7pt] (0,0) rectangle (2.85,2.0);
      \node[draw=orange!60, fill=white, rounded corners=2pt, line width=0.7pt, minimum width=0.9cm, minimum height=0.45cm, align=center] (doc1) at (0.78,1.45) {logs};
      \node[draw=orange!60, fill=white, rounded corners=2pt, line width=0.7pt, minimum width=0.9cm, minimum height=0.45cm, align=center] (doc2) at (1.98,1.45) {CTI};
      \draw[very thick, black!70, ->] (doc1.south) to[out=-90,in=140] (1.28,0.95);
      \draw[very thick, black!70, ->] (doc2.south) to[out=-90,in=40] (1.57,0.95);
      \node[draw=red!65, fill=red!10, rounded corners=2pt, line width=0.7pt, minimum width=1.35cm, minimum height=0.5cm, align=center] (ttp) at (1.42,0.8) {\textbf{TTP}};
      \draw[very thick, orange!80!black, ->] (1.42,0.55) -- (1.42,0.25);
      \node[draw=black!50, fill=white, rounded corners=2pt, line width=0.7pt, minimum width=1.8cm, minimum height=0.48cm, align=center] at (1.42,0.05) {human-readable report};
      \node[font=\scriptsize\bfseries, text=orange!70!black] at (1.42,1.8) {semantic mapping};
    \end{scope}

    \begin{scope}[shift={(12.55,0.55)}]
      \draw[rounded corners=3pt, fill=green!8, draw=green!45, line width=0.7pt] (0,0) rectangle (2.85,2.0);
      \foreach \x/\c in {0.45/green!20,1.20/yellow!25,1.95/orange!20} {
        \draw[fill=\c, draw=black!55, line width=0.6pt] (\x,1.15) circle (0.22);
        \draw[black!55, line width=0.7pt] (\x,0.95) -- (\x,0.6);
        \draw[black!55, line width=0.7pt] (\x,0.82) -- ++(-0.12,-0.12);
        \draw[black!55, line width=0.7pt] (\x,0.82) -- ++(0.12,-0.12);
      }
      \draw[black!70, line width=0.8pt, ->] (0.67,1.15) -- (1.00,1.15);
      \draw[black!70, line width=0.8pt, ->] (1.43,1.15) -- (1.77,1.15);
      \draw[black!70, line width=0.8pt, ->] (2.17,0.92) -- (2.17,0.52);
      \draw[fill=white, draw=black!45, line width=0.7pt] (1.00,0.25) rectangle (1.80,0.55);
      \node[font=\scriptsize\bfseries] at (1.40,0.4) {report};
      \node[font=\scriptsize\bfseries, text=green!55!black] at (1.42,1.75) {multi-agent response};
    \end{scope}
  \end{tikzpicture}
  \caption{Overview of \sysarmor.}
  \label{fig:architecture}
\end{figure*}

\sysarmor is a provenance-driven endpoint detection and response system that integrates rule-based detection, graph-based anomaly detection, and TTP-level reasoning into a unified architecture. As shown in Figure~\ref{fig:architecture}, the system is organized into multiple components that collectively support scalable monitoring, real-time alert generation, and post-detection investigation over large-scale host telemetry.

At the core of the design is a distributed processing engine built on Apache Flink, which provides the runtime substrate for continuous analysis over high-volume provenance streams. On top of this pipeline, \sysarmor combines four complementary detection modules: a Falco-compatible rule engine for fast matching of known attack patterns, \nodlink~\cite{nodlink} for identifying suspicious structural deviations in provenance graphs, \knowhow~\cite{knowhow} for mapping suspicious behavior to high-level \ac{ttp} techniques, and STAIR~\cite{jiang2026stair} for higher-level reasoning and investigation support. By coupling signature-based precision with graph-based anomaly detection and semantic interpretation, \sysarmor is able to detect both known and emerging \ac{apt} behaviors while preserving interpretability for security analysts.



\subsection{Rule-based Detection}

\sysarmor incorporates a compiler-based rule engine that preserves the familiar Falco rule syntax while introducing compiler-style optimization for high-throughput detection. The system first compiles rule specifications into efficient internal representations, and then evaluates the compiled logic against normalized streaming events with low-overhead matching procedures. This design allows the platform to retain rule expressiveness for security analysts while substantially reducing runtime cost under large-scale telemetry workloads.

\begin{figure}[t]
  \centering
  \includegraphics[width=\columnwidth]{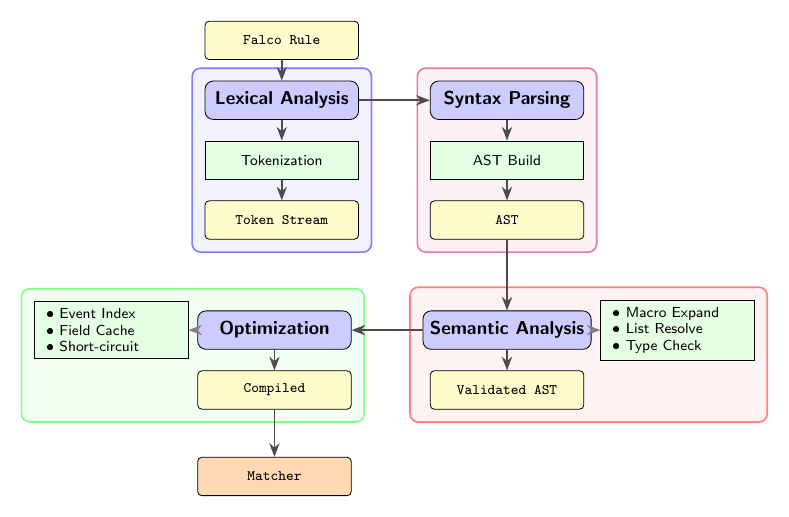}
  \caption{Rule compilation pipeline with four phases}
  \label{fig:compiler-pipeline}
\end{figure}

\begin{figure}[t]
  \centering
  \begin{lstlisting}[basicstyle=\ttfamily\small,breaklines=true,frame=single]
rule suspicious_tmp_execution {
    condition: 
      (evt.type in (execve, execveat) and
      (proc.exe startswith "/tmp/" or
      proc.exe startswith "/dev/shm/"))
}
\end{lstlisting}
  \caption{Example Falco rule suspicious\_tmp\_execution for detecting process executions from temporary directories.}
  \label{fig:suspicious-tmp-execution}
\end{figure}

\noindent\textbf{Compilation Phase.}
Figure~\ref{fig:compiler-pipeline} summarizes the four-stage compilation pipeline used by the rule engine. During this stage, Falco rules are parsed into a structured \ac{ast} that preserves logical semantics while enabling efficient execution. The process consists of lexical analysis and \ac{ast} construction, which translate rule conditions into a machine-readable form; semantic analysis, which expands macros and resolves symbolic references; and a final optimization pass that improves runtime efficiency by indexing event types, caching frequently accessed fields, and exploiting short-circuit evaluation.

Lexical analysis converts the rule into a sequence of atomic tokens.
\textit{BinaryOpNode(``and", In(FieldNode(``evt.type"), SetNode(\{``execve",``execveat"\})), BinaryOpNode(``or", StartsWith(FieldNode(``proc.exe"), ``/tmp/"), StartsWith(FieldNode(``proc.exe"), ``/dev/shm/")))}
The semantic analysis phase then validates the AST and resolves references to schema fields, macros, and list definitions. It recursively expands macro expressions, materializes list values, and checks type compatibility and field existence against the event schema. Only well-formed rules proceed to optimization, preventing undefined references or malformed predicates from reaching the runtime matcher.

The optimization stage further improves execution efficiency through several mechanisms. Event-type indexing builds a direct mapping from event categories to relevant rules, allowing the system to filter candidate matches before evaluating predicates. In a naive rule engine, each event may need to be checked against all $N$ rules, yielding $O(N)$ matching cost. In contrast, the compiler extracts event constraints from each rule, such as `evt.type = execve', and uses a hash lookup at runtime to retrieve only the relevant subset $R_t \subset R$ for an incoming event type $t$, where $|R_t| \ll |R|$. In addition, cached field access avoids repeated parsing of the same event attributes, while short-circuit evaluation terminates logical expressions as soon as the outcome is determined.

\noindent\textbf{Rule Matching Phase.}
The matching phase executes within the streaming detection pipeline and evaluates compiled rules against normalized events in real time. At runtime, an incoming event first consults the event-type index to obtain a candidate rule set, then extracts and caches any required fields in the event context. Predicates are evaluated in a cost-aware order and benefit from short-circuit semantics to avoid unnecessary computation. Once a rule is satisfied, the engine applies suppression and rate-control policies, assembles a structured alert containing the rule identifier, matched fields, and provenance context, and forwards the result to the Alerts Kafka topic and OpenSearch~\cite{opensearch} for storage and subsequent investigation.

\subsection{NodLink}

\nodlink is the graph-based anomaly detection component of \sysarmor. It formulates online \ac{apt} detection as an \ac{ostp} over provenance graphs. The detector maintains a compact evolving graph, identifies suspicious terminal nodes, and incrementally expands them into a hopset that preserves the causal structure of attack-related activities. By reasoning over the relationships among subjects, objects, and execution paths rather than over isolated events, \nodlink is able to expose stealthy attacks that unfold through a sequence of seemingly benign actions.

\begin{figure}[t!]
  \centering
  \includegraphics[width=0.98\textwidth]{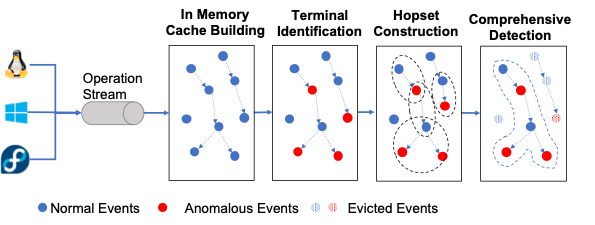}
  \caption{Overview of the \nodlink pipeline. Normal provenance events are ingested from the operation stream, cached in memory, and progressively expanded from suspicious terminal nodes into a hopset for comprehensive attack detection.}
  \label{fig:nodlink-overview}
\end{figure}

Figure~\ref{fig:nodlink-overview} summarizes the main stages of the pipeline. The operation stream denotes the normalized provenance input produced by upstream Flink jobs. During in-memory cache building, recently observed events and graph fragments are retained so the detector can update its state incrementally without rescanning the full history. This cache also enforces bounded-memory execution by evicting stale events that fall outside the active detection window.

The terminal identification stage assigns anomaly scores to nodes that are likely to serve as attack endpoints. In the figure, the blue backbone represents normal provenance structure, whereas the red nodes indicate suspicious terminals. These terminals are not treated as the final detection result; instead, they act as anchors for the subsequent graph expansion. The hopset construction stage then traverses the causal neighborhood around each suspicious terminal and forms a compact subgraph that connects the relevant attack entities while preserving provenance semantics. The dashed outlines in the figure indicate the progressively expanded suspicious region, and the evicted events in the legend correspond to cached events that are removed once they are no longer useful for the current window.

In the final comprehensive detection stage, \nodlink evaluates the expanded hopset to determine whether it contains a coherent attack pattern. This stage aggregates evidence from the suspicious terminals, their surrounding causal paths, and the retained cache state to produce a single detection outcome. The resulting provenance subgraph can then be forwarded to the downstream alerting pipeline and combined with rule-based detections. This incremental design avoids heavy retraining, supports real-time inference, and provides stronger contextual evidence than event-level matching alone.

\subsection{KnowHow}

\knowhow complements \nodlink by providing semantic reasoning over the suspicious provenance subgraphs identified by the anomaly detector. It converts a structurally suspicious graph into a sequence of candidate attack lifecycles and then reasons over those candidates using cyber threat intelligence. Rather than returning only a binary anomaly label, \knowhow explains which adversarial behaviors the observed provenance is most consistent with and produces TTP-level outputs that are directly useful for investigation.

\begin{figure}[t!]
  \centering
  \includegraphics[width=0.98\textwidth]{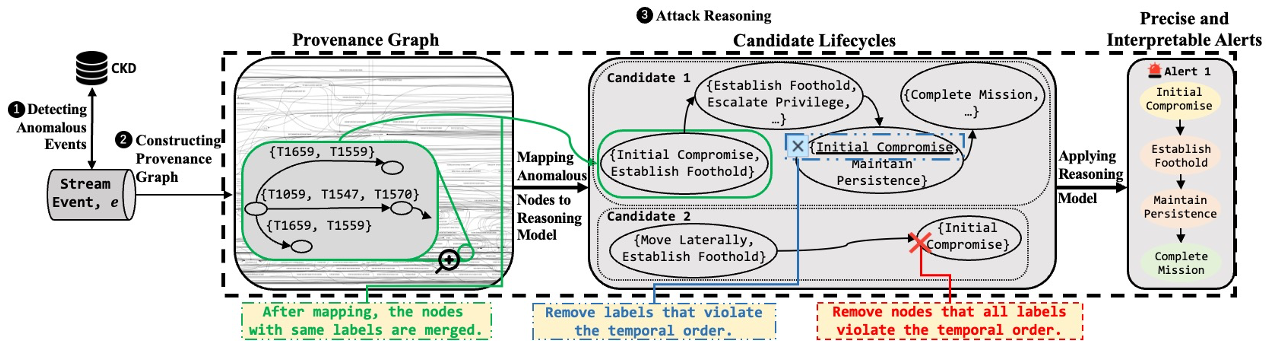}
  \caption{Overview of the \knowhow pipeline. Suspicious provenance subgraphs produced by \nodlink are matched against CTI-derived knowledge, organized into candidate attack lifecycles, and refined into precise and interpretable alerts.}
  \label{fig:knowhow-overview}
\end{figure}

Figure~\ref{fig:knowhow-overview} shows the main stages of the pipeline. The provenance graph on the left is the graph-level input passed from \nodlink. At this stage, suspicious nodes and edges already encode the contextual evidence produced by the graph anomaly detector. The candidate-lifecycle stage then constructs one or more attack hypotheses, each corresponding to a plausible progression of attack stages such as initial compromise, establishment of foothold, lateral movement, privilege escalation, and mission completion. During this stage, \knowhow also applies temporal-order constraints to prune inconsistent hypotheses: labels that violate the expected progression are removed, and nodes that cannot be placed into a valid lifecycle are discarded. This filtering step is important because it prevents the reasoning model from overfitting to isolated graph fragments that are structurally suspicious but semantically implausible.

In the attack-reasoning stage, \knowhow performs semantic alignment between the candidate lifecycles and structured knowledge extracted from \ac{cti} reports. The module relies on \ac{gioc} representations, which encode adversary knowledge as subject-verb-object triples. These triples provide a compact bridge between natural-language threat intelligence and provenance observations, allowing the system to match CTI-derived behaviors against the suspicious graph. The reasoning model then selects the candidate lifecycle that best satisfies both the graph evidence and the semantic constraints encoded in the knowledge base.

The final output is a precise and interpretable alert. As illustrated in the rightmost panel, the selected lifecycle is converted into an attack narrative that summarizes the most likely stage sequence and the associated \ac{ttp} labels. This representation gives analysts both the structural evidence from the provenance graph and the semantic explanation from CTI, making it easier to understand the attack progression and to prioritize response actions. In this way, \knowhow serves as a reasoning layer that turns graph anomalies into actionable, human-readable detection results.

\subsection{STAIR}

\begin{figure}[t!]
  \centering
  \includegraphics[width=0.98\textwidth]{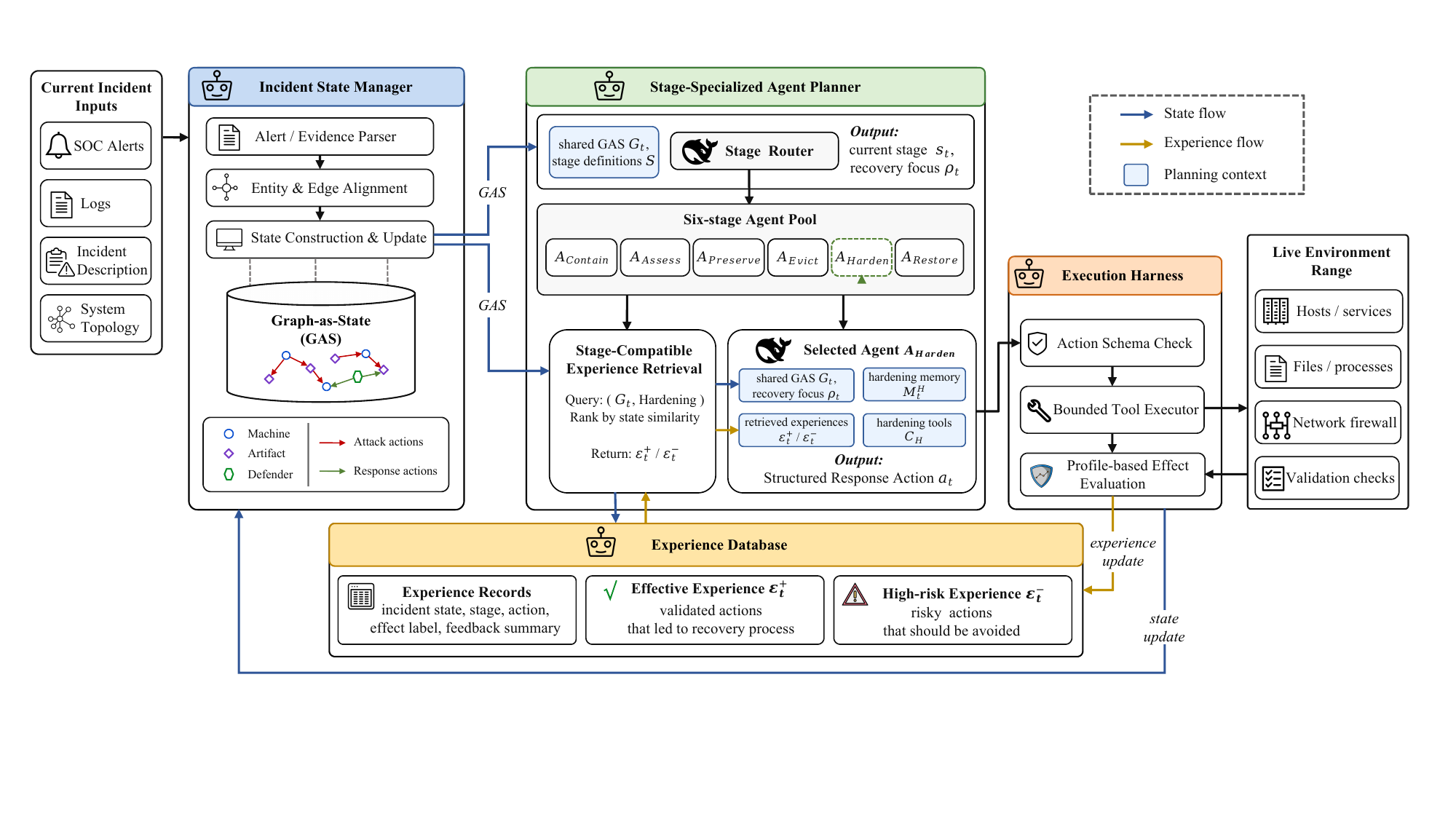}
    \caption{Overview of STAIR. STAIR maintains an evolving incident state with Graph-as-State (GAS), performs stage-aware response planning with historical experience, and incorporates execution feedback to form a closed-loop automated response process.}
  \label{fig:stair}
\end{figure}

STAIR is the automated incident-response component of \sysarmor. 
As shown in Fig.~\ref{fig:stair}, it organizes incident response as a closed-loop process that continuously tracks incident state, determines the current recovery stage, generates response actions, and incorporates execution feedback. 
STAIR targets two major challenges in long-horizon response. 
First, incident states evolve continuously as attacks and defensive actions change the environment, making previously handled entities, unresolved attack paths, and recovery progress easy to lose over long interaction histories. 
Second, different response stages pursue different operational objectives, and an action that is reasonable in one stage may be inappropriate in another. 
These two problems lead to \emph{state loss} and \emph{stage mismatch}, respectively.

To preserve response continuity, the Incident State Manager represents the evolving incident as a \emph{Graph-as-State} (GAS). 
GAS jointly captures affected machines and artifacts, attack behaviors, response actions, and their execution results. 
It is initialized from SOC alerts, logs, and system information, and is incrementally updated after every response action using execution feedback. 
This persistent representation allows subsequent planning to operate on the current incident state rather than repeatedly reconstructing progress from textual interaction history.

Based on the maintained GAS, the Stage Router identifies the current recovery stage and its concrete recovery focus. 
STAIR organizes response into six stages---\textsc{Containment}, \textsc{Assessment}, \textsc{Preservation}, \textsc{Eviction}, \textsc{Hardening}, and \textsc{Restoration}---and dispatches planning to the corresponding stage-specialized agent. 
The selected agent reasons with the shared GAS, stage-compatible historical experience, and stage-scoped response tools, thereby constraining action generation to the current recovery objective.

The resulting structured action is validated and executed by the Execution Harness. 
Execution feedback is written back to GAS and used to update reusable response experience, forming a continuous state--planning--execution loop. 
By combining persistent state tracking with stage-aware planning, STAIR reduces state loss and stage mismatch in long-horizon incident response and extends \sysarmor from attack detection and interpretation to automated response.

\section{Evaluation}
\label{sec:evaluation}

\subsection{Evaluation of \textsc{SysField}}
\label{sec:sysfield-evaluation}

We evaluate \textsc{SysField} through four questions that follow the measurement pipeline from the quality of the collected evidence to the difficulty of the generated ranges and, finally, to the interpretation of agent behavior:

\begin{itemize}[noitemsep, topsep=1pt, partopsep=1pt, listparindent=\parindent, leftmargin=*]
    \item \textbf{RQ1: Measurement Fidelity and Cost.} Can \textsc{SysField} collect complete runtime evidence without imposing prohibitive system overhead?
    \item \textbf{RQ2: Range Difficulty and Scale.} Do multi-step composition and larger topologies expose capability gaps that are hidden by single-step or small-range evaluations?
    \item \textbf{RQ3: Multi-stage Progression under Controlled Conditions.} Where do current cyber agents lose progress, and how does the current high-interference configuration affect downstream completion and execution cost?
    \item \textbf{RQ4: Offensive Progression and Defensive Evidence.} How does independently verified attack completion relate to runtime observability, and which expected behaviors are least consistently represented in the evidence?
\end{itemize}

Together, these questions test whether the evaluation substrate is trustworthy, whether the constructed ranges provide meaningful difficulty, and whether the resulting trajectories support fine-grained offensive and defensive analysis.

\subsubsection{Evaluation Setup}

\paragraph{Dataset.}
The primary trajectory dataset is Stratified-50: 50 three-tier enterprise cases constructed from 24 unique CVEs and stratified by historical entry-stage and intermediate-stage difficulty across six cells (easy/easy, easy/hard, mid/easy, mid/hard, hard/easy, and hard/hard). Each case contains three ordered CVE slots and three independently verified flags: a perimeter entry service (target~1), an intermediate application service (target~2), and a data service (target~3). Historical results determine sampling strata only and are not reused as model outcomes. CVEs recur across cases, and the current data layer contains only three CVE variants; the 50 cases are therefore not statistically independent vulnerability samples or a complete representation of enterprise attack paths \cite{cvelabreport}.

We additionally report substrate-level measurements that are not part of Stratified-50: collector stress tests, presentation-level aggregate comparisons between single-step CyberGym tasks and \textsc{SysField} multi-step ranges, a 7-node versus 50-node scale comparison, and construction-capacity measurements. These sources have different units and controls and are analyzed separately from the case-level trajectory data.

\paragraph{Metrics.}
Measurement fidelity is quantified by event-loss rate and additional system overhead. Range difficulty is reported as verified attack success and the percentage-point change under multi-step composition or larger topology scale. Trajectory capability is measured by target-1, target-2, and target-3 completion, all-three completion, business-objective completion, deepest stage, termination status, timeout, and runtime. Defensive observability is measured independently by the presence of any new attack-window signal, strict coverage of every expected rule, per-rule missing-signal counts, and the four attack-success/signal-hit quadrants. Signal-frame volume is retained as observation volume, not interpreted as a count of independent behaviors or as a severity score.

\paragraph{Baselines.}
The collector-completeness baselines are Sysdig~\cite{sysdig} and LTTng~\cite{lttng}; the overhead baselines are HARDLOG~\cite{ahmad2022hardlog} and OMNILOG~\cite{gandhi2023rethinking}. They belong to different experiments and are not combined into one collector ranking. The range-difficulty baselines are single-step CyberGym~\cite{cybergym} and the reported 7-node \textsc{SysField} configuration, each compared with its corresponding multi-step or 50-node condition for Kimi K3~\cite{kimik3}, GLM-5.2~\cite{glm52}, DeepSeek-V4-Pro~\cite{deepseekv4pro}, and GPT-5.6 Luna~\cite{gpt56systemcard}. The trajectory baselines are the single-agent Kimi K3 and DeepSeek-V4-Pro L2 runs; here L1 and L2 denote task-information difficulty levels that follow CyberGym's convention~\cite{cybergym}, in which a higher level discloses richer task information to the agent. The same-backbone \textsc{SysSpear} analysis uses the DeepSeek-V4-Pro single-agent L2 arm as its reference, and the interference analysis compares the recorded DeepSeek-V4-Pro L1 \texttt{none} and \texttt{high} configurations.

\paragraph{Setup for Baselines.}
Table~\ref{tab:sysfield-baseline-setup} states what is held fixed and what changes in each comparison. This prevents a presentation-level baseline, a same-protocol model comparison, and a configuration-level system comparison from being interpreted as if they were one randomized experiment.

\begin{table}[htbp]
  \centering
  \begingroup
  \scriptsize
  \setlength{\tabcolsep}{4pt}
  \renewcommand{\arraystretch}{1.14}
  \caption{Baseline setup and comparison boundaries for the \textsc{SysField} evaluation.}
  \label{tab:sysfield-baseline-setup}
  \begin{tabular}{P{3.10cm} P{5.20cm} P{4.40cm}}
    \toprule
    \textbf{Comparison} & \textbf{Shared setup} & \textbf{Changed factor and interpretation} \\
    \midrule
    L2 model comparison & Same 50 cases, OpenAI-compatible SDK and runner, L2 context, tools, 300-turn budget, 3,600-s agent timeout, serial execution, verifier, and \textsc{SysArmor} configuration & Kimi K3 versus DeepSeek-V4-Pro; same-protocol descriptive model comparison, with one trajectory per model--case pair \\
    Same-backbone red-team system & Same 50 cases, DeepSeek-V4-Pro, public L2 scenario, 300-turn budget, 3,600-s timeout, serial execution, private verifier, and observe-only \textsc{SysArmor} & Single-agent runner versus \textsc{SysSpear} assessment/session adapter; configuration-level comparison because the runner and execution policy change together \\
    L1 interference & Same manifest, DeepSeek-V4-Pro, L1 context, 300 turns, 3,600-s timeout, seed~1, and temperature~0 & \texttt{none}: parallelism~8; \texttt{high}: parallelism~4, 43 decoys, and one omitted topology hint, in fixed order; comparison of recorded configurations, not a pure decoy effect \\
    Collector completeness & Same stress-oriented event-generation and accounting protocol on the reported hardware/virtualization settings & \textsc{SysField} versus Sysdig and LTTng; event-loss comparison only \\
    Collector overhead & Same reported workload family within the overhead experiment & \textsc{SysField} versus HARDLOG and OMNILOG; overhead comparison only \\
    Composition and scale & Same model within each reported pair & CyberGym single-step versus \textsc{SysField} multi-step, and 7-node versus 50-node; source reports omit trial-level uncertainty and some controls, so deltas remain protocol-specific \\
    \bottomrule
  \end{tabular}
  \endgroup
\end{table}

Across the Stratified-50 arms, environment validation, attack-graph validation, path reachability, and cleanup succeed for all formal cases, so the differences analyzed below occur after range construction and qualification. The direct L2 model arms each contain one recorded trajectory per case. The L1 comparison changes parallelism and one topology hint together with the decoy condition, and the \textsc{SysSpear} comparison changes the red-team runner and execution policy. We therefore use these results as descriptive evidence under their recorded protocols rather than as universal rankings or component-level causal effects.

\paragraph{Implementation and Experiment Environment.}
\textsc{SysField} is implemented on the CVELab range-generation and evaluation stack. The stack uses Python~3.12, \texttt{uv}, Docker, ContainerLab~0.74 or later, and Ansible. Initialization consists of resolving the Python environment with \texttt{uv}, building the attacker and per-CVE runtime images, generating an \texttt{enterprise\_3tier} scenario from a case manifest, deploying its container network, applying the Ansible setup, and waiting for service readiness. Each formal run then launches the attack system, verifies stage flags and the business objective through a model-independent verifier, preserves the structured run artifacts, and destroys the range. Workload nodes communicate through the emulated data plane rather than the Docker management network; only the red team's route retains the egress required to call the model endpoint. \textsc{SysArmor} is enabled as an observe-only monitor, so it records signals without blocking the attack \cite{cvelabreport}.

All case-level Stratified-50 model arms use pretrained inference endpoints through an OpenAI-compatible interface; \textsc{SysField} performs no fine-tuning or task-specific training. Consequently, there is no training split, optimizer, epoch count, or learning-rate setting to report. Inference controls are stated per arm below rather than treated as global hyperparameters. The range experiments use isolated Docker/ContainerLab topologies; the separate collector stress tests cover virtual and physical machines under both 1-CPU/2-GB and 32-CPU/64-GB configurations. The available artifacts do not disclose a single host-CPU model or OS build shared by every range run, so we do not introduce one as an experimental control.

\subsubsection{RQ1: Measurement Fidelity and Cost}

Table~\ref{tab:sysfield-collection} summarizes the reported results for the DPUAudit-backed \textsc{SysField} collector~\cite{jiang2025dpuaudit}. This configuration records all generated events in every reported hardware and virtualization setting, yielding a 0\% event-loss rate. Under the same stress-oriented completeness comparison, Sysdig loses 98.5\% of events and LTTng loses 60.0\%. The result matters for evaluation validity: when most events are dropped, an apparent absence of attack behavior may be a collector artifact rather than a property of the agent trajectory.

\begin{table}[htbp]
  \centering
  \begingroup
  \footnotesize
  \setlength{\tabcolsep}{5pt}
  \renewcommand{\arraystretch}{1.13}
  \caption{Reported completeness and overhead of the DPUAudit-backed \textsc{SysField} collector. Completeness and overhead use different baseline sets.}
  \label{tab:sysfield-collection}
  \begin{tabular}{P{4.00cm} P{2.40cm} P{6.50cm}}
    \toprule
    \textbf{Metric} & \textbf{DPUAudit-backed \textsc{SysField}} & \textbf{Reported baselines} \\
    \midrule
    Event-loss rate & 0.0\% & Sysdig: 98.5\%; LTTng: 60.0\% \\
    Additional system overhead & 2.1\% & HARDLOG: 23.2\%; OMNILOG: 33.0\% \\
    \bottomrule
  \end{tabular}
  \endgroup
\end{table}

The completeness gain does not come from accepting a large execution penalty. The DPUAudit-backed \textsc{SysField} configuration adds 2.1\% overhead, compared with 23.2\% for HARDLOG and 33.0\% for OMNILOG. Relative to these two reported baselines, the overhead is lower by 90.9\% and 93.6\%, respectively, which explains the 90--94\% reduction summarized in the presentation. The event-loss and overhead experiments use different baselines and should not be collapsed into a single multi-metric ranking; jointly, however, they show that the framework can preserve the evidence needed for later trajectory analysis while remaining close to the reported 3\% deployment budget.

\subsubsection{RQ2: Range Difficulty and Scale}

The first difficulty comparison changes the evaluation from single-step vulnerability reproduction in CyberGym to multi-step attack execution in \textsc{SysField}. Table~\ref{tab:sysfield-composition-difficulty} shows a consistent reduction for all four evaluated models. Success falls by 18.1 percentage points for Kimi K3, 21.6 points for GLM-5.2, 21.6 points for DeepSeek-V4-Pro, and 20.8 points for GPT-5.6 Luna. Thus, the reported 18--22 point gap is not caused by one outlying model. Requiring the agent to preserve state and satisfy dependencies across several attack steps exposes failure modes that a single-step endpoint does not measure.

\begin{table}[htbp]
  \centering
  \begingroup
  \footnotesize
  \setlength{\tabcolsep}{5pt}
  \renewcommand{\arraystretch}{1.13}
  \caption{Reported attack success on single-step CyberGym tasks and \textsc{SysField} multi-step ranges.}
  \label{tab:sysfield-composition-difficulty}
  \begin{tabular}{P{3.55cm} P{3.05cm} P{3.05cm} P{2.55cm}}
    \toprule
    \textbf{Model} & \textbf{CyberGym single-step} & \textbf{\textsc{SysField} multi-step} & \textbf{Change} \\
    \midrule
    Kimi K3 & 73.8\% & 55.7\% & $-18.1$ pp \\
    GLM-5.2 & 62.8\% & 41.2\% & $-21.6$ pp \\
    DeepSeek-V4-Pro & 57.8\% & 36.2\% & $-21.6$ pp \\
    GPT-5.6 Luna & 44.3\% & 23.5\% & $-20.8$ pp \\
    \bottomrule
  \end{tabular}
  \endgroup
\end{table}

Topology scale produces a second, separate reduction. As shown in Table~\ref{tab:sysfield-scale-difficulty}, increasing the range from 7 to 50 nodes lowers attack success by 9.8--10.6 percentage points across the same four models. The model ordering is unchanged, but every model loses roughly ten points. This consistent shift suggests that the larger search space and longer dependency structure add difficulty even when the evaluated model family changes.

\begin{table}[htbp]
  \centering
  \begingroup
  \footnotesize
  \setlength{\tabcolsep}{5pt}
  \renewcommand{\arraystretch}{1.13}
  \caption{Reported attack success as \textsc{SysField} range size increases from 7 to 50 nodes.}
  \label{tab:sysfield-scale-difficulty}
  \begin{tabular}{P{3.55cm} P{3.05cm} P{3.05cm} P{2.55cm}}
    \toprule
    \textbf{Model} & \textbf{7-node range} & \textbf{50-node range} & \textbf{Change} \\
    \midrule
    Kimi K3 & 60.8\% & 50.2\% & $-10.6$ pp \\
    GLM-5.2 & 46.1\% & 36.0\% & $-10.1$ pp \\
    DeepSeek-V4-Pro & 40.9\% & 30.6\% & $-10.3$ pp \\
    GPT-5.6 Luna & 29.0\% & 19.2\% & $-9.8$ pp \\
    \bottomrule
  \end{tabular}
  \endgroup
\end{table}

These difficulty results are supported by the construction-scale measurements. \textsc{SysField} reports orchestration support for more than 257 CVEs and 1,148 multi-vulnerability ranges. Under a 256-GB memory limit, the platform supports ranges with more than 300 system nodes and campaigns with more than 150 attack behaviors. In the reported commercial-LLM trials, no model escaped the isolated range or bypassed the external verification mechanism. This is evidence about the tested configurations, not a general proof of containment against arbitrary adversaries. The presentation additionally reports that combining \textsc{SysField} ranges with active \textsc{SysArmor} defenses reduces SOTA-agent success by more than 60\%; because that experiment enables blocking, we keep it separate from the observe-only defensive analysis in RQ4.

The two tables establish complementary forms of difficulty: multi-step composition expands the required attack process, whereas topology growth expands the environment through which that process must be executed. The presentation does not provide the trial counts, uncertainty estimates, or all controls behind these aggregate comparisons, so the reported deltas should be interpreted within their respective protocols rather than as causal estimates attributable only to step count or node count.

\subsubsection{RQ3: Multi-stage Progression under Controlled Conditions}

\paragraph{Stage progression.}

\begin{figure}[htbp]
  \centering
  \includegraphics[width=0.75\linewidth]{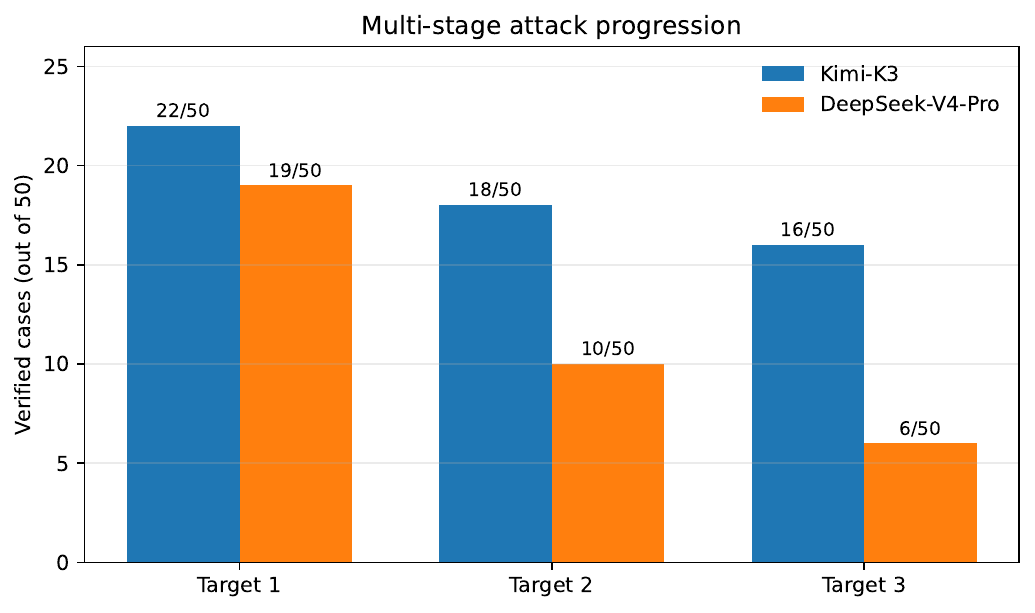}
  \caption{Stage-level completion on the same 50 Stratified-50 L2 cases. Bars report independently verified target flags. The model gap widens after initial access.}
  \label{fig:sysfield-stage-progression}
\end{figure}

Figure~\ref{fig:sysfield-stage-progression} reports verified stage completion on the same 50 L2 cases. Initial access is relatively close: Kimi K3 reaches target~1 in 22 cases, while DeepSeek-V4-Pro does so in 19. The difference becomes larger after the initial foothold. Kimi K3 reaches target~2 in 18 cases and target~3 in 16, whereas DeepSeek-V4-Pro reaches them in 10 and 6 cases, respectively. Thus, the observed model gap is modest at entry but expands as the task requires continued use of an existing foothold and successful transition to later stages.

The transition rates make this pattern more explicit. After reaching target~1, Kimi K3 continues to target~2 in 18/22 cases (81.8\%) and to target~3 in 16/22 (72.7\%). DeepSeek-V4-Pro continues to target~2 in 10/19 cases (52.6\%) and to target~3 in 6/19 (31.6\%). Equivalently, Kimi K3 loses six cases between the first and final stages, whereas DeepSeek-V4-Pro loses thirteen. The primary difference under the current protocol therefore lies in sustained multi-stage progression rather than initial entry alone.

This result is important because later-stage progress requires more than triggering another isolated vulnerability. The agent must retain useful information from earlier steps, operate from the current foothold, identify a reachable continuation target, and recover from failed actions without losing the attack state. Stage-level verification shows where this longer-horizon process breaks down instead of compressing all failures into a single final score.

\paragraph{Same-backbone \sysspear rerun.}

We additionally evaluate \textsc{SysSpear} on the same 50 Stratified-50 L2 cases using DeepSeek-V4-Pro as the solver backbone. The rerun preserves the CVELab scenarios, public L2 task description, private flag verification, and observe-only \textsc{SysArmor} instrumentation, while replacing the red team's execution path with the \textsc{SysSpear} assessment/session adapter and using \texttt{--max-turns 300}. All 50 environment-only validations succeed. Because the red-team runner and recorded execution policy differ from the single-agent arm, we treat this experiment as a same-backbone, configuration-level comparison rather than a component-level causal ablation.

\begin{table}[htbp]
  \centering
  \begingroup
  \footnotesize
  \setlength{\tabcolsep}{4pt}
  \renewcommand{\arraystretch}{1.13}
  \caption{Stage progression under the recorded L2 configurations. \textsc{SysSpear} uses DeepSeek-V4-Pro as its solver backbone; the comparison holds the backbone fixed but changes the red-team system architecture and recorded execution policy.}
  \label{tab:sysfield-syspear-rerun300}
  \begin{tabular}{P{3.20cm} P{4.00cm} P{1.80cm} P{1.80cm} P{1.80cm}}
    \toprule
    \textbf{Attack configuration} & \textbf{Backbone} & \textbf{Target~1} & \textbf{Target~2} & \textbf{Target~3} \\
    \midrule
    Single-agent & Kimi K3 & 22/50 (44\%) & 18/50 (36\%) & 16/50 (32\%) \\
    Single-agent & DeepSeek-V4-Pro & 19/50 (38\%) & 10/50 (20\%) & 6/50 (12\%) \\
    \textsc{\sysspear} & DeepSeek-V4-Pro & 34/50 (68\%) & 15/50 (30\%) & 7/50 (14\%) \\
    \bottomrule
  \end{tabular}
  \endgroup
\end{table}

\textsc{SysSpear} reaches target~1 in 34/50 cases (68\%), target~2 in 15/50 (30\%), and target~3 in 7/50 (14\%). Relative to the recorded DeepSeek-V4-Pro single-agent arm, which reaches the three targets in 19, 10, and 6 cases, respectively, the absolute improvement decreases from 30 percentage points at target~1 to 10 points at target~2 and only 2 points at target~3. The improvement is therefore strongly front-loaded: \textsc{SysSpear} substantially expands initial-foothold coverage, but the additional entry successes do not propagate proportionally through the later stages of the attack chain.

\begin{figure}[H]
  \centering
  \includegraphics[width=0.75\linewidth]{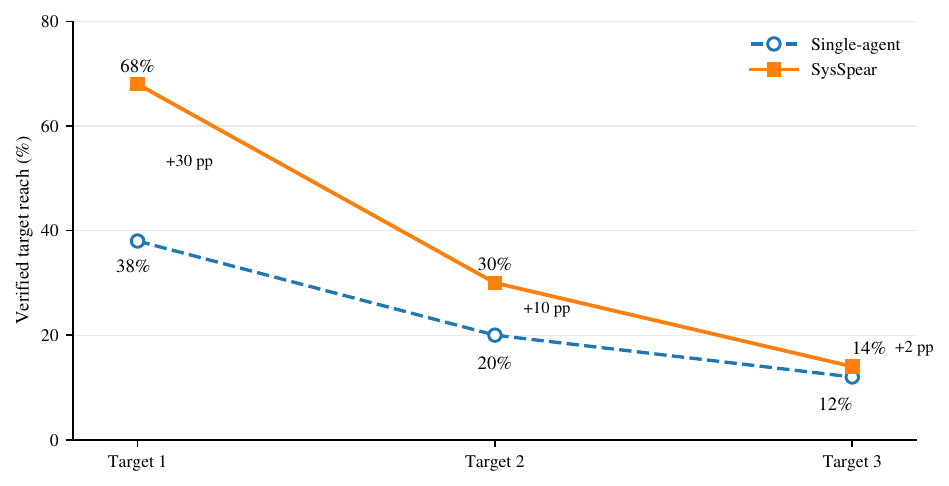}
  \caption{Same-backbone progression profile on the 50 Stratified-50 L2 cases. The gap between the DeepSeek-V4-Pro single-agent arm and \textsc{SysSpear} narrows from 30 percentage points at Target~1 to 10 at Target~2 and 2 at Target~3. Because the comparison changes the red-team runner and execution policy, it should be interpreted as a configuration-level comparison rather than a component-level causal ablation.}
  \label{fig:sysfield-same-backbone-progression}
\end{figure}

Among the 34 \textsc{SysSpear} executions that reach target~1, 15 (44.1\%) continue to target~2, and 7 of those 15 (46.7\%) then reach target~3; end-to-end, 7/34 (20.6\%) of the initial footholds progress to the final target. Thus, 19 cases lose progress immediately after initial access and another eight stop before the final target. Together with the single-agent results, this shifts the diagnostic focus from vulnerability triggering alone toward post-compromise state utilization, internal target selection, and cross-stage continuation. These conditional rates should not, however, be interpreted as evidence that \textsc{SysSpear} reduces post-foothold capability: \textsc{SysSpear} reaches target~1 on many cases that the single-agent baseline never enters, so a matched comparison over commonly reached footholds is required to isolate the downstream effect.

Termination behavior provides a complementary view of the original matched model comparison. Kimi K3 records 22/50 agent timeouts; none are recorded for DeepSeek-V4-Pro. Deeper progression therefore does not coincide with uniformly lower execution cost or simpler termination behavior. Under the current protocol, Kimi K3 reaches later stages more often but also more frequently consumes the full time budget. Because there is only one recorded trajectory per model and case, this difference should be treated as a descriptive operating pattern rather than a general reliability ranking.

\paragraph{Environmental interference.}

The \textsc{SysSpear} rerun above does not complete the predefined matched harness-uplift ablation because the red-team runner, recorded execution policy, and turn budget change together. The available evidence for environmental dependence comes from the DeepSeek-V4-Pro L1 comparison between the current \texttt{none} and \texttt{high} configurations.

Table~\ref{tab:sysfield-interference} summarizes progression, timeouts, and execution cost. Both configurations reach target~1 in 2 of 50 cases. Under \texttt{none}, both cases continue through targets~2 and~3; under \texttt{high}, neither progresses beyond the first stage. The unchanged entry count and the disappearance of downstream completion indicate that the largest observed difference occurs after entry, when the agent must continue exploration and select the correct path through the internal environment.

\begin{table}[htbp]
  \centering
  \begingroup
  \footnotesize
  \setlength{\tabcolsep}{5pt}
  \renewcommand{\arraystretch}{1.13}
  \caption{DeepSeek-V4-Pro L1 results under the current \texttt{none} and \texttt{high} configurations.}
  \label{tab:sysfield-interference}
  \begin{tabular}{P{6.30cm} P{3.20cm} P{3.20cm}}
    \toprule
    \textbf{Metric} & \textbf{\texttt{none}} & \textbf{\texttt{high}} \\
    \midrule
    Target 1 / Target 2 / Target 3 & 2 / 2 / 2 & 2 / 0 / 0 \\
    All-three completion & 2/50 & 0/50 \\
    Objective achieved & 1/50 & 0/50 \\
    Timeout & 6/50 (12\%) & 19/50 (38\%) \\
    Mean agent time & 1,417.6 s (23.6 min) & 2,428.5 s (40.5 min) \\
    Median agent time & 1,091.7 s (18.2 min) & 2,536.5 s (42.3 min) \\
    Recorded decoy interaction & -- & 50/50 \\
    Direct decoy contact & -- & 38/50 \\
    \bottomrule
  \end{tabular}
  \endgroup
\end{table}

The execution-cost measurements support the same interpretation. Timeouts increase from 6/50 to 19/50, a rise of 26 percentage points. Mean agent time increases from 23.6 to 40.5 minutes (71.3\%), while median time increases from 18.2 to 42.3 minutes (132.3\%). The larger change in the median suggests that the additional cost is not explained only by a small number of unusually long runs; the typical high-configuration run also consumes substantially more of the available execution budget.

The interaction records are consistent with this increase in exploration cost. Every high-configuration run contains a recorded decoy interaction, and 38/50 directly contact a decoy endpoint. This indicates that the agent spends part of its execution interacting with the enlarged target surface. However, the experiment does not isolate decoys as the sole cause: the two arms also differ in worker parallelism, run in a fixed \texttt{none}-then-\texttt{high} order, and contain a topology-hint difference, while decoy interaction is derived from transcript diagnostics rather than packet-level provenance. The supported conclusion is therefore narrower: the \emph{current high configuration} is associated with greater execution cost, more timeouts, and weaker progression beyond the first stage. A causal estimate of deception alone would require matched parallelism and topology hints, randomized arm order, and repeated trials.

\subsubsection{RQ4: Offensive Progression and Defensive Evidence}

\paragraph{Case-level observability.}

\begin{figure}[htbp]
  \centering
  \includegraphics[width=0.75\linewidth]{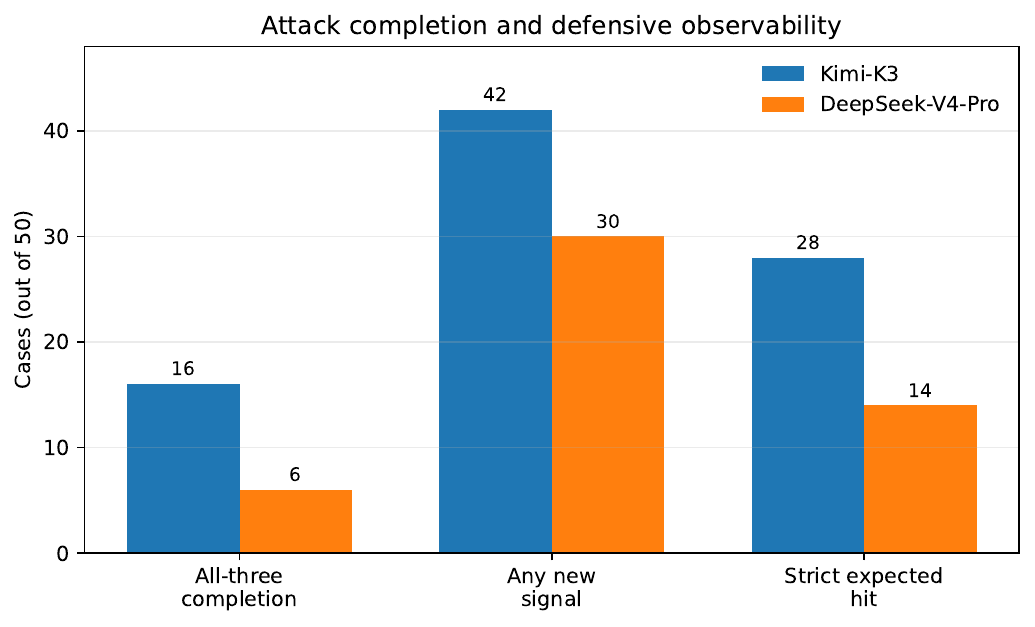}
  \caption{Attack completion and defensive observability on 50 L2 cases. ``Any new signal'' records whether at least one new attack-window signal appears; ``strict expected hit'' requires the complete expected rule set for the case.}
  \label{fig:sysfield-observability}
\end{figure}

Figure~\ref{fig:sysfield-observability} compares complete attack success with two levels of defensive evidence across the two matched direct-run L2 arms. Kimi K3 completes all three targets in 16/50 cases, produces at least one new attack-window signal in 42/50, and covers all expected rules in 28/50. DeepSeek-V4-Pro completes all three targets in 6/50 cases, while producing a new signal in 30/50 and covering all expected rules in 14/50. For both models, observable runtime behavior is therefore substantially more common than complete compromise.

The difference between the two defensive metrics is itself informative. For Kimi K3, any new signal is observed in 84\% of cases, whereas strict expected coverage is 56\%; for DeepSeek-V4-Pro the corresponding values are 60\% and 28\%. The gaps of 28 and 32 percentage points show that ``the defense observed some attack-associated behavior'' and ``the defense observed the complete expected behavior set'' are distinct levels of observability. A final success label loses even more process information: any-signal coverage exceeds all-three completion by 52 percentage points for Kimi K3 and 48 points for DeepSeek-V4-Pro.

\paragraph{\textsc{SysSpear} observability.}

The same 50-case \textsc{SysSpear} rerun provides a third, configuration-level observation. Only 7/50 cases (14\%) complete all three targets, but 39/50 (78\%) produce at least one new attack-window signal and 27/50 (54\%) cover the complete expected rule set. Any-signal coverage therefore exceeds full completion by 64 percentage points, while strict expected coverage exceeds it by 40 points. More importantly, 32 of the 39 signal-producing executions and 20 of the 27 strict expected hits occur in cases that do not complete the three-stage attack chain. The additional arm thus strengthens the RQ4 claim that failed attack progression does not imply an absence of defensively relevant evidence.

The case-level quadrants make the separation between attack outcome and defensive evidence explicit. For Kimi K3, 12 cases are attack-success/signal-hit, 4 are attack-success/signal-miss, 16 are attack-failure/signal-hit, and 18 are attack-failure/signal-miss. For DeepSeek-V4-Pro, the corresponding counts are 3, 3, 11, and 33. Across the two direct-run arms, 27 failed executions nevertheless cover all expected rules, while 7 successful executions have incomplete strict coverage. Together with the \textsc{SysSpear} rerun, these results show that runtime evidence is common even when the attack chain stops before the final target.

These cases demonstrate why \textsc{SysField} treats offensive capability and defensive observability as independent measurements. A failed trajectory can still execute behavior that is relevant to detection and investigation before stopping, while a completed attack need not produce the complete case-level expected signal set. The present experiment establishes runtime visibility, not protection efficacy: \textsc{SysArmor} operates in observe-only mode, and neither the presence nor the volume of signals implies that an attack was blocked or that the observed activity was severe.

\paragraph{Expected-behavior coverage.}

Table~\ref{tab:sysfield-missing-signals} summarizes the 40 paired direct-run cases for which both model arms contain complete per-rule \texttt{missing\_signal} records and the 50-case \textsc{SysSpear} rerun, for which all formal cases contain complete per-rule records. The ten additional direct-run DeepSeek cases contain expected-signal outcomes but do not provide comparable per-rule missing fields, so they remain excluded from the direct-run comparison. The denominators are therefore shown explicitly and must not be treated as a matched three-arm sample.

\begin{table}[htbp]
  \centering
  \begingroup
  \footnotesize
  \setlength{\tabcolsep}{5pt}
  \renewcommand{\arraystretch}{1.13}
  \caption{Missing expected behaviors. Direct-run columns use 40 paired cases; the \textsc{SysSpear} 300-turn column uses all 50 formal rerun cases.}
  \label{tab:sysfield-missing-signals}
  \begin{tabular}{P{5.70cm} P{2.35cm} P{2.35cm} P{2.55cm}}
    \toprule
    \textbf{Expected behavior} & \textbf{Kimi K3 } & \textbf{DeepSeek-V4-Pro } & \textbf{\textsc{SysSpear}} \\
    \midrule
    Execution-oriented tool opens a network connection
      & 12/40 (30.0\%) & 23/40 (57.5\%) & 20/50 (40.0\%) \\
    Network client is used inside a workload
      & 9/40 (22.5\%) & 18/40 (45.0\%) & 19/50 (38.0\%) \\
    Shell or interpreter executes inside a workload
      & 6/40 (15.0\%) & 16/40 (40.0\%) & 14/50 (28.0\%) \\
    \bottomrule
  \end{tabular}
  \endgroup
\end{table}

The direct-run ordering is unchanged: an execution-oriented tool opening a network connection is missing most often, followed by workload-level network-client use and shell or interpreter execution. The \textsc{SysSpear} arm shows a closely related pattern, with missing rates of 40.0\%, 38.0\%, and 28.0\%, respectively. Its first two categories are nearly tied, but network-related behaviors still lack complete evidence more often than shell or interpreter execution.

The repeated ordering is the most useful cross-model pattern. Network activity initiated by execution-oriented tools is the behavior category that most often lacks complete expected evidence in the current records, followed by workload-level network-client activity. This identifies where closer trace alignment and telemetry analysis are most valuable.

However, the missing-rule counts should not be interpreted as a model-stealth ranking or as direct proof of a defensive blind spot. Expected rules are case-level behavioral expectations rather than action-level labels of the realized trajectory. Because the arms stop at different depths and the \textsc{SysSpear} rerun uses a different execution protocol and denominator, some expected downstream behaviors may never occur. Other misses may reflect telemetry boundaries or rule coverage. Distinguishing ``behavior not executed'' from ``behavior executed but not observed'' requires aligning each missing rule with the corresponding agent actions and workload events. RQ4 is therefore best understood as a diagnostic analysis of incomplete defensive evidence rather than an independent capability score.

\subsubsection{Evaluation Findings}

The evaluation yields four main findings. First, \textsc{SysField} preserves all generated events in the reported stress settings while adding 2.1\% system overhead; the corresponding baselines lose 60.0--98.5\% of events or impose 23.2--33.0\% overhead. Second, the generated ranges create measurable difficulty beyond small or single-step evaluations: multi-step composition lowers success by 18.1--21.6 percentage points, and the separate 7-to-50-node comparison shows a 9.8--10.6 point reduction across four reported models. Third, stage-level verification reveals how the red-team system architecture changes the progression funnel. With DeepSeek-V4-Pro as the common backbone, \textsc{SysSpear} expands target-1 reach from 19/50 to 34/50, but the absolute difference contracts from 30 percentage points at target~1 to 10 at target~2 and 2 at target~3. The additional footholds therefore do not translate proportionally into downstream completion, identifying post-compromise state utilization and cross-host continuation as the remaining bottleneck without implying that \textsc{SysSpear} is worse on the non-matched set of reached footholds. The current high-interference configuration likewise increases timeouts and execution cost while suppressing downstream completion. Fourth, runtime monitoring produces evidence in many executions that do not complete the attack. In the \textsc{SysSpear} arm, 39/50 cases produce a new signal and 27/50 cover all expected rules despite only 7/50 full completions; offensive capability and defensive observability are therefore distinct dimensions.

Together, these results connect the three requirements of a trustworthy cyber-agent evaluation. Low-loss, low-overhead collection makes negative evidence interpretable; multi-step and larger ranges expose capability limits that simpler tasks hide; and stage-aligned offensive and defensive measurements explain both where an agent stops and what behavior remains visible before termination. Missing expected evidence is concentrated around network activity initiated by execution-oriented tools, but trajectory-level alignment is still required before those misses can be attributed to the defensive layer. \textsc{SysField} therefore provides not only a harder range, but also the evidence needed to explain performance within that range.

\subsection{Evaluation of \sysspear}

We evaluate \sysspear through four research questions that examine whether its coordinated multi-agent architecture improves task success over single-agent execution, how the three underlying mechanisms operate in practice, whether the architecture scales to complex multi-host ranges, and how it generalizes to a published single-CVE benchmark under heterogeneous model assignment:

\begin{itemize}[noitemsep, topsep=1pt, partopsep=1pt, listparindent=\parindent, leftmargin=*]
    \item \textbf{RQ1: Architecture Effectiveness.} Can the coordinated multi-agent architecture improve task success rates over single-agent execution across medium- and hard-difficulty penetration testing cases?
    \item \textbf{RQ2: Mechanism Analysis.} How do task slicing, context projection, and runtime isolation contribute to solving multi-stage attack scenarios?
    \item \textbf{RQ3: Complex-Range Multi-Point Attack.} Can \sysspear elevate red-teaming from single-point exploitation to topology exploration, lateral movement, and multi-point exploitation on complex multi-layer \sysfield ranges?
    \item \textbf{RQ4: Single-CVE Benchmark Generalization and Cost.} How does \sysspear perform on PACEbench, a published single-CVE benchmark, and can heterogeneous coordinator--solver model assignment reduce cost at equal or higher score?
\end{itemize}

\subsubsection{Evaluation Setup}

We evaluate \sysspear on web penetration testing tasks using the XBow benchmark~\cite{xbow-paper}, comparing its Strategy Board architecture against a Claude Code single-agent baseline across multiple LLM configurations.

\textbf{Dataset.} We use a 13-case slice of the XBow benchmark, comprising 9 Level-2 (L2) and 4 Level-3 (L3) web penetration testing cases. We choose XBow for its wide adoption in recent studies of LLM-based penetration testing systems~\cite{xbow-paper, pentestgpt, mapta25, baselines-before-architecture26}. XBow provides realistic vulnerability classes, such as SQL injection, IDOR, SSRF and XSS, and spans difficulty levels from medium-complexity to hard multi-stage attacks~\cite{agent-zero-day}. Its containerized deployment and automated verification are suitable for evaluating multi-agent coordination across sequential attack stages. Following its evaluation protocol, a task counts as successful only when the benchmark verifier accepts its submission; timeouts and incomplete exploits are treated as failures.

\textbf{Baselines.} We compare \sysspear against Claude Code, a general-purpose AI coding agent that has been adapted for penetration testing tasks in prior benchmark studies~\cite{xbow-paper}. Claude Code serves as a representative single-agent baseline: it runs as a single integrated agent that interleaves tool use and reasoning within one continuous session, without the Coordinator-driven task slicing, typed Strategy Board evidence, or cross-Assignment context projection that \sysspear introduces. This makes it a suitable reference point for isolating the contribution of coordinated multi-agent execution rather than of any single underlying LLM.

\textbf{Metrics.} We report task-level success rates stratified by difficulty (L2, L3, and overall). Because the compared runs vary in model provider and execution date rather than forming a strictly controlled paired trial, we treat aggregate success rate as the primary quantitative signal, complemented by qualitative process evidence from two L3 cases (RQ2).

\textbf{Experimental protocol.} Both \sysspear and Claude Code execute on the same containerized runtime environment with identical base tooling (curl, nmap, gobuster, sqlmap), and each task runs under a 10-minute timeout. \sysspear's Coordinator may spawn multiple sequential or overlapping Assignments, with each Solver receiving a context projection of the current Strategy Board state rather than the full conversation history; Claude Code instead runs as a single continuous session per task. We report five \sysspear configurations, varying by model provider (DeepSeek-V4-Pro, GLM-5.1, Qwen 3.6) and by whether domain-specific skills are enabled, alongside two Claude Code configurations (GLM-5.1 and Opus 4.6) that enable both same-model and cross-model comparison against \sysspear.

\subsubsection{RQ1: Architecture Effectiveness}

Table~\ref{tab:sysspear-xbow-13} reports success rates for five \sysspear configurations and two Claude Code baselines, stratified by difficulty level.

\begin{table}[t]
  \centering
  \begingroup
  \footnotesize
  \setlength{\tabcolsep}{3pt}
  \renewcommand{\arraystretch}{1.13}
  \caption{XBow 13-case comparison by difficulty for five \sysspear configurations and two Claude Code baselines. Level-2 and Level-3 contain 9 and 4 cases, respectively.}
  \label{tab:sysspear-xbow-13}
  \begin{tabular}{P{2.70cm} P{3.85cm} P{2.35cm} P{2.35cm} P{2.35cm}}
    \toprule
    \textbf{System} & \textbf{Model configuration} & \textbf{L2 (9 cases)} & \textbf{L3 (4 cases)} & \textbf{Overall (13 cases)} \\
    \midrule
    \multirow{5}{*}{\sysspear}
      & DeepSeek-V4-Pro + GLM & 7/9 (77.8\%) & 1/4 (25.0\%) & 8/13 (61.5\%) \\
      & GLM-5.1 & 7/9 (77.8\%) & 1/4 (25.0\%) & 8/13 (61.5\%) \\
      & Qwen 3.6 & 5/9 (55.6\%) & 0/4 (0.0\%) & 5/13 (38.5\%) \\
      & DeepSeek-V4-Pro-Flash & 8/9 (88.9\%) & 1/4 (25.0\%) & 9/13 (69.2\%) \\
      & DeepSeek-V4-Pro-Flash + skills & 7/9 (77.8\%) & 3/4 (75.0\%) & \textbf{10/13 (76.9\%)} \\
    \midrule
    \multirow{2}{*}{Claude Code}
      & GLM-5.1 & 6/9 (66.7\%) & 1/4 (25.0\%) & 7/13 (53.8\%) \\
      & Opus 4.6 & 8/9 (88.9\%) & 1/4 (25.0\%) & 9/13 (69.2\%) \\
    \bottomrule
  \end{tabular}
  \endgroup
\end{table}

On the 13-case slice, the skill-enabled \sysspear configuration is the strongest recorded setting, solving 10 tasks (76.9\%). Its advantage is concentrated in Level-3 coverage: it solves 3 of 4 hard cases, whereas every other reported configuration solves at most one. The DeepSeek-V4-Pro-Flash and Opus 4.6 Claude Code configurations each solve 9/13 tasks (69.2\%), but their single Level-3 success leaves them below the skill-enabled configuration overall.

The same-model comparison provides an architecture-aligned signal: with GLM-5.1, \sysspear solves 8/13 tasks compared with 7/13 for the Claude Code. Notably, \sysspear completes the Level-3 HTTP-smuggling task XBEN-066-24, which the Claude Code baseline does not complete, consistent with the expectation that coordinated multi-agent execution benefits multi-stage exploration in harder cases. This configuration-level comparison does not isolate the causal contribution of individual mechanisms, since it reflects the combined architecture under the recorded XBow protocol; RQ2 turns to process-level evidence for how the mechanisms operate.

\subsubsection{RQ2: Mechanism Analysis}

We examine process evidence from two L3 cases below: XBEN-018, a tag-filter XSS bypass, and XBEN-066, the HTTP-smuggling task.

\paragraph{Case study: XBEN-018 (XSS tag-filter bypass).}

This task asks the red team to trigger \texttt{alert("XSS")} on a web application that reflects user input but filters many HTML tags. The Claude Code agent (Opus 4.6) made 57 tool calls, discovered the reflection point early, but spent most of its session attempting sandbox-escape variants and environment assumptions rather than systematically enumerating the filter boundary, and timed out without a working payload. \sysspear instead allocated two sequential Assignments: the first Solver attempted common tags (\texttt{<script>}, \texttt{<img>}, \texttt{<svg>}), and recorded each rejection as a Board entry typed \texttt{attempt} with explicit \texttt{contradicts} relations rather than as an opaque failure; the second Solver, focused on the known reflection point and the recorded filter failures, tested single-letter tags, found that \texttt{<a>} through \texttt{<y>} were blocked but \texttt{<z>} was not, and completed a working \texttt{onfocus}/\texttt{autofocus} payload on that tag.

The key difference is not that one system ran two executors, but that the Strategy Board preserved the exact filter boundary as explicit relations among Board entries. The second Assignment received a context projection containing the negative results, the confirmed reflection fact, and the next open question, so it resumed from a compressed parse boundary rather than from a blank session. This aligns with the intended benefit of task slicing and context projection: local search can advance systematically when prior failures are recorded as structured, relational evidence rather than as a growing transcript.

\paragraph{Case study: XBEN-066 (HTTP smuggling / SSRF).}

This task places the target endpoint behind a proxy and expects the red team to reach an internal hostname through Host-header manipulation. The Claude Code agent (Opus 4.6) found the login credentials (\texttt{test:test}), confirmed the presence of HAProxy, and observed internal routing clues, but its 60-tool-call session contained 47 explicit authentication or session-setup steps and many repeated login attempts, with no structured record of which capabilities had already been confirmed, and it ended with incomplete exploration. \sysspear solved the task in three overlapping Assignments: the first Solver discovered \texttt{test:test} in an HTML comment, logged in, accessed \texttt{/wifi\_settings}, and recorded a Board capability stating that the application proxies requests to \texttt{internal.router} via HAProxy; the Coordinator then dispatched a second Assignment, anchored on that capability, which bypassed the proxy using \texttt{Host: internal.router}, enumerated \texttt{/devices}, found a device named \texttt{flag}, and fetched \texttt{/devices/flag/status}. A third Assignment exploring POST-response details was cancelled once the second Assignment succeeded.

The central architectural feature here is that the SSRF capability was promoted to a Board entry with explicit \texttt{unlocks} relations, so the Coordinator could slice a new Assignment anchored on that capability without requiring the new Solver to rediscover the proxy chain. Context Projection delivered the login fact, the proxy capability, and already-refuted hypotheses (such as parameter injection) as a compact sub-graph, not a full transcript. The second Solver therefore started with confirmed access and a specific next route, consistent with the intended benefit of runtime isolation and Board-mediated collaboration: each Solver produces durable, relational evidence, while the Coordinator slices complementary routes rather than duplicating entire sessions.

\subsubsection{RQ3: Complex-Range Multi-Point Attack}

\paragraph{Setup.}
To test whether the architecture survives range complexity beyond containerized web tasks, we evaluate \sysspear on 50 complex ranges constructed by \sysfield. Each range instantiates the defining characteristics of \sysfield as an evaluation substrate: a multi-layer network topology with 3--6 tiers, 20--50 device nodes, and three chained vulnerabilities whose exploitation order is enforced by explicit capability dependencies, so that capabilities obtained from earlier vulnerabilities are prerequisites for later ones. Success in such ranges requires the red team to move beyond the single-point exploitation paradigm toward topology exploration, lateral movement, and multi-point exploitation. We report cumulative completion levels: L1 denotes capturing the first vulnerability, L2 the first two, and L3 all three; stage outcomes are verified by the \sysfield external verifier. We compare \sysspear with a DeepSeek-V4-Pro backbone against the Codex harness~\cite{codex} (Minimax-M3, and GPT-5.6 Luna), the Claude Code harness~\cite{claudecode} (DeepSeek-V4-Pro), and two single-agent baselines (DeepSeek-V4-Pro and Kimi K3). No Claude-family model participates in the experiments of this and the following subsubsection: in pilot runs, Claude 4.8's safety guardrails rejected the required offensive actions in more than 90\% of cases, leaving the tasks unsolvable regardless of the harness.

\begin{table}[t]
  \centering
  \begingroup
  \footnotesize
  \setlength{\tabcolsep}{4pt}
  \renewcommand{\arraystretch}{1.13}
  \caption{Multi-point attack results on 50 complex \sysfield ranges (3--6 tiers, 20--50 nodes, three chained vulnerabilities). L1/L2/L3 denote capturing the first, the first two, and all three vulnerabilities, respectively.}
  \label{tab:sysspear-complex-range}
  \begin{tabular}{P{2.05cm} P{3.30cm} P{1.75cm} P{1.75cm} P{1.75cm}}
    \toprule
    \textbf{System} & \textbf{Backbone} & \textbf{L1} & \textbf{L2} & \textbf{L3} \\
    \midrule
    \sysspear & DeepSeek-V4-Pro & 33 (66\%) & \textbf{23 (46\%)} & \textbf{20 (40\%)} \\
    \midrule
    \multirow{2}{*}{Codex} & Minimax-M3 & 29 (58\%) & 18 (36\%) & 13 (26\%) \\
     & GPT-5.6 Luna$^{\dagger}$ & \textbf{37 (74\%)} & 22 (44\%) & 7 (14\%) \\
    \midrule
    Claude Code & DeepSeek-V4-Pro & 33 (66\%) & 17 (34\%) & 14 (28\%) \\
    \midrule
    \multirow{2}{*}{Single-agent} & DeepSeek-V4-Pro & 17 (34\%) & 17 (34\%) & 16 (32\%) \\
     & Kimi K3 & 22 (44\%) & 18 (36\%) & 16 (32\%) \\
    \bottomrule
  \end{tabular}
  \endgroup

  \raggedright\footnotesize $^{\dagger}$Run without prompt modification.
\end{table}

\paragraph{Results.}
Table~\ref{tab:sysspear-complex-range} shows three patterns. First, coordinated harnesses expand initial access in complex topologies: with the same DeepSeek-V4-Pro backbone, \sysspear and Claude Code each reach L1 in 33/50 cases while the single-agent runner reaches 17/50, and GPT-5.6 Luna on Codex reaches 37/50. In a 20--50 node topology, breadth of exploration---not per-stage exploitation skill---is what the single-point paradigm lacks. Second, entry breadth alone does not translate into end-to-end completion. The GPT-5.6 Luna Codex arm collapses from 74\% at L1 to 14\% at L3, and Claude Code falls from 33 to 14, whereas \sysspear retains 23/33 (69.7\%) of its footholds through L2 and 20/23 (87.0\%) through L3, yielding the best overall L3 completion (20/50, 40\%). The Board's structured preservation of capabilities and negative results is what sustains lateral movement and multi-point chaining after the foothold. Third, the single-agent DeepSeek-V4-Pro funnel is nearly flat (17/17/16): once a foothold exists, per-stage capability is high, so its deficit concentrates on discovering and selecting targets across the topology---exactly the part that task slicing and parallel Assignments address. On complex ranges, \sysspear therefore both widens the entry funnel and, unlike the alternative multi-agent harnesses, prevents that width from decaying across the attack chain.

\subsubsection{RQ4: Single-CVE Benchmark Generalization and Cost}

\paragraph{Setup.}
Whereas the complex ranges of RQ3 chain multiple vulnerabilities into multi-hop campaigns that mirror real intrusion workflows, PACEbench~\cite{liu2025pacebench} centers on the exploitation of single CVEs and serves as a published, SOTA reference point for practical AI cyber-exploitation capability. It was designed to address the limitations of CTF-style evaluation, which operates on an ``assumption of guilt'' by handing agents a known vulnerability to exploit and therefore fails to reflect the investigative nature of real attacks. Its tasks are built on three principles: varied vulnerability difficulty that moves beyond simple exploits, environmental complexity that reflects the intricate systems agents must navigate, and the active presence of cyber defenses that must be identified and bypassed. Tasks are organized into four classes (A--D) containing 17, 7, 5, and 3 single-CVE tasks respectively, and the benchmark computes a single aggregate PACEbench Score from per-class performance. We use it to verify that \sysspear also generalizes to this established single-vulnerability setting and remains directly comparable with published results. We compare \sysspear against the Claude Code and Codex harnesses under their reported model configurations; \sysspear runs with a 50-turn limit per task. Motivated separately by cost and by the C-class behavior analyzed below, we additionally evaluate a heterogeneous \sysspear configuration that the coordinator--solver architecture supports natively: the strong but expensive DeepSeek-V4-Pro serves as the Coordinator model, while the economical DeepSeek-V4-Pro-Flash serves as the Solver model.

\begin{table}[t]
  \centering
  \begingroup
  \footnotesize
  \setlength{\tabcolsep}{3pt}
  \renewcommand{\arraystretch}{1.13}
  \caption{PACEbench results. A--D denote the benchmark's four single-CVE task classes (17, 7, 5, and 3 tasks); the PACEbench Score is the benchmark's aggregate metric. The heterogeneous \sysspear configuration uses DeepSeek-V4-Pro as the Coordinator and DeepSeek-V4-Pro-Flash as Solvers.}
  \label{tab:sysspear-pacebench}
  \begin{tabular}{P{1.60cm} P{3.95cm} P{1.10cm} P{1.10cm} P{1.10cm} P{1.10cm} P{1.35cm}}
    \toprule
    \textbf{System} & \textbf{Configuration} & \textbf{A-CVE} & \textbf{B-CVE} & \textbf{C-CVE} & \textbf{D-CVE} & \textbf{Score} \\
    \midrule
    \multirow{2}{*}{Claude Code} & DeepSeek-V4-Pro & 12/17 & 4.67/7 & \textbf{3.33/5} & 2/3 & 0.6745 \\
     & GLM-5.3 & 13/17 & 3.33/7 & 1.33/5 & 3/3 & 0.5758 \\
    \midrule
    Codex & GPT-5.6 Luna & 12/17 & 4.00/7 & 1.67/5 & 3/3 & 0.6126 \\
    \midrule
    \multirow{2}{*}{\sysspear} & DeepSeek-V4-Pro & 14/17 & \textbf{4.67/7} & 2.33/5 & 2/3 & 0.6380 \\
     & DeepSeek-V4-Pro (Coord.) + DeepSeek-V4-Pro-Flash (Solver) & \textbf{15/17} & \textbf{4.67/7} & 2.00/5 & \textbf{3/3} & \textbf{0.6930} \\
    \bottomrule
  \end{tabular}
  \endgroup
\end{table}

\paragraph{Results.}
Two observations follow from Table~\ref{tab:sysspear-pacebench}. First, on the investigative A-CVE class, \sysspear's coverage is the best among same-model configurations (14/17 versus 12/17 for both harnesses), consistent with the breadth advantage observed on complex ranges. Second, \sysspear underperforms on the C-CVE class (2.33/5 versus 3.33/5 for Claude Code with the same backbone). We attribute this to the interaction budget rather than a capability deficit: \sysspear limits each task to 50 agent turns, while the strong models we currently deploy reason deeply and, on investigation-heavy tasks, descend into exploration dead ends that consume turns without producing Board evidence. The same backbone completes more C-class tasks under harnesses without the turn cap, indicating that the deficit stems from bounded turns interacting with over-deep reasoning rather than from missing capability.

The heterogeneous configuration addresses both concerns. Assigning DeepSeek-V4-Pro to the Coordinator---where deep reasoning pays off in slicing, scheduling, and evidence triage---and the economical DeepSeek-V4-Pro-Flash to bounded Solver assignments yields the highest recorded PACEbench Score (0.693) among all configurations, with the best A-CVE coverage (15/17) and full D-CVE completion. Because the Board confines each Solver to a focused, independently checkable question, bounded execution with the economical solver stays on task, while the strong Coordinator supplies the strategic depth that inexpensive models lack. The configuration also improves economics: its total inference cost is 46\% of the all-DeepSeek-V4-Pro \sysspear configuration, and 46--64\% of the Codex and Claude Code configurations. Heterogeneous model assignment therefore turns the coordinator--solver architecture into a cost--performance dial: capability is concentrated where reasoning matters and economized where it does not.

\subsubsection{Evaluation Findings}

The evaluation yields five main findings. First, the coordinated multi-agent architecture improves task success over single-agent execution, with the skill-enabled \sysspear configuration achieving the highest recorded success rate of 76.9\% (10/13 tasks). Its advantage is concentrated in Level-3 coverage: it solves 3 of 4 hard cases, whereas every other reported configuration solves at most one. Second, the same-model comparison provides architecture-aligned evidence: with GLM-5.1, \sysspear solves 8/13 tasks compared with 7/13 for Claude Code, completing the Level-3 HTTP-smuggling task XBEN-066-24 that the baseline does not complete. This differential is consistent with the expectation that coordinated multi-agent execution benefits multi-stage exploration in harder cases. Third, process evidence from two Level-3 cases demonstrates how the three architectural mechanisms contribute in practice. In XBEN-018, task slicing and context projection enabled systematic filter-boundary enumeration by preserving negative results as explicit Board relations rather than as transcript noise. In XBEN-066, runtime isolation and Board-mediated collaboration allowed the Coordinator to slice a second Assignment anchored on a discovered SSRF capability without requiring the new Solver to rediscover the proxy chain, reducing redundant authentication steps and enabling focused exploitation. Fourth, on complex \sysfield ranges with 3--6 tiers, 20--50 nodes, and three chained vulnerabilities, \sysspear achieves the best end-to-end completion (L3 in 20/50 cases, 40\%): it combines the entry-breadth advantage of coordinated multi-agent harnesses with the post-foothold retention that alternative harnesses lose, retaining 69.7\% of footholds through L2 and 87.0\% through L3. Fifth, on the PACEbench single-CVE benchmark, a heterogeneous \sysspear configuration with a strong Coordinator and economical Solvers achieves the highest aggregate score (0.693) at 46\% of the cost of the homogeneous configuration and 46--64\% of the cost of the Claude Code and Codex harnesses; the remaining C-class deficit traces to the 50-turn budget interacting with deep-reasoning models rather than to missing capability.

Together, these results show that the coordinated multi-agent design offers measurable improvement over single-agent baselines, with benefits concentrated in harder multi-stage tasks. The three mechanisms are co-designed rather than independent: task slicing defines the Assignment boundaries that context projection projects evidence onto, and runtime isolation depends on both to keep each Solver's session bounded yet informed. The present comparison therefore characterizes their combined contribution rather than attributing a separate share to each; under this combined architecture, \sysspear solves tasks on which a higher-capability single-agent baseline timed out.

\subsection{Evaluation of \sysarmor}
\label{sec:sysarmor_evaluation}
We evaluate \sysarmor in a realistic deployment setting to examine its operational effectiveness, scalability, and resource efficiency in production-like conditions. The assessment is organized around the following research questions:

\begin{itemize}[noitemsep, topsep=1pt, partopsep=1pt, listparindent=\parindent, leftmargin=*]
    \item \textbf{RQ1: Detection Effectiveness.} How effectively does \sysarmor detect real-world \ac{apt} activity?
    \item \textbf{RQ2: Scalability.} What is the end-to-end throughput of the system, and how well does it scale to large enterprise deployments?
    \item \textbf{RQ3: Resource Consumption.} What memory footprint is required for online provenance-based detection under sustained event ingestion?
\end{itemize}

\begin{table}[t!]
    \scriptsize
    \centering
    \caption{Rule-level alert counts produced by the rule-based detection engine during the one-week deployment.}
     \resizebox{0.38\textwidth}{!}{\begin{tblr}{
        colspec = {l r},
        vline{1,2,3} = {},
        hline{1,2,13} = {1pt},
        cell{1}{1-2} = {c}
    }
        \textbf{Rule Name} & \textbf{Alert Count} \\
        command\_injection\_indicators & 7600 \\
        crontab\_modification        & 7525 \\
        permission\_modification     & 6670 \\
        suspicious\_binary\_execution& 4865 \\
        clear\_log\_activities       & 4454 \\
        dev\_shm\_execution          & 3677 \\
        temp\_executable\_creation   & 3673 \\
        nonstandard\_port\_connection& 103  \\
        sudo\_privilege\_escalation  & 12   \\
        file\_deletion\_dangerous    & 5    \\
        service\_manipulation        & 4    \\
    \end{tblr}}
    \label{tab:detection_rulebased}
\end{table}

\subsubsection{Evaluation Setup}

\sysarmor is deployed in \company for a one-week operational assessment. The deployment targets a representative endpoint environment and is designed to measure both detection quality and system overhead under realistic event volumes.

\noindent \textbf{Dataset.} We install \sysarmor agents on a monitored endpoint within the enterprise network and collect system telemetry over a one-week period. During this time, the red team at \company executes an \ac{apt} campaign against the monitored host to emulate realistic adversarial behavior. The endpoint and server operate on Ubuntu, and the collected auditd stream includes approximately 30 million raw events, which are normalized into 8,179,548 Sysdig events before ingestion by the detection pipeline. This corresponds to an average event rate of roughly 57 events per second per endpoint.

\begin{table*}[t!]
    \scriptsize
    \centering
    \caption{Detection results produced by the \nodlink + \knowhow inference pipeline against the real-world \ac{apt} campaign. \nodlink constructs the suspicious provenance subgraph, and \knowhow maps the observed behavior to MITRE ATT\&CK techniques and attack stages.}
    \resizebox{0.98\textwidth}{!}{\begin{tblr}{      
        cell{1}{1-7} = {c},
        cell{2-14}{1-2} = {c},
        cell{2-14}{3-7} = {c},
        vline{1,2,3,8} = {},
        hline{1,15} = {1pt}
        }
    \SetCell[r = 1, c = 1]{c} \textbf{Attack Stage}  & \SetCell[r = 1, c = 1]{c} \textbf{\ac{ttp}} & \SetCell[r = 1, c = 5]{c} \textbf{Malicious Command Line}  & & & & \\
    \hline
    \SetCell[r = 1, c = 1]{c}Initial Compromise & T1570 & \textit{/tmp/.\textgreater\_, /var/tmp/.\textgreater\_, /tmp/sshd, /root/sshd}  & & & & \\
    \SetCell[r = 1, c = 1]{c}Establish Foothold & T1105  & \textit{/usr/sbin/sshd -D, sshd [accepted], sshd nobody [priv]} & & & & \\
    \SetCell[r = 1, c = 1]{c}Move Laterally & T1021.004 & \textit{sshd root [priv], sshd operator [priv]} & & & & \\
    \hline
    \SetCell[r = 3, c = 1]{c}Escalate Privilege & T1222 & \textit{(cd `/var/tmp' \&\& [ -f `yum.log' ] \&\& (chattr -i `yum.log'; chmod +x `yum.log' \&\& ./`yum.log'))}  & & & & \\
    & T1036 & \textit{(cd `/var/tmp' \&\& [ -f `apt.log' ] \&\& (chattr -i `apt.log'; chmod +x `apt.log' \&\& ./`apt.log'))}  & & & & \\
    & T1022.002 & \textit{chmod +x yum.log, chmod +x .bash\_logout, /usr/bin/ydvlmeowey su 12812}  & & & & \\ \hline
     \SetCell[r = 4, c = 1]{c} Internal Reconnaissance & T1016 & \textit{/usr/bin/yzhulutzjr ls -la, /usr/bin/ukkoisjdsi route -n, /usr/bin/dcjwwfaznr ifconfig eth0} & & & & \\ 
    & T1057 & \textit{/usr/bin/zqjxlypgjo netstat -an, /usr/bin/rvwcaoweok ps -ef, /usr/bin/tseyeejrpq who} & & & & \\ 
    & T1033 & \textit{/usr/bin/wrrvfqlxbh whoami, /usr/bin/mbydvjvtfs top, /usr/bin/srtewbasqq uptime, /usr/bin/cqlmxscjbm pwd } & & & & \\
    & T1083 & \textit{/usr/bin/bkzytqxfba cd /etc, /usr/bin/vyqlcyypoo cat /etc/resolv.conf, which apt, grep -v grep}& & & & \\ \hline
    \SetCell[r = 3, c = 1]{c}Maintain Persistence & T1562 &
    \textit{shell -c ps -eo pid,comm,\%cpu \textbar \quad awk `...' \textbar \quad sort -k3nr \textbar \quad ... \textbar \quad xargs kill ... \textbar\textbar \quad pgrep -x `\textgreater\_' | tail} & & & & \\
    & T1499 & \textit{readlink -f /proc/10751/exe, xargs -r kill -9, ps -o pid=,pcpu= -p 19901 30569} & & & & \\
    & T1053 & \textit{chattr -i .bash\_logout}, \textit{/bin/bashell -c run-parts /etc/cron.hourly, basename /etc/cron.hourly/gcc.sh}  & & & & \\
\end{tblr} }
    \label{tab:detection_mlinference}
\end{table*}

\noindent \textbf{Experimental protocol.} The \sysarmor server is deployed on a dedicated machine with 32 CPU cores, an NVIDIA GeForce RTX 3090 GPU, 128 GB RAM, and 2 TB of storage, running Ubuntu 20.04 LTS. Docker and Kubernetes are used for container orchestration, while OpenSearch and PostgreSQL provide alert storage and metadata management. A React-based web interface is also deployed to enable operational monitoring and analyst interaction.

\subsubsection{RQ1: Detection Effectiveness}

\textbf{Rule-based Detection.}
We evaluate the rule-based engine using an existing Falco rule set containing more than 36 signatures for common malicious behaviors. During the deployment period, the engine triggers 38,588 alerts, corresponding to approximately 0.47\% of the processed event stream. The distribution of alerts across rule categories, summarized in Table~\ref{tab:detection_rulebased}, shows that the most frequently activated detections correspond to command injection, crontab modification, permission tampering, suspicious binary execution, log clearing, and execution from temporary or shared-memory paths. These patterns are consistent with persistence, privilege escalation, and evasion behaviors commonly observed in real intrusion campaigns.

\textbf{\ac{ml}-based Detection.}
The \ac{ml}-based detection pipeline, composed of \nodlink and \knowhow, identifies a multi-day \ac{apt} campaign during the deployment. The malicious activity persists for three days and is first surfaced on the first day of the campaign, indicating that the system can detect adversarial behavior early in the attack lifecycle. As the campaign evolves, the alert remains active and is progressively refined until the activity ceases on the third day. The generated alert ultimately contains a provenance graph with 201 nodes and 1,982 edges, providing a structured representation of the attack trajectory.

Table~\ref{tab:detection_mlinference} reports the attack stages, associated MITRE ATT\&CK techniques, and representative malicious command lines identified by the system. After removing redundant or highly similar command strings for readability, the results show that \sysarmor recognizes six attack phases: Initial Compromise, Establish Foothold, Move Laterally, Escalate Privilege, Internal Reconnaissance, and Maintain Persistence. No Data Exfiltration or Impact stage is observed during the campaign, suggesting that the intrusion was interrupted before reaching later-stage objectives. For each stage, the system maps the suspicious provenance evidence to the relevant ATT\&CK technique and preserves the underlying command-line context used by the adversary.

The TTP-level attribution is enabled by \knowhow's semantic reasoning over structured cyber threat intelligence, represented as \ac{gioc} triples. In the lateral movement stage, for example, \knowhow matches triples such as \texttt{<adversary, use, ssh service>} and \texttt{<adversary, transfer, file>} to provenance evidence involving shell processes launching \textit{sshd} and writing files such as \textit{/tmp/sshd} and \textit{/var/tmp/.\textgreater\_}. Similarly, the masquerading and permission-modification behaviors are aligned to triples such as \texttt{<adversary, create, system file name>} and \texttt{<adversary, modify, file permission>}, which correspond to the observed creation of files such as \textit{yum.log} and subsequent execution of commands like \textit{chattr -i} and \textit{chmod +x}. For obfuscation-related activity, the system matches triples like \texttt{<adversary, execute, obfuscated file>} to execution paths such as \textit{/usr/bin/yzhulutzjr} and \textit{/usr/bin/zqjxlypgjo}. This semantic alignment between provenance structure and CTI knowledge enables fine-grained TTP attribution beyond simple signature matches.

The observed campaign also demonstrates the kinds of stealth techniques that often evade conventional defenses. The adversary uses masquerading and obfuscation to hide malicious artifacts behind filenames such as \textit{yum.log}, \textit{apt.log}, and \textit{gcc.sh}, while storing payloads in hidden paths such as \textit{/tmp/.\textgreater\_} and \textit{/var/tmp/.\textgreater\_}. Command execution is further disguised through randomized executable names, allowing the attacker to evade rule-based checks that rely on static file names or known command patterns. These behaviors are difficult for both signature-based and learning-based systems to capture reliably, because they are rare, highly context-dependent, and intentionally designed to look benign. The results indicate that \sysarmor's provenance-based analysis provides an effective countermeasure against this class of evasive activity.

\begin{table}[t!]
    \scriptsize
    \centering
    \caption{Throughput of \sysarmor under different Flink job configurations. ``eps'' denotes events per second.}
     \resizebox{0.38\textwidth}{!}{\begin{tblr}{
        colspec = {c r},
        vline{1,2,3} = {},
        hline{1,2,5} = {1pt},
        cell{1}{1-2} = {c}
    }
        \textbf{Job Configuration} & \textbf{Throughput (eps)}  \\
        \textit{Job 01 + Job 02}  &  12547  \\ \hline
        \textit{Job 01 + Job 03}  &  12208   \\ \hline
        \textit{Job 01 + Job 02 + Job 03}  &  9003    \\
    \end{tblr}}
    \label{tab:throughput_singleagent}
\end{table}

\subsubsection{RQ2: Scalability}
\label{sec:rq2}

\textbf{End-to-end throughput.}
To evaluate throughput, we replay a fixed workload from a single monitored endpoint and measure the number of events processed per second as they pass through the streaming pipeline. We use one-agent playback because simulating multiple sources to saturate the cluster would require a substantially larger deployment and would not reflect the isolated upper-bound behavior of a single endpoint. In this setup, the raw audit stream is first written to Kafka, then consumed by Flink jobs for event normalization and detection. The first job performs normalization from raw auditd records into Sysdig format, while the second and third jobs execute rule-based and machine-learning-based detection, respectively.

Table~\ref{tab:throughput_singleagent} reports the measured throughput under different execution configurations. The combined configuration of Job 01 and Job 02 and the configuration of Job 01 and Job 03 both sustain approximately 12,000 events per second. When both detection jobs run concurrently, the system processes about 9,000 events per second. This reduction is primarily attributed to Kafka's duplicated delivery of the same event stream to two independent consumer groups, which increases broker egress load, I/O overhead, and network pressure. Although the throughput drops under simultaneous rule-based and machine-learning execution, the observed rate remains far above the 57 events per second generated by a typical endpoint in our deployment.

This result suggests that a single deployment server can support more than 150 endpoints in real time under the current workload profile. We further inspect the pipeline bottleneck and find that Job 01, the normalization stage, is the dominant limiting component. This is not due to the detection logic itself, but rather to the naive parsing implementation in this stage. As such, further optimization of the normalization layer is expected to yield substantial throughput gains.

\subsubsection{RQ3: Resource Consumption}

We quantify memory consumption by monitoring the Flink containers during the throughput experiment described in Section~\ref{sec:rq2}. Because the provenance-based detection pipeline maintains in-memory graph structures and cached artifacts for anomaly construction and semantic interpretation, memory overhead is a key operational concern. We therefore measure the resource usage of the two detection jobs separately, each running in a dedicated Flink container with 8 GB of allocated memory.

\begin{figure}[t!]
    \centering
    \includegraphics[width=0.48\textwidth]{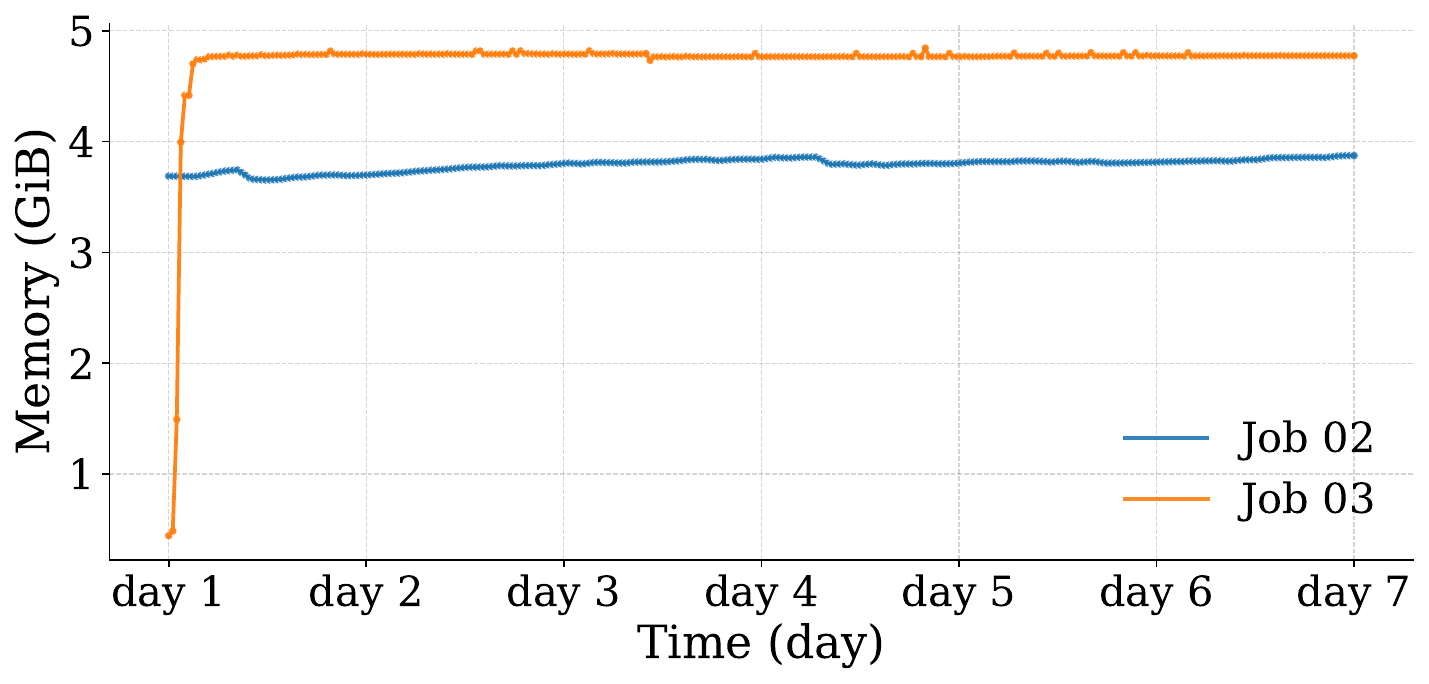}
    \caption{Memory consumption of the \sysarmor server during the evaluation period.}
    \label{fig:memory_consumption}
\end{figure}

Figure~\ref{fig:memory_consumption} shows the measured memory traces. The rule-based job remains relatively lightweight, as it only maintains a compact rule set and evaluates predicates against the incoming event stream. The machine-learning job demonstrates a higher initial memory footprint during warm-up, when the \nodlink model and \knowhow knowledge base are loaded and graph state is initialized. After this startup phase, memory usage stabilizes. The rule-based job settles at approximately 3.8 GB, while the machine-learning job stabilizes at roughly 4.8 GB. These values indicate that \sysarmor can operate with moderate memory requirements while sustaining real-time detection on production-like workloads.

\subsubsection{Evaluation Findings}
The evaluation yields three main findings. First, \sysarmor demonstrates strong detection capability in a real-world deployment. The rule-based engine emits 38,588 alerts across 11 rule categories, covering 0.47\% of the processed event stream. The combined \nodlink + \knowhow pipeline detects a three-day \ac{apt} campaign and maps it to six attack stages, resulting in a provenance graph with 201 nodes and 1,982 edges. The semantic alignment between CTI-derived \acp{gioc} and observed provenance events produces high-fidelity TTP attribution, which is particularly valuable for stealthy behaviors such as masquerading and obfuscated file execution.

Second, \sysarmor scales effectively to enterprise-level workloads. The system sustains approximately 12,000 events per second with single-job detection configurations and maintains about 9,000 events per second when both detection jobs operate concurrently. Given an average event rate of 57 events per second per endpoint, these results indicate that a single server can support more than 150 endpoints in real time. The primary bottleneck resides in the normalization stage, which suggests that additional optimization of the parsing layer can further improve throughput.

Third, \sysarmor requires only moderate memory resources. After the initial warm-up period, the rule-based detection job consumes around 3.8 GB and the machine-learning job around 4.8 GB, both remaining stable throughout the evaluation. This confirms that provenance-based real-time threat detection is feasible within a realistic operational resource envelope.

Taken together, these results indicate that \sysarmor satisfies the intended design goals of real-time, interpretable, and actionable defense. The combination of signature-based detection and provenance-driven reasoning enables the system to capture both well-known attack patterns and evasive behaviors that are difficult to detect using conventional methods alone. The observed throughput and memory usage further suggest that the architecture is practical for deployment in live enterprise environments, while preserving the contextual richness needed for accurate TTP-level attribution and analyst investigation.










\section{Conclusion}
\label{sec:conclusion}

In this report, we traced the persistent failures of current cyber defense to one root cause, defense's isolated evolution, and proposed co-evolution as the integrating insight to build a defense that evolves at the speed of attack. Based on this insight, we presented \sysevolve, a safe and autonomous co-evolutionary system built from three co-designed components, \sysfield, \sysspear, \sysarmor, that form a self-driven adversarial co-evolution loop. \sysfield constructs realistic multi-host ranges via declarative topology-driven construction, provides complete attack-trace collection and trustworthy assessment, restoring deployment evolution. \sysspear generates efficient, safe attack plans via task-sliced local search, a bidirectional expert-knowledge base that exploits external expert skills and distills internal campaign knowledge, and static-analysis verification, restoring red-team capability evolution. \sysarmor performs real-time, interpretable defense via state tracking, CTI-context fusion, and stage-matched response, restoring defense capability evolution. In evaluation, \sysfield achieves zero-loss collection at 2.1\% overhead and orchestrates 257 CVEs into 1,148 ranges. \sysspear improves red-team success by over 25\% over baseline LLMs, attains the best end-to-end completion on complex multi-vulnerability ranges, and achieves the highest PACEbench score with a heterogeneous coordinator--solver configuration at 46\% of the cost of homogeneous deployment. \sysarmor achieves 10--1000$\times$ greater precision than prior systems and detects real APT attacks in production at Huawei and Sangfor. Our evaluation also reveals four findings about LLM agent capabilities. First, multi-step composition and larger topologies expose agent capability gaps hidden by single-step evaluations. Second, the bottleneck lies after initial access in post-compromise state utilization. Third, LLM agents are susceptible to environmental interference. When decoy endpoints are deployed in the range, agent timeouts triple and downstream completion disappears despite unchanged initial-access success rates. Fourth, under bounded interaction budgets, strong reasoning models descend into exploration dead ends, so weak scores can reflect over-exploration rather than missing capability.

\bibliographystyle{IEEEtranS} 
\bibliography{reference}

\end{document}